\documentclass[fleqn,usenatbib]{mnras}

\usepackage{newtxtext,newtxmath}
\usepackage[T1]{fontenc}

\DeclareRobustCommand{\VAN}[3]{#2}
\let\VANthebibliography\thebibliography
\def\thebibliography{\DeclareRobustCommand{\VAN}[3]{##3}\VANthebibliography}

\usepackage[english]{babel}
\usepackage{float}
\usepackage{graphicx,epsfig}
\usepackage{multicol}
\usepackage{amsmath}
\usepackage{amsfonts}
\usepackage{dsfont}
\usepackage{amsxtra}
\usepackage{hyperref}
 
\usepackage{amssymb}
\usepackage{upgreek} 
\usepackage{changes}
\usepackage{multirow}
\usepackage{url}

\usepackage{dcolumn}

\usepackage{xspace} 

\usepackage{color}
\usepackage{xcolor}

\usepackage[switch]{lineno}
\usepackage{lipsum}

\usepackage{graphicx}	
\usepackage{amsmath}	
\usepackage{amssymb}	

\definecolor{darkgreen}{rgb}{0.0, 0.7, 0.0}

\title[Gamma-rays from young radio galaxies and quasars]{Gamma-ray emission from young radio galaxies and quasars}

 \author[G. Principe et al.]{G. Principe,$^{1,2,3}$ \thanks{E-mail:giacomo.principe@inaf.it}
L.~Di~Venere $^{4,5}$
M. Orienti, $^{3}$
G. Migliori, $^{3}$
F. D'Ammando, $^{3}$  
M. N.~Mazziotta, $^{5}$
\newauthor and M.~Giroletti $^{3}$
\\
$^{1}$Dipartimento di Fisica, Universit\'a di Trieste, I-34127 Trieste, Italy\\
$^{2}$ Istituto Nazionale di Fisica Nucleare, Sezione di Trieste, I-34127 Trieste, Italy\\
$^{3}$ INAF - Istituto di Radioastronomia, I-40129 Bologna, Italy\\
$^{4}$Dipartimento di Fisica “M. Merlin” dell’Universita` e del Politecnico di Bari, I-70126 Bari, Italy\\
$^{5}$ Istituto Nazionale di Fisica Nucleare, Sezione di Bari, I-70126 Bari, Italy\\
}

\date{Accepted August 11, 2021. Received April 26, 2021.}

\pubyear{2021}

\begin{document}
\label{firstpage}
\pagerange{\pageref{firstpage}--\pageref{lastpage}}
\maketitle

\begin{abstract}
According to radiative models, radio galaxies and quasars are predicted to produce gamma rays from the earliest stages of their evolution. Exploring their high-energy emission is crucial for providing information on the most energetic processes, the origin and the structure of the newly born radio jets.
Taking advantage of more than 11 years of \textit{Fermi}-LAT data, we investigate the gamma-ray emission of 162 young radio sources (103 galaxies and 59 quasars), the largest sample of young radio sources used so far for such a gamma-ray study. 
We separately analyze each source and perform the first stacking analysis of this class of sources to investigate the gamma-ray emission of the undetected sources.
We detect significant gamma-ray emission from 11 young radio sources, four galaxies and seven quasars, including the discovery of significant gamma-ray emission from the compact radio galaxy PKS\,1007+142 (z=0.213). 
The cumulative signal of below-threshold young radio sources is not significantly detected. 
However, it is about one order of magnitude lower than those derived from the individual sources, providing stringent upper limits on the gamma-ray emission from young radio galaxies ($F_{\gamma}< 4.6 \times 10^{-11}$ ph cm$^{-2}$ s$^{-1}$) and quasars ($F_{\gamma}< 10.1 \times 10^{-11}$ ph cm$^{-2}$ s$^{-1}$), and enabling a comparison with the models proposed.
With this analysis of more than a decade of \textit{Fermi}-LAT observations, we can conclude that while individual young radio sources can be bright gamma-ray emitters, the collective gamma-ray emission of this class of sources is not bright enough to be detected by \textit{Fermi}-LAT.

\end{abstract}


\section{Introduction}
\label{sec:intro}
Understanding the origin and evolution of the high-energy emission in extragalactic radio galaxies and quasars is one of the greatest challenges faced by modern astrophysics.
The extragalactic gamma-ray sky is dominated by blazars \citep{2015A&ARv..24....2M,2020ApJS..247...33A}, which are a sub-class of active galactic nuclei (AGN) for which the gamma-ray emission is favoured by their small jet inclination to the line of sight and by relativistic beaming. 
The increasing amount of data collected by the Large Area Telescope (LAT) on board the \textit{Fermi Gamma-ray Space Telescope} \citep{2009ApJ...697.1071A} allows us to investigate many classes of objects in the gamma-ray sky \citep{2018A&A...614A...6S}.
A small percentage, $\sim2\%$, of the fourth catalog of gamma-ray AGN \citep[4LAC,][]{2020ApJ...892..105A}, are radio galaxies (or misaligned AGN), which have larger jet inclination angles ($> 10 ^{\circ}$) and a smaller Doppler factor ($\delta \leq2-3$) than blazars. With their misaligned jets, they offer a unique tool to probe some of the non-thermal processes at work in unbeamed regions in AGN, which are usually overwhelmed by beamed emission from the jet in blazars.

In the evolutionary scenario, the size of a radio galaxy is strictly related to its age \citep{1995A&A...302..317F}.
Extragalactic compact radio objects, with a projected linear size (LS) smaller than about 20 kpc, i.e. they reside within the host galaxy, are important objects because they are expected to be the progenitors of extended radio galaxies \citep{1996ApJ...460..612R}.
The young nature of these objects is strongly supported by the determination of kinematic and radiative ages in some of the most compact sources, which were found to be $t \sim 10^2 - 10^5$ years \citep{1982A&A...106...21P,1999A&A...345..769M,2005ApJ...622..136G,2009AN....330..193G}, while large-size objects have an age of $t \sim 10 ^7 - 10^8$ years \citep{2005A&A...433..467J,2017MNRAS.469..639H}.

Extragalactic compact radio objects can be classified into GHz-peaked spectrum (GPS) and compact steep spectrum (CSS) sources depending on their radio spectra. GPS objects are powerful radio sources whose spectra present peak frequencies $\nu_{p} > 0.5$ GHz \citep{1998PASP..110..493O}.
CSS objects are similarly powerful inverted-spectrum radio sources but with peak frequencies in a lower frequency range when compared to the GPS population, $\nu_{p} \lesssim 0.5$ GHz. 
Morphologically, GPS/CSS sources may be reminiscent of a smaller version of classical doubles (Fanaroff-Riley type-II radio galaxies), with pairs of symmetric lobes present on opposite sides of a weak radio nucleus. In such cases they are called compact symmetric objects (CSOs) if LS $\lesssim$ 1 kpc, medium symmetric objects (MSOs) if LS $\sim 1 - 20$ kpc, and large symmetric objects (LSOs) if LS $>20$ kpc \citep{1990A&A...231..333F,2021A&ARv..29....3O}. As shown by \citet{1997AJ....113..148O}, there seems to be a relatively tight correlation between the peak frequency and the source’s linear size.
This unifies the GPS and CSS populations and suggests that they are both manifestations of the same physical phenomenon.

GPS and CSS objects reside either in galaxies or quasars. 
For the quasars, gamma-ray emission is favored by the smaller jet-inclination angle and beaming effects, while the origin of gamma-ray emission in galaxies is still a matter of debate.
Young radio sources were predicted to constitute a relatively numerous class of extragalactic objects detectable by \textit{Fermi}-LAT \citep{2008ApJ...680..911S}.
They are entirely located within the innermost region of the host galaxies, surrounded by dense and inhomogeneous interstellar medium, which may be a rich source of UV/optical/IR photons. 
In compact radio sources associated with quasars the high-energy emission could be due to inverse Compton (IC) of the synchrotron photons by a dominant jet component, producing an emission that can be strongly beamed \citep[e.g.][]{2012ApJ...749..107M,2014ApJ...780..165M}.
However, the most compact and powerful radio galaxies are expected to produce isotropic $\gamma$-ray emission up to the GeV band through IC scattering of the UV/optical/IR photons by the electrons in the compact radio lobes \citep[][see the latter for a discussion of hadronic models]{2008ApJ...680..911S,2011MNRAS.412L..20K}. In these models the high-energy luminosity of radio galaxies strictly depends on different source parameters such as linear size, jet power, UV/optical/IR photon density and the equipartition condition in the lobes. 

The search for gamma-ray emission from young radio sources is crucial for providing information on the physical conditions in the central region of the host
galaxy, the energetic processes possibly at work in such regions, as well as the origin and the structure of the newly born radio jets.
However, systematic searches for young radio sources at gamma-ray energies have so far been unsuccessful \citep{2016AN....337...59D}.
Dedicated studies have reported a handful of detections. \citet{2016ApJ...821L..31M} reported the first association of a gamma-ray source with a GPS radio galaxy, NGC\,6328 ($z$ = 0.014).
Beyond the confirmation of this object in the fourth \textit{Fermi}-LAT catalog \citep[4FGL,][]{2020ApJS..247...33A}, detections have been reported for only five CSS sources (3C\,138, 3C\,216, 3C\,286, 3C\,380, and 3C\,309.1), all associated with quasars, and for a second GPS radio galaxy (NGC\,3894, $z$  = 0.0108) \citep{2020A&A...635A.185P}. The five CSS quasars have high gamma-ray luminosity ($>10^{46}$ erg s$^{-1}$) and show flaring activity at high energies, suggesting (mildly-)beamed emission.


Taking advantage of the increased exposure provided by more than eleven years of LAT data, we investigate the gamma-ray properties of a sample of 162 young radio sources (103 galaxies and 59 quasars).
In addition to the gamma-ray analysis of each young radio object, we perform the first stacking analysis of this class of sources in order to investigate the gamma-ray emission of the young radio sources still below the detection threshold in the high-energy regime.

Throughout this article, we assume $H_{0} = 70$ km s$^{-1}$ Mpc$^{-1}$ , $\Omega_{M} = 0.3$, and $\Omega_{\Lambda}=0.7$ in a flat Universe.
\section{Sample of young radio sources}
\label{sec:sample}
To select young radio sources, we base our sample on the following resources which contain radio galaxies and quasars with projected linear size below 50 kpc:
\begin{itemize}
    \item the sample of 51 \textit{bona-fide} young radio sources created by \citet{2014MNRAS.438..463O} selected on the basis of the detection of the core as a crucial requirement for classifying a source as a genuine CSO/MSO/LSO. It contains 32 galaxies and 19 quasars with linear sizes from a few pc to tens of kpc. In particular, for half of them, their youth is strongly supported by the determination of kinematic and radiative ages which were found to be a few thousand years or less.
    \item The sample of 25 nearby ($z$ < 0.25) and compact ($\theta$ < 2'') radio galaxies selected by \citet{2009A&A...498..641D} from the COmpact RAdio sources at Low Redshift (CORALZ) sample \citep{2004MNRAS.348..227S}. For this sample the authors investigated the size and morphological classification by means of VLBI observations, reporting sizes spanning a couple of pc to a few kpc. 
    \item The list of sources used for the investigation of the optical properties of young radio AGNs based on the Sloan Digital Sky Survey (SDSS) spectroscopy created by \citet{2020MNRAS.491...92L}, where they excluded all possible blazar-like objects with optical spectral variability. This list contains 126 (54 galaxies and 72 quasars) sources with different physical properties: a redshift value between 0.001 and 3.5 and linear sizes between 0.4 pc and 30 kpc.
    \item The sample of 17 GPS and/or CSOs with measured redshifts below 1 and linear sizes below 1 kpc from \citet{2020ApJ...892..116W}.
\end{itemize}
Since several sources are present in more than one of the above-listed samples, we removed all the repetitions. In addition to these samples, we included a handful of objects selected from the following resources:
\begin{itemize}
    \item three sources, the galaxies NGC\,3894, TXS\,0128+554, and the quasar 3C\,380, have been specifically selected since they have been detected at high energy and investigated in \citet{2020A&A...635A.185P}, \citet{2020ApJ...899..141L} and \citet{2020ApJ...899....2Z}, respectively. 
    \item The young radio galaxy 0402$+$379 ($z$ = 0.0545, LS =7.3 pc) was selected because it has been proposed, together with other sources already included in our sample, as a promising candidate
gamma-ray source in the study by \citet{2020ApJ...897..164K}, where the X-ray emission of 29 GPS and CSO objects ($t \lesssim$ 3 kyr, LS $<$ 300 pc) was investigated using high-angular resolution X-ray telescopes.
\end{itemize}

Our final sample consists of 162 young radio sources with known position, redshift, linear size, radio luminosity and peak frequency (see Table \ref{table_sample} in the Appendix \ref{appendix_tables}). Among them, 103 are classified as galaxies and 59 as quasars.

\begin{figure}
\centering
\includegraphics[width=\columnwidth]{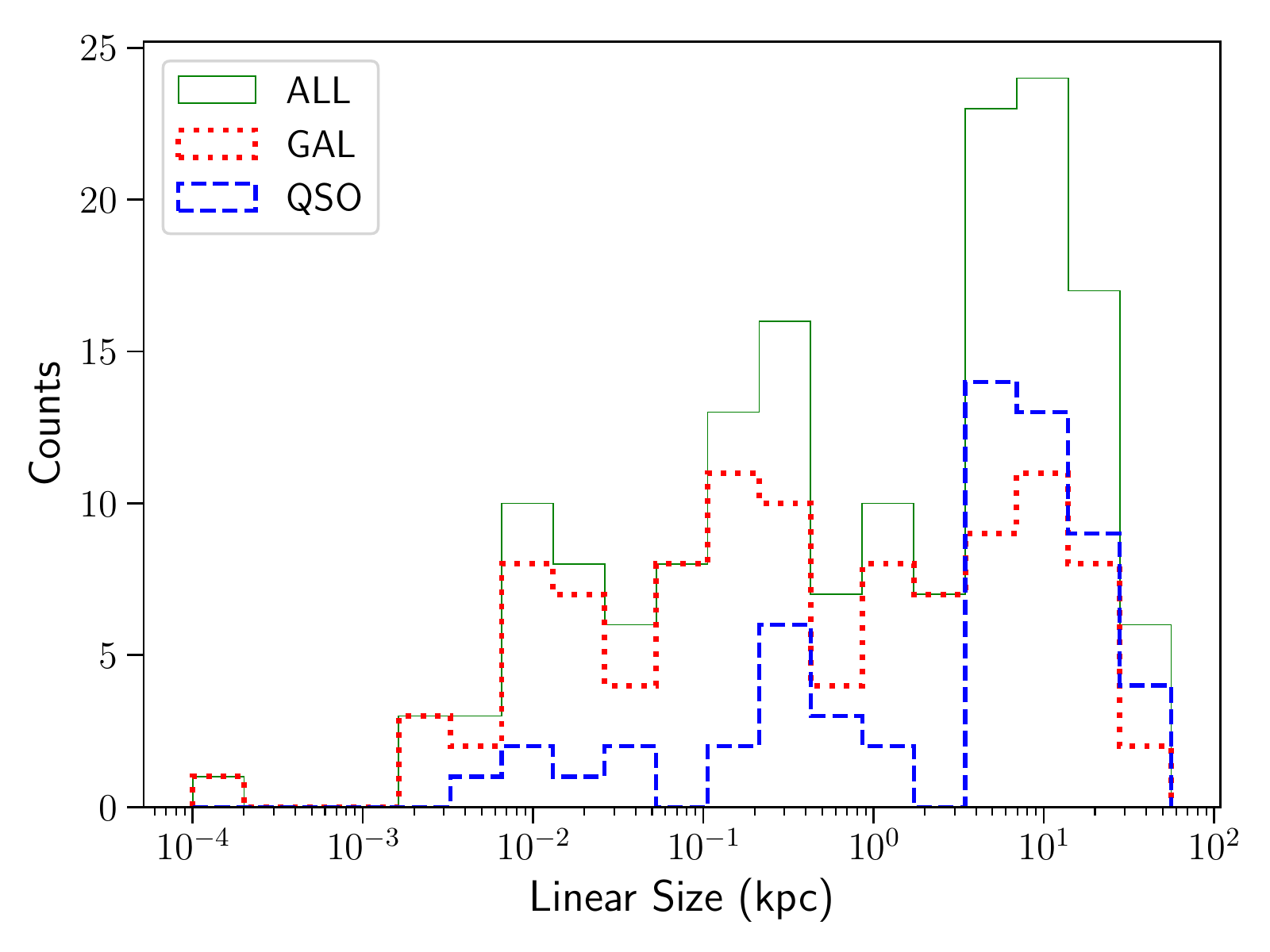}
\caption{\small \label{histo_linear_size}
Projected linear size for the full sample of young radio sources described in Sect. \ref{sec:sample}. Red dotted line, blue dashed line and green solid line represent galaxies, quasars and all the sources, respectively.}
\end{figure}

The selected sources have redshift values between 0.001 and 3.5 and linear sizes spanning from less than 1 pc up to a few tens of kpc.
Most (129) of the sources have redshift below 1, with seven sources located in the local Universe ($z$ < 0.05, $D_L \lesssim$200 Mpc). 
In Fig. \ref{histo_linear_size} we show the distribution of the linear size for our sample of sources, discriminating between galaxies and quasars.  

\noindent Considering the morphological classification, about half (79) of the sources are classified as CSOs (LS $<1$ kpc), 70 sources as MSO (LS $\sim 1-20$ kpc), and 13 LSOs with LS between 20 and 50 kpc.

\begin{figure}
\centering
\includegraphics[width=\columnwidth]{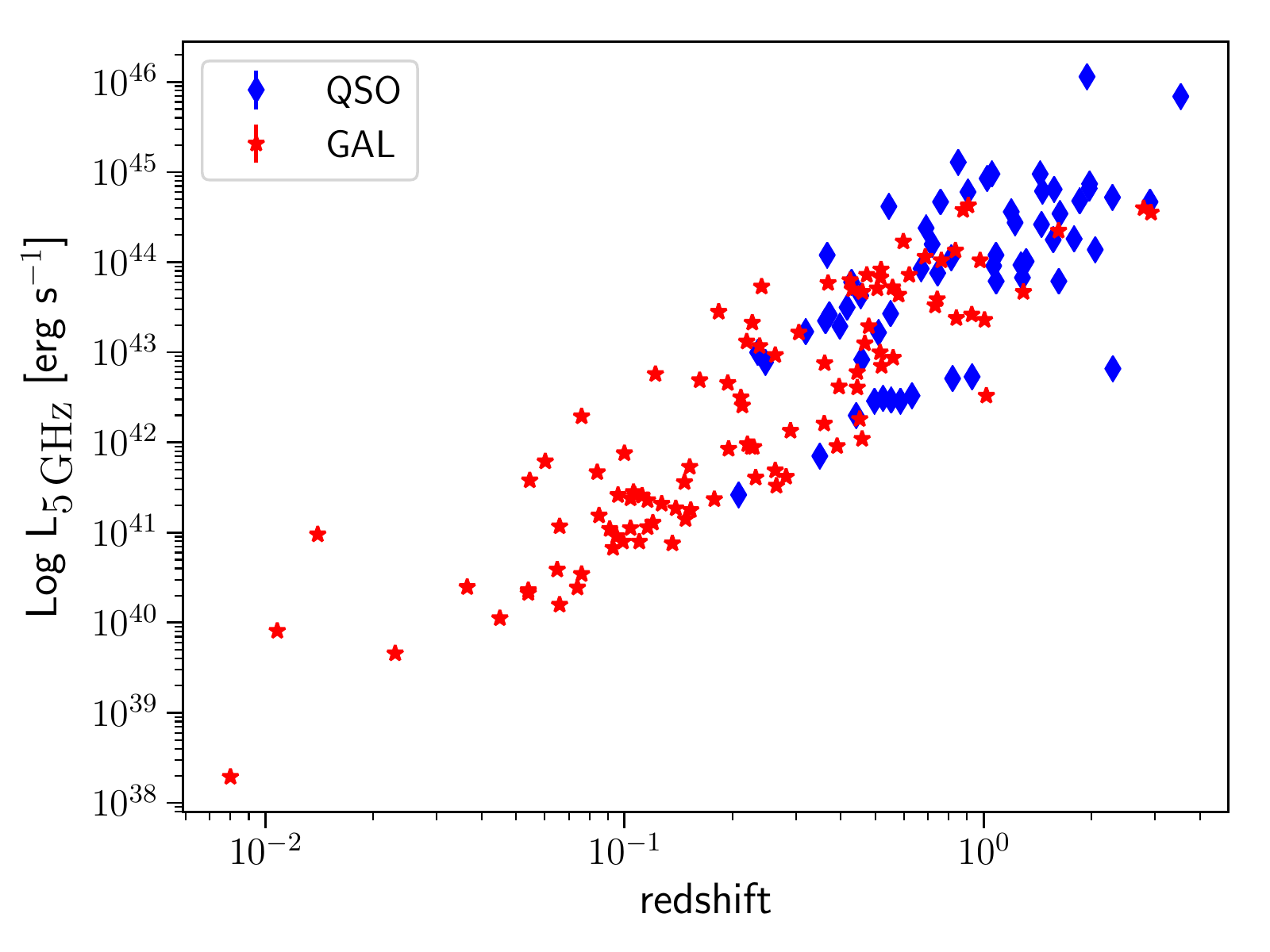}
\caption{\small \label{fig:lum_radio_redshift}
Radio luminosity vs redshift for the galaxies (red) and quasars (blue) contained in our sample.}
\end{figure}

Concerning their radio spectra and peak frequency ($\nu_p$), 52 sources are classified as GPS ($\nu_p > 0.5$ GHz), with the remaining 110 being classified as CSS sources ($\nu_p < 0.5$ GHz).
For several CSS sources only upper limits on the peak frequency have been found in the literature.
The radio luminosity ($\nu L_{\nu \textrm{=5\,GHz}}$) of the sources contained in our sample varies by more than 8 orders of magnitude ($\nu L_{\nu \textrm{= 5\,GHz}}  \sim 10^{38} - 10^{46}$ erg s$^{-1}$).
Fig. \ref{fig:lum_radio_redshift} shows the distribution of the radio luminosity vs redshift for all the sources selected in this work.

\section{Analysis Description}
\label{sec:analysis_desription}
The LAT is a gamma-ray telescope that detects photons by conversion into electron-positron pairs and has an operational energy range from 20\,MeV to 2\,TeV. It comprises a high-resolution converter tracker (for direction measurement of the incident gamma rays), a CsI(Tl) crystal calorimeter (for energy measurement) and an anti-coincidence detector to identify the background of charged particles \citep{2009ApJ...697.1071A}.
The analysis procedure applied in this work is mainly based on two steps. First, we investigate the gamma-ray data of each individual source with a standard likelihood analysis \citep[see e.g.][as described in detail in the next subsection]{2020A&A...635A.185P,2021arXiv210406855A}. Subsequently we performed a stacking analysis of the sources which were not significantly detected in the individual study, in order to investigate the general properties of the population of young radio galaxies and quasars.

\subsection{Analysis of individual sources}
\label{sec:analysis_single_source}
We analysed the \textit{Fermi}-LAT data of each individual source in our sample in order to determine whether it is detected or not (using a Test Statistic TS\footnote{The test statistic (TS) is the logarithmic ratio of the likelihood $\mathcal{L}$ of a model with the source being at a given position in a grid to the likelihood of the model without the source, TS=$2 \log \frac{\mathcal{L}_\mathrm{src}}{\mathcal{L}_\mathrm{null}}$ \citep{1996ApJ...461..396M}.} $>25$ as a threshold).
We analysed more than 11 years of \textit{Fermi}-LAT data between August 5, 2008, and November 1, 2019 (MJD 54683 -- 58788). We selected events which have been reprocessed with the P8R3\_Source\_V2 instrument response functions (IRFs) \citep[IRFs,][]{2018arXiv181011394B}, in the energy range between 100\,MeV and 1\,TeV.
The low energy threshold is motivated by the large uncertainties in the arrival directions of the photons below 100 MeV, leading to a possible confusion between point-like sources and the Galactic diffuse component \citep[see ][for a different analysis implementation to solve this and other issues at low energies with \textit{Fermi}-LAT]{2017AIPC.1792g0016P,2018A&A...618A..22P, 2019RLSFN.tmp....7P}.

We reduced the contamination from the low-energy Earth limb emission \citep{2009PhRvD..80l2004A} by applying a zenith angle cut to the data.
We made a harder cut at low energies by selecting event types with the best point spread function\footnote{A measure of the quality of the direction reconstruction is used to assign events to four quartiles. Gamma rays in Pass 8 data can be separated into 4 PSF event types: 0, 1, 2, 3, where PSF0 has the largest point spread function and PSF3 has the best one.} (PSF).
For energies below 300 MeV we excluded events with zenith angle larger than 85$^{\circ}$, as well as photons from PSF0 and PSF1 event types, while between 300 MeV and 1 GeV  we excluded events with zenith angle larger than 95$^{\circ}$, as well as photons from the PSF0 event type. Above 1 GeV we use all events with zenith angles less than 105$^{\circ}$. 

The binned likelihood analysis (which consists of model optimisation, and localisation, spectrum and variability analyses) was performed with Fermipy\footnote{http://fermipy.readthedocs.io/en/latest/} \citep{2017arXiv170709551W}, a python package that facilitates the analysis of LAT data with the \textit{Fermi} Science Tools, of which the version 11-07-00 was used. 
For each source in our sample we considered a region of interest (ROI) of about 15$^{\circ}$ radius centred on the source position, and each ROI is analysed separately. In each ROI we binned the data with a pixel size of $0.1^{\circ}$ and 8 energy bins per decade.
The model used to describe the sky includes all point-like and extended LAT sources located at a distance $<20^{\circ}$ from the source position and listed in the 4FGL \citep{2020ApJS..247...33A}, as well as the Galactic diffuse and isotropic emission.
For these two latter contributions, we made use of the same templates\footnote{https://fermi.gsfc.nasa.gov/ssc/data/access/lat/ \\BackgroundModels.html} adopted to compile the 4FGL.

For the analysis we first optimised the model for the ROI, then we searched for possible additional faint sources in each ROI, not included in 4FGL, by generating TS maps (significance maps). Subsequently, we re-localised the sources of our sample with TS $>$ 4 ($\sim 2\sigma$).
We performed the spectral analysis during which we left free to vary the diffuse background and the spectral parameters of the sources within 5$^{\circ}$ of our targets.
For the sources in a radius between 5$^{\circ}$ and 10$^{\circ}$ only the normalisation was fit, while we fixed the parameters of all the sources within the ROI at larger angular distances from our targets.
For the spectral energy distribution (SED) plot of the detected sources, we repeated the spectral analysis dividing the photons into seven energy bands: six logarithmically spaced bands between 100 MeV and 100 GeV and one band between 100 GeV and 1 TeV.
We modeled the spectrum of each source with a power-law (PL) function
\begin{equation} \label{eq:power_law}
  \dfrac{dN}{dE} = N_{0} \times (\frac{E}{E_{b}})^{- \Gamma} ;  
\end{equation}
\noindent using $E_b=1$ GeV.
Upper limits at 95\% confidence level are reported in the Appendix \ref{appendix_tables}, for the sources with no significant gamma-ray emission (TS < 10). In order to derive the upper limits we repeated the spectral analysis fixing the photon index ($\Gamma=2$).

Finally we extracted a light curve for each source using time bins of 1 year. For the brightest and variable sources, the lightcurve analysis was repeated using time intervals of 3 months in order to better characterise the emission variability.
The fluxes in each interval were obtained by leaving only the normalisation free to vary and freezing the other spectral parameters to the best fit values obtained from the full range analysis. Using the same method applied in \citet{2020ApJS..247...33A}, for all the detected sources we computed the variability index $TS_{var}$. Variability is considered probable when $TS_{var} >23$ ($>$68), corresponding to 99\% confidence in a $\chi^{2}$ distribution with $N_{int} - 1 = 10$ (44) degrees of freedom, where $N_{int}$ is the number of intervals corresponding in our case to the 11 yearly-length periods (to the 45 three-months long periods).
Properties of the detected young radio sources, including SEDs and lightcurves, are described in Sect. \ref{sec:detected_sources}, while the results for all the individual sources are reported in the Appendix (Table \ref{table_sample}).

\subsection{Stacking analysis}
\label{sec:stacking_analysis}


As discussed in Sect. \ref{sec:detected_sources}, only a few young radio sources have been significantly detected. 
We analysed in detail the population of the undetected sources looking for collective emission from these objects.

To reach this goal we performed a stacking analysis of the sources using the spectrum results of each object as described in Sect. \ref{sec:analysis_single_source}. 
For each source and each energy bin a log-likelihood profile $\log \mathcal{L}_{i,k}$ was calculated, i.e. the log-likelihood value as a function of the photon flux. The indices $i$ and $k$ represent the source and the energy bin, respectively.
We assumed a spectral shape common to all sources $dN/dE$ and calculated the corresponding log-likelihood value at a given energy. The total log-likelihood was obtained by summing over all the energy bins and sources:

\begin{equation}
    \log \mathcal{L} = \sum_i \sum_k \log \mathcal{L}_{i,k}|_{dN/dE(E_k)}
\end{equation}

\noindent We assumed a simple power-law spectrum (see Eq. \ref{eq:power_law}) for the entire population. We varied the normalisation $N_0$ and photon index $\Gamma$ to create a 2-dimensional likelihood profile in order to search for the parameter values which maximise the log-likelihood.
The significance of the potential detection was checked by comparing the maximum log-likelihood value with the one of the null hypothesis ($\log \mathcal{L}_{null}$), i.e., the hypothesis in which the flux of the gamma-ray emitter is zero. We obtained the ($\log \mathcal{L}_{null}$) from the 2-dimensional profile by setting $k=0$ and defined the TS = $2(\log \mathcal{L} - \log \mathcal{L}_{null}$).
If the log likelihood distribution in the null hypothesis is asymptotically Gaussian, then we expect values of $\Delta \log \mathcal{L}>4.61$ to occur by chance only 5\% of the time.  We performed MCMC simulations (see Appendix \ref{verification_stacking_simulation}) and verified that this threshold corresponds to a 5\% false positive detection rate.

In the case of no detection, we estimated the upper limit at 95\% confidence level on the photon flux deriving the 2-dimensional contour corresponding to a $\Delta \log \mathcal{L}=4.61/2$, having two additional free parameters in the model (Ciprini, Di Venere and Mazziotta, in preparation).
A similar method was applied in \citet{2020ApJ...894...88A} to study a different class of celestial objects.


To verify the robustness of our stacking method, we simulated 11 years of Pass 8 data for 100 sources in random positions on the sky with the same spectrum. The flux value was chosen such that the sources are below the LAT detection threshold. The simulation confirmed the robustness of the method to detect cumulative gamma-ray emission from sources of the same population, assuming they have similar spectral properties (see Appendix \ref{verification_stacking_simulation} for details).



As discussed in detail in Sect. \ref{sec:stacking_results}, the stacking procedure was applied to several sub-samples of the population available. The selection was based on different parameters (e.g. linear size, distance).
Since the photon flux upper limit depends on the number of sources included in the stacking procedure, we repeated the analysis changing the number of sources $N$ included in the stacking procedure. 
See Appendix \ref{verification_stacking_background} for more details.



\section{Results}
\label{sec:results}
In this section we present the results of the analysis of each individual source as well as the cumulative results obtained from the stacking analysis of all the undetected (TS<25) young radio sources. Our analysis expands the 4FGL study to more than 11 years of \textit{Fermi}-LAT data.
For the individual sources we report here the results of those with a TS value $\geq$ 25 \footnote{TS = 25 with 2 degrees of freedom, as in the case of a simple power-law model, corresponds to an estimated statistical significance of $\sim 4.6\sigma$ assuming that the null-hypothesis TS distribution follows a ${\chi }^{2}$ distribution (see \citet{1996ApJ...461..396M}).}, and those with a marginal detection, 10 $\leq$ TS < 25. The high-energy properties of all the sources of our sample are listed in Appendix \ref{appendix_tables}.

\subsection{Young radio sources: individual detections}
\label{sec:detected_sources}
\begin{figure*}
\centering
\includegraphics[trim=1.5cm 2.5cm 1.5cm 3.3cm,clip,width=14cm]{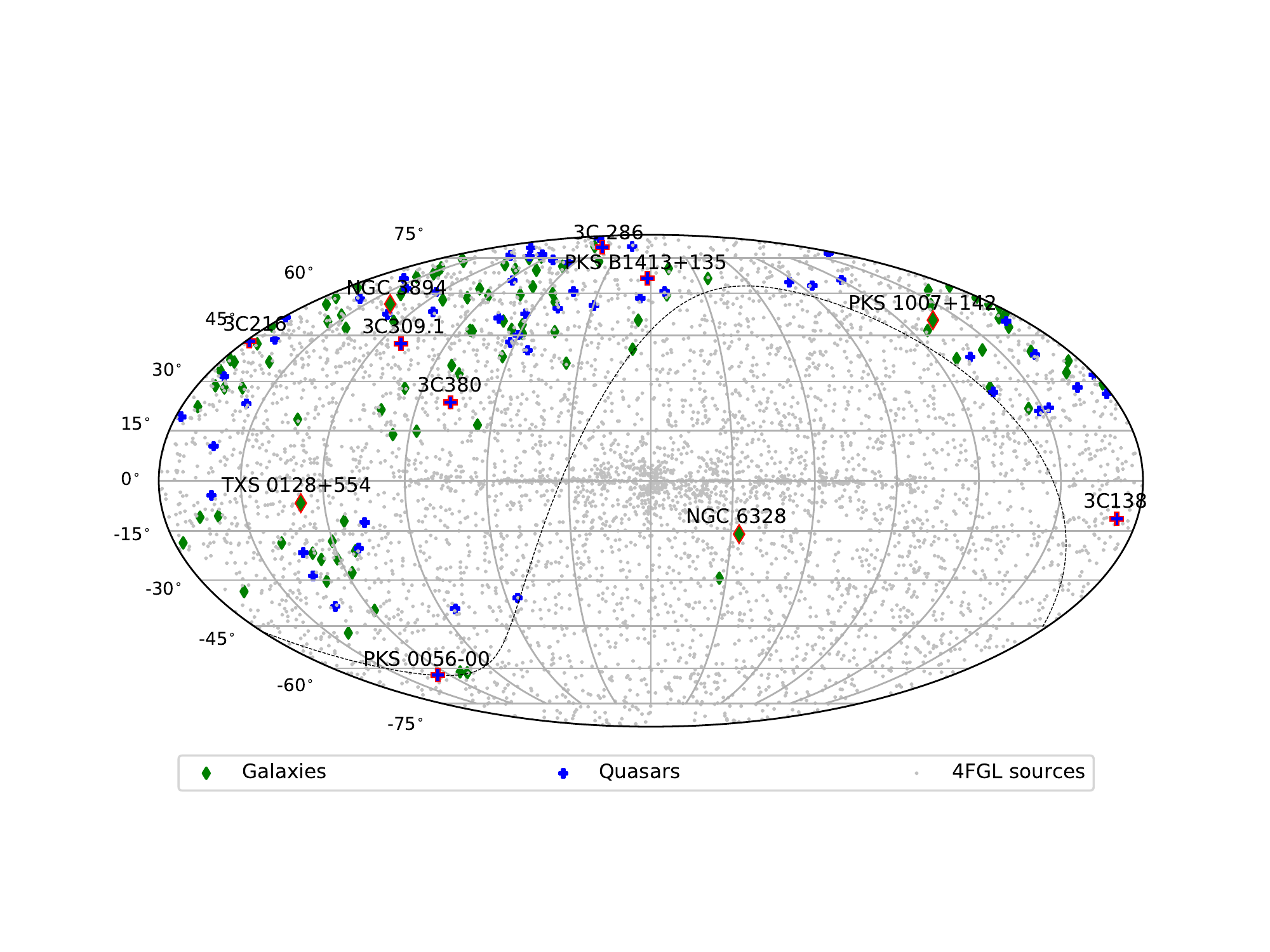}
\caption{\small \label{fig:sky_map}
Sky map, in Galactic coordinates and Mollweide projection, showing the young radio sources in our sample. The detected sources are labelled in the plot. All the 4FGL sources \citep{2020ApJS..247...33A} are also plotted, with grey points, for comparison.}
\end{figure*}

From our analysis we detect significant 
gamma-ray emission (TS $>$ 25) at the positions of 11
young radio sources (see Fig. \ref{fig:sky_map}), four galaxies and seven quasars, whose characteristics are reported in Table \ref{table_detected}. 

\begin{table*}
\caption{\small \label{table_detected} List of young radio sources detected in our analysis. We report the name, morphological/spectral type, redshift, projected linear size (LS) [kpc], peak frequency ($\nu_p$) [GHz], radio power at 5 GHz [W Hz$^{-1}$], gamma-ray significance (TS), gamma-ray flux ({0.1--1000 GeV}) in units of 10$^{-9}$ ph cm$^{-2}$ s$^{-1}$, power-law photon index, gamma-ray luminosity [10$^{44}$ erg s$^{-1}$] and gamma-ray variability (TS$_{var}$) estimated with the one-year time intervals. Sources with variable gamma-ray emission are marked with $^{\dag}$. 
The table is divided into two blocks: the upper one lists the galaxies, while the lower part reports the quasars. The parameter TS$_{var}$ indicates the significance of the variability. 
The newly detected source, PKS\,1007+142 is marked with `*'.
}
\small
\hspace*{-0.5cm}
\centering
\begin{tabular}{c|ccccc|ccccc}
\hline \hline
Name & type & $z$ & LS & $\nu_p$ & log L$_{5\, \textrm{GHz}}$& TS & F$_{\gamma}$ & $\Gamma$ & L$_{\gamma}$ & TS$_{var}$\\
 & & & kpc & GHz & W Hz$^{-1}$  & & 10$^{-9}$ cm$^{-2}$ s$^{-1}$ & & 10$^{44}$ erg s$^{-1}$ & \\
\hline
\multicolumn{11}{c}{Galaxies}\\
\hline
NGC 6328 & CSO/GPS & 0.014 & 0.002 & 4 & 24.28 & 36 & 5.30$\pm$1.45 & 2.60$\pm$0.14 & 0.011 & 5\\
NGC 3894 & CSO/GPS & 0.0108 & 0.010 & 5 & 24.60 & 95 & 2.03$\pm$0.48 & 2.05$\pm$0.09 & 0.006 & 11\\
TXS 0128+554 & CSO/GPS & 0.0365 & 0.012 & 0.66 & 23.69 & 178 & 8.03$\pm$1.46 & 2.20$\pm$0.07 & 0.19 & 9\\ 
PKS 1007+142* & MSO/GPS & 0.213 & 3.3 & 0.5-2 & 25.71 & 31 & 4.65$\pm$1.55 & 2.56$\pm$0.18 & 2.8 & 4\\ 
\hline 
\multicolumn{11}{c}{Quasars}\\
\hline
3C 138$^{\dag}$ & MSO/CSS & 0.759 & 5.9 & 0.176 & 27.97 & 34 & 2.09$\pm$0.89 & 2.05$\pm$0.12 & 64 & 68\\
3C 216$^{\dag}$ & LSO/CSS & 0.6702 & 56 & 0.066 & 27.23 & 153 & 7.78$\pm$0.98 & 2.60$\pm$0.09 & 97 & 24\\
3C 286 & LSO/CSS & 0.85 & 25 & <0.05 & 28.41 & 67 & 5.60$\pm$1.10 & 2.52$\pm$0.12 & 110 & 8\\ 
3C 309.1$^{\dag}$ & MSO/CSS & 0.905 & 17 & <0.076 & 28.08 & 207 & 6.33$\pm$0.74 & 2.47$\pm$0.07 & 180 & 215\\
3C 380$^{\dag}$ & MSO/CSS & 0.692 & 11 & <0.05 &27.68 & 2274 & 36.44$\pm$1.48 & 2.41$\pm$0.03 & 510 & 68\\
PKS 0056-00 & MSO/CSS & 0.719 & 15 & <0.14 &27.50 & 52 & 5.21$\pm$1.48 & 2.30$\pm$0.15 & 74 & 11\\ 
PKS B1413+135$^{\dag}$ & CSO/GPS & 0.247 & 0.03 & 8.4-15 & 26.19 & 1198 & 14.72$\pm$1.02 & 2.10$\pm$0.03 & 28 & 321\\ 
\hline
\end{tabular}
\end{table*}
\noindent Nine out of the 11 detected sources were present in previous \textit{Fermi}-LAT catalogs, while PKS\,0056-00 has been recently reported in the latest
release of LAT sources 4FGL-DR2\footnote{https://fermi.gsfc.nasa.gov/ssc/data/access/
lat/10yr\_catalog/} \citep{2020ApJS..247...33A}.

\begin{figure*}
\begin{center}
\rotatebox{0}{\resizebox{!}{65mm}{\includegraphics{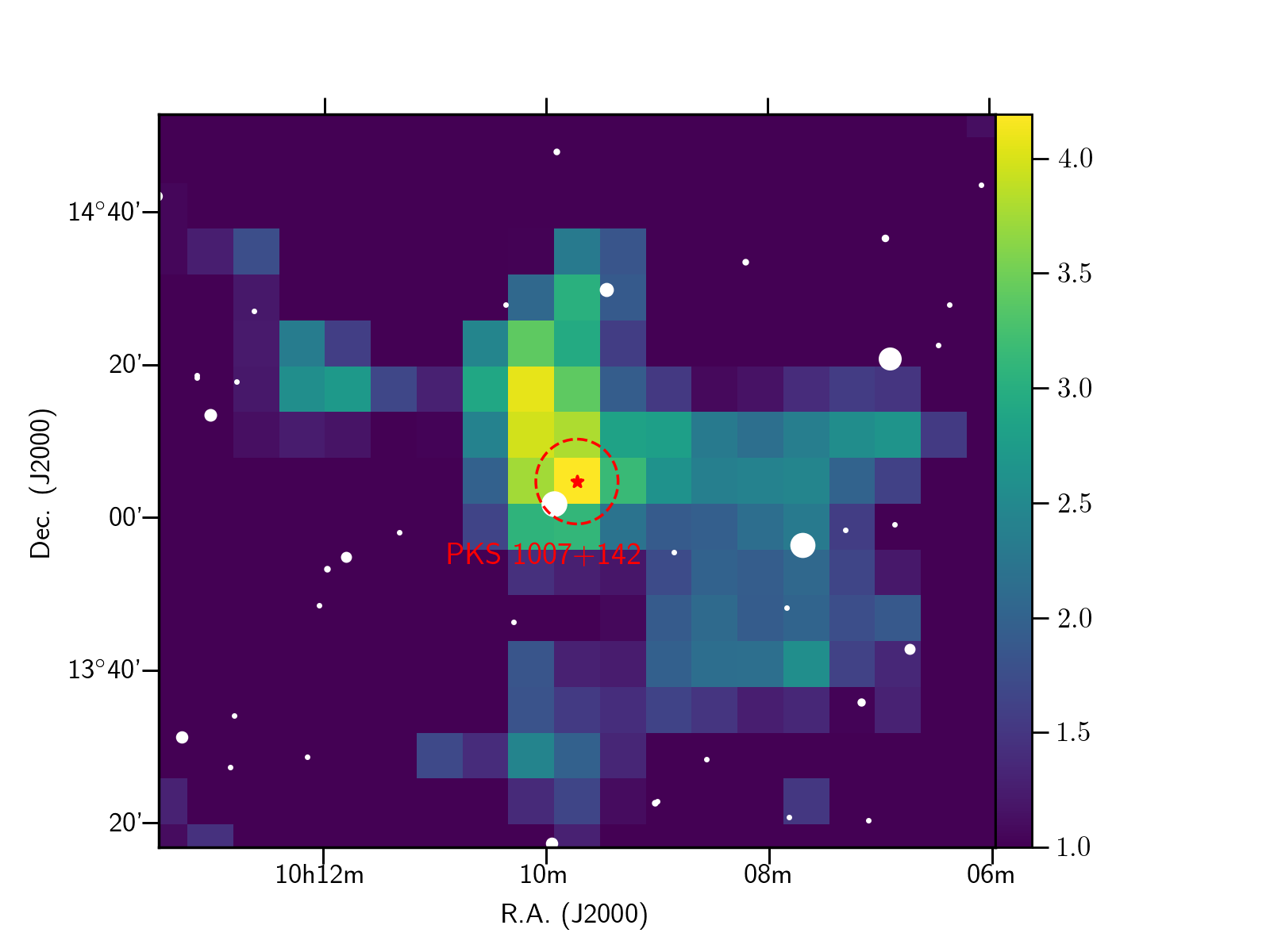}}}
\hspace{0.1cm}
\rotatebox{0}{\resizebox{!}{60mm}{\includegraphics{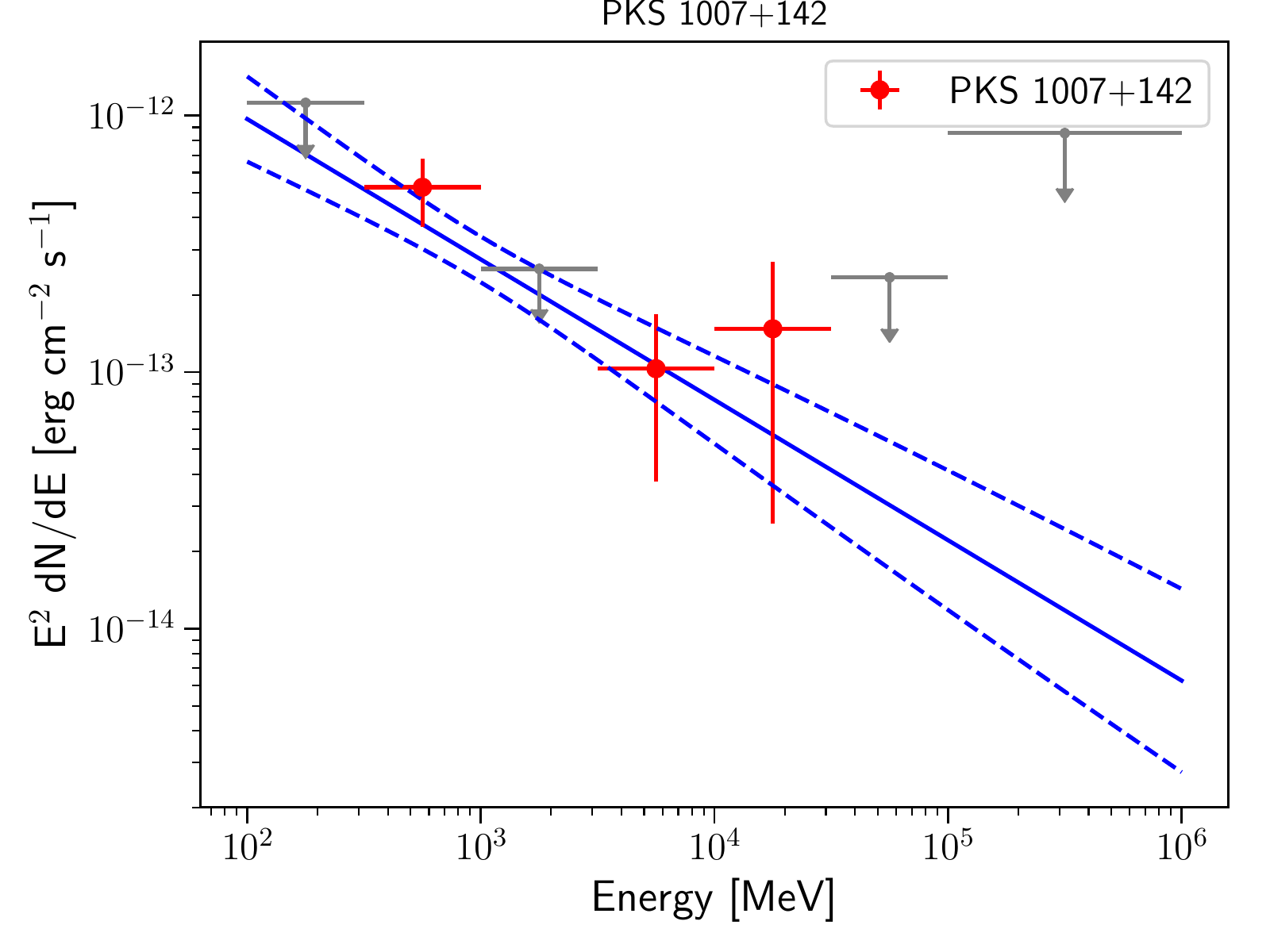}}}
\caption{\small {\it Left}: \textit{Fermi}-LAT TS map (in sigma units) above 100 MeV of the region around PKS\,1007$+$142. The red star and circle represent the central position and the 68\% confidence-level uncertainty R$_{68}=0.08^{\circ}$ of the gamma-ray source, respectively. White dots show radio sources from the NVSS survey whose sizes are arbitrarily scaled depending on their radio flux density. {\it Right}: \textit{Fermi}-LAT SED of the galaxy PKS\,1007$+$142. The SED has been fit with a PL (blue line). The $1\sigma$ upper limit is reported when TS $<4$. }
\label{fig:map_sed_pks1007}
\end{center}
\end{figure*}

In addition to the sources already included in the 4FGL-DR2, we report here the discovery of
gamma-ray emission from the young radio galaxy PKS\,1007+142 ($z$ = 0.213).
We significantly (TS = 31) detected gamma-ray emission from the compact radio galaxy PKS\,1007$+$242. The LAT best-fit position of PKS\,1007$+$142 (R.A., Dec. (J2000)= 152.43$^{\circ} \pm 0.05^{\circ}$, 14.08$^{\circ} \pm 0.06^{\circ}$), 68\% confidence-level uncertainty $R_{68}= 0.08^{\circ}$, is compatible with its radio counterpart (see left panel of Fig. \ref{fig:map_sed_pks1007}). 
We note that the 4FGL-DR2 catalog reports the detection of the source 4FGL J1010.0+1416 (TS$=31$, $\Gamma=2.73\pm0.19$) located in the vicinity of PKS\,1007+142 and less than 0.2$^{\circ}$ away from the source we report here. The source, detected in the 4FGL-DR2 catalog which is based on 10 years of data, presents a similar significance, however it does not present a clear association with PKS\,1007+142. For a crosscheck, we repeated the analysis using 11.3 years starting from the position of the 4FGL-DR2 source obtaining a position compatible, considering the statistical error, with the one reported above. This can indicate a possible better localization achieved in this work thanks to the dedicated analysis and the longer exposure.

\noindent PKS\,1007$+$142 presents a soft gamma-ray spectrum (see right panel of Fig. \ref{fig:map_sed_pks1007}) with best-fit results $\Gamma = 2.56 \pm 0.18$ and $F =(4.65 \pm 1.55) \times 10^{-9}$ ph cm$^{-2}$ s$^{-1}$.

In order to verify the association of the newly detected gamma-ray source with its radio counterpart we used a Bayesian method \citep{2012ApJS..199...31N} which is based only on spatial coincidence between the gamma-ray source and its potential counterpart. A uniform threshold of P $> 0.8$ is applied to the posterior probability for the association to be retained. The resulting association probability confirms the association of the new \textit{Fermi}-LAT galaxy with the radio counterpart PKS\,1007+142 (P = 0.92).

Considering the detected sources, all the galaxies are GPS ($\nu_p$ > 0.5 GHz), while all the quasars are CSS, with the exception of the peculiar GPS PKS\,B1413$+$135.

\subsubsection{Galaxies}
In our analysis we observe significant gamma-ray emission from four young radio galaxies: NGC\,3894, NGC\,6328, TXS\,0128$+$554 and the newly detected source PKS\,1007$+$142. 
Compared to galaxies previously reported in the 4FGL catalog, NGC\,3894 presents a relatively flat spectrum ($\Gamma = 2.05 \pm 0.05$), while NGC\,6328 has a softer spectrum ($\Gamma = 2.60 \pm 0.14$). The spectral results obtained for NGC\,3894 and NGC\,6328 are compatible with those found in the 4FGL as well as those obtained in dedicated studies \citep{2020A&A...635A.185P,2016ApJ...821L..31M}.

A recent multi-frequency radio radio very long baseline array (VLBA) study of the galaxy TXS\,0128$+$554 ($z$ = 0.0365), presented in \citet{2020ApJ...899..141L}, provided new information on this radio source, which was previously classified as blazar candidate of uncertain type (BCU) in the 4FGL catalog. 
They measured the compact size (LS $\sim$ 12 pc), misaligned nature (43$^{\circ}<\theta<59^{\circ}$) and the advanced speed of the jet separation ($v = 0.32 \pm \,0.07c$) of the source, classifying it as a young radio galaxy with kinematic age of only 82 $\pm$ 17 years.
In our analysis, the source is significantly detected with a TS = 178, and presents spectral results ($\Gamma = 2.20 \pm 0.07$, $F_{\gamma}=8.03 \pm 1.46 \, \times 10^{-9}$ ph cm$^{-2}$ s$^{-1}$) which are compatible with the 4FGL results.
No significant gamma-ray emission variability has been found for the detected young radio galaxies ($TS_{var}<23$; see  Appendix \ref{appendix_detected_sources} for the light curve plots). 

\subsubsection{Quasars}
In our analysis we detect significant gamma-ray emission from seven quasars already reported in earlier works: 3C\,138, 3C\,216, 3C\,286, 3C\,309.1, 3C\,380, PKS\,0056-00, PKS\,B1413$+$135. 
For all the detected quasars, with the exception of 3C\,138, the spectral parameters obtained in this work are compatible with those reported in the 4FGL catalog. Five quasars have a relatively soft photon index ($\Gamma >$ 2.3), while 3C\,138 and PKS\,B1413$+$135 present a flatter spectrum with photon indexes $\Gamma \sim$ 2 (see Table \ref{table_detected}, and the SED plots in the Appendix \ref{appendix_detected_sources}).

Five quasars (3C\,138, 3C\,216, 3C\,309.1, 3C\,380 and PKS\,B1413$+$135) present significant variability (TS$_{var}$ > 23) of the gamma-ray emission. As can be seen from their light curves (see the Appendix \ref{appendix_detected_sources} for the light curve plots) the sources underwent strong flares during the LAT observations.
In particular, the quasar 3C\,138 underwent a strong gamma-ray flare in 2012, during which the emission spectrum was quite soft $\Gamma_{2012}\gtrsim 2.5$. Thereafter, the activity quickly decreased: the source is only marginally detected until 2016, after which the flux falls below the detection threshold. Similarly the measured flux and photon index also decrease with the time interval considered for the analysis, reaching a flux of $F=2.1\pm0.9 \times 10^{-9}$ ph cm$^{-2}$ s$^{-1}$ and a photon index $\Gamma= 2.05 \pm 0.12$, as obtained in this work.

\begin{figure}
\includegraphics[width=0.97\columnwidth]{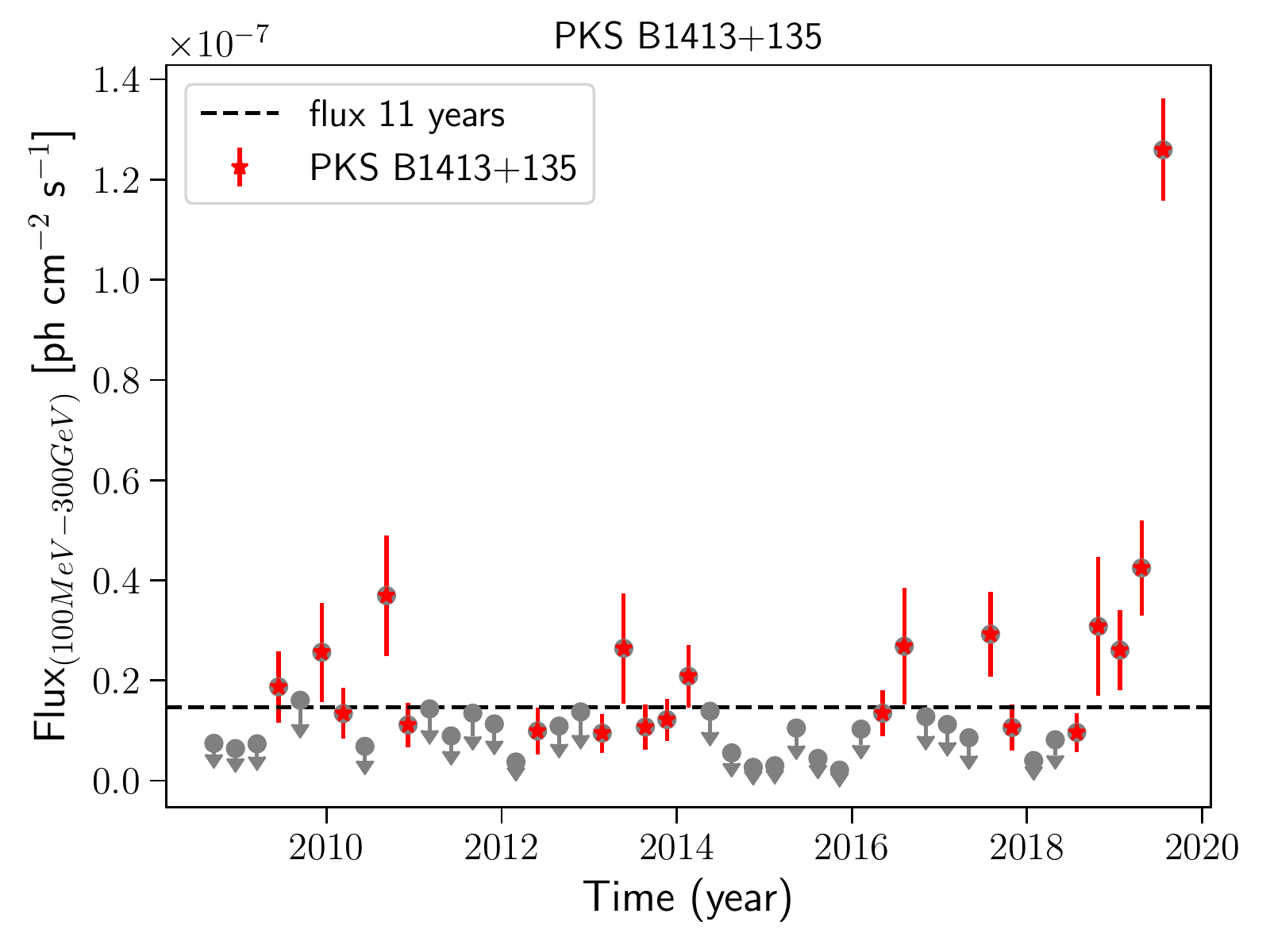}
\caption{\small \textit{Fermi}-LAT three-month binned light curve of PKS\,B1413$+$135. The $1\sigma$ upper limit is reported when TS $<4$. The dashed line represents the averaged flux for the entire period. The flux values have been estimated for the energy range 100 MeV -- 300 GeV.}
\label{fig:sed_lightcurve_pks1413}
\end{figure}

PKS\,B1413$+$135 is significantly detected (TS = 1198) in our analysis, showing extremely bright gamma-ray emission, with an average flux $F = (14.7 \pm 1.0) \times 10^{-9}$ ph cm$^{-2}$ s$^{-1}$. Its light curve reveals a strong gamma-ray flare in the latest period of observation considered for this work (August -- November 2019), when the source reached a flux $F = (126 \pm 10) \times 10^{-9}$ ph cm$^{-2}$ s$^{-1}$, an order of magnitude above the averaged one (see Fig. \ref{fig:sed_lightcurve_pks1413} for the light curve of the source). 
\noindent We analysed the gamma-ray emission at the time of the flare with daily time-bins. The flux reached the peak on August 29, 2019, when the source was detected with a significance of TS = 67. We measured a daily flux of $F_{E > 100 \,\textrm{MeV}} = (5.4 \pm 1.9) \times 10^{-7}$ ph cm$^{-2}$ s$^{-1}$ and a significant hardening of the spectrum: $\Gamma = 2.0 \pm 0.2$ ($\Gamma_{\rm\,4FGL} = 2.41 \pm 0.07$), in agreement with the preliminary results reported by \citet{2019ATel13049....1A}. 
PKS\,B1413$+$135 has long been considered an unusual object, with a BL-Lac-like AGN hosted in a spiral galaxy at redshift $z$ = 0.247 \citep{1992ApJ...400L..13C,2017ApJ...845...90V}. The detection of the flaring activity supports the idea that the gamma-ray emission is beamed and produced by a relativistic jet at a relatively small viewing angle, similar to the case of the AGN PKS\,0521$-$36 \citep{2015MNRAS.450.3975D}. Alternatively, it can represent a particular case of high-activity episode of a misaligned AGN as seen in 3C84 \citep{2011MNRAS.413.2785B,2018A&A...614A...6S,2018ApJ...855...93F}. A recent study by \citet{2021ApJ...907...61R} argues that the association with the spiral host galaxy is just due to a chance alignment, and instead supports the hypothesis that PKS\,B1413+135 is a background blazar-like object lying in the redshift range  0.247 $<$ z $<$ 0.5.


The presence of gamma-ray flares in the other quasars suggests that their high-energy emission is due to a relativistic jet and beaming effect, confirming their non-misaligned nature.

\subsection{Results on individual sources}
\begin{figure}
\centering
\includegraphics[width=\columnwidth]{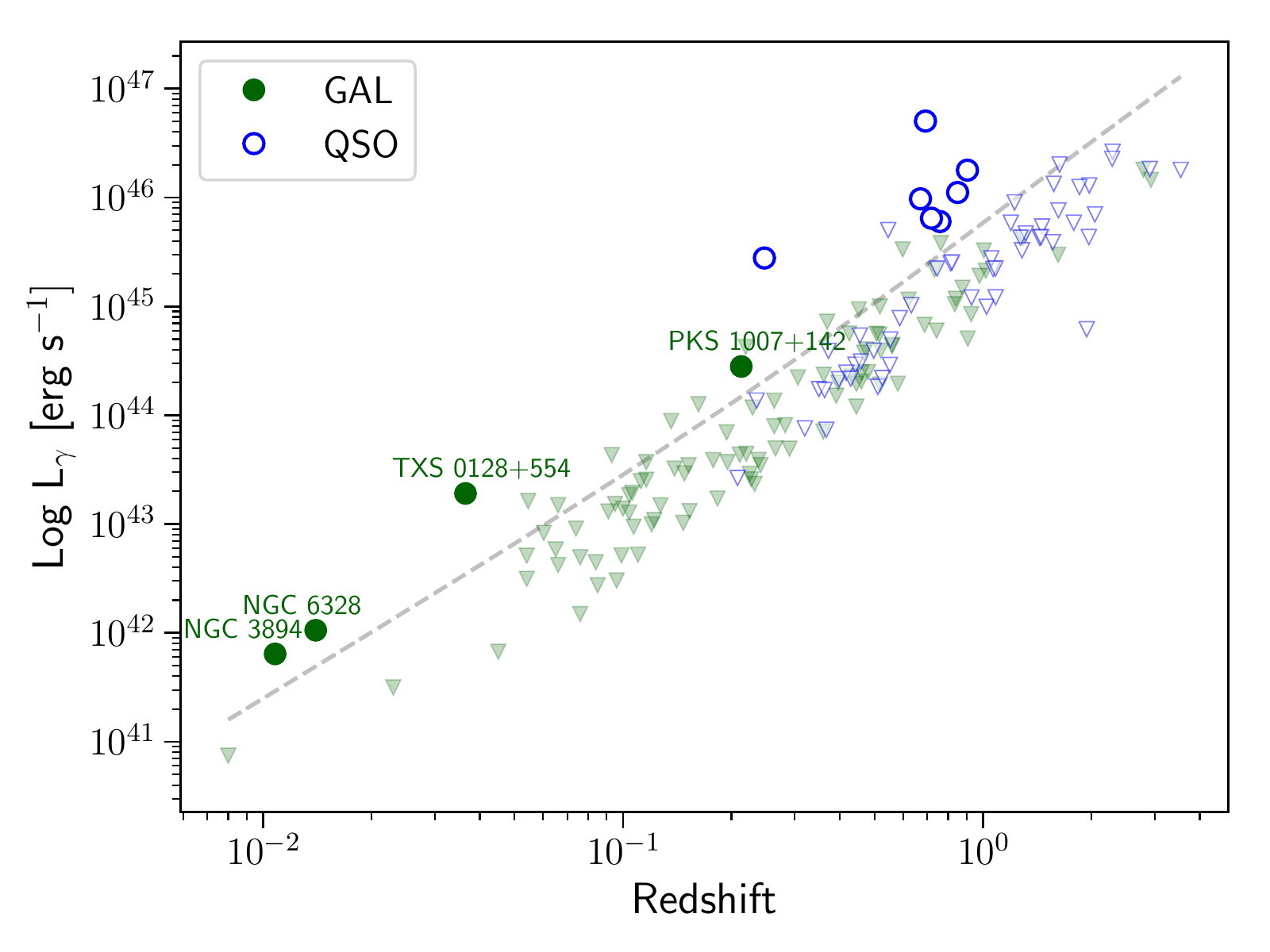}
\caption{\small \label{fig:lum_vs_redshift}
Gamma-ray luminosity vs redshift for all the sources in our sample. The circle (triangles) represents the detected (undetected) sources, distinguishing between galaxies (in green) and quasars (blue). The dashed line represents the averaged \textit{Fermi}-LAT ten-year sensitivity ($F_{E >100\,\textrm{MeV}}=1.8\times 10^{-9}$ MeV cm$^{-2}$ s${-1}$, $\Gamma = 2.0$) for an isolated point source outside the Galactic plane ($\mid b \mid > 25^{\circ}$), as a function of the redshift.}
\end{figure}

In order to investigate the possible origin of the high-energy emission in the population of young radio sources, we compared the radio and gamma-ray properties of the investigated sources. 
\noindent Figure \ref{fig:lum_vs_redshift} shows the distribution of the gamma-ray luminosity vs redshift for all the sources in our sample. 

\noindent
The dashed line represents the averaged \textit{Fermi}-LAT ten-year sensitivity as a function of the redshift ($F_{E >100\,\textrm{MeV}}=1.5\times 10^{-9}$ MeV cm$^{-2}$ s$^{-1}$, $\Gamma = 2.0$) for an isolated point source outside the Galactic plane ($\mid b \mid > 25^{\circ}$).
All the detected sources have a gamma-ray luminosity above, or in the proximity, of the \textit{Fermi}-LAT sensitivity, while we derived upper limits mainly below the \textit{Fermi}-LAT sensitivity for the remaining ones. In addition, nine sources, eight galaxies and one quasar (3C\,147), were marginally detected. This means that, although below the TS $=$ 25 threshold to formally claim a detection, they present a non-negligible gamma-ray emission (TS > 10, corresponding to a significance > 3$\sigma$, see Table \ref{table_ts_10_20}). Of these, the radio galaxies are at redshift between 0.093 and 0.763, i.e. more distant than the radio galaxies detected by \textit{Fermi}-LAT so far, with the exception of PKS\,1007$+$142. Their gamma-ray luminosity is between $\sim$ 2$\times$10$^{43}$ and 2.2$\times$10$^{45}$ erg s$^{-1}$, with the quasar 3C\,147 reaching a luminosity of about 5$\times$10$^{45}$ erg s$^{-1}$.
It is likely that the detection of these sources, or a fraction of them, will be confirmed with the increase of statistics in the coming years.

\begin{table*}
\caption{\small \label{table_ts_10_20} List of the young radio sources with marginal detection (TS > 10). We report  name, type, redshift, projected linear size (LS) [kpc], radio peak frequency ($\nu_p$) [GHz], radio luminosity at 5 GHz [W Hz$^{-1}$], gamma-ray significance (TS), gamma-ray flux ({0.1--1000 GeV}) in units of 10$^{-9}$ ph cm$^{-2}$ s$^{-1}$, power-law photon index ($\gamma$) and gamma-ray luminosity [10$^{44}$ erg s$^{-1}$]. 
}
\small
\hspace*{-0.8cm}
\centering
\begin{tabular}{c|ccccc|cccc}
\hline \hline
Name & type & $z$ & LS & $\nu_p$ & log L$_{5\, \textrm{GHz}}$& TS & Flux$_{\gamma}$ & $\Gamma$ & Lum$_{\gamma}$\\
 & & & kpc & GHz & W Hz$^{-1}$  & & 10$^{-9}$ cm$^{-2}$ s$^{-1}$ & & 10$^{44}$ erg s$^{-1}$\\
\hline
\multicolumn{10}{c}{Galaxies}\\
\hline
0404+768 & CSO/GPS & 0.598 & 0.866 & 0.55 & 27.53 & 12 & 2.70$\pm$0.81 & 2.61$\pm$0.29 & 22.2\\
1323+321 & CSO/GPS & 0.369 & 0.305 & 0.68 & 27.07 & 19 & 1.36$\pm$0.41 & 2.15$\pm$0.23 & 4.0\\
3C346 & LSO/CSS & 0.162 & 22.056 & <0.045 & 25.99 & 13 & 1.23$\pm$0.43 & 2.07$\pm$0.20 & 0.82\\
1843+356 & CSO/GPS & 0.763 & 0.022 & 2 & 27.32 & 11 & 0.59$\pm$0.24 & 1.93$\pm$0.24 & 22.6\\
J140051+521606 & CSO/CSS & 0.116 & 0.32 & <0.15 & 24.36 &  17 & 0.12$\pm$0.05 & 1.64$\pm$0.32 & 0.20\\
J083411.09+580321.4 & CSO/CSS & 0.093 & 0.0086 & <0.4 & 24.13 & 15 & 3.53$\pm$0.96 & 2.66$\pm$0.20 & 0.30\\
J092405.30+141021.4 & CSO/CSS & 0.136 & 0.74 & <0.4 & 24.18  & 13 & 2.15$\pm$0.69 & 2.33$\pm$0.24 & 0.58\\
J155235.38+441905.9 & MSO/CSS & 0.452 & 6.93 & <0.4 & 25.56  & 17 & 0.78$\pm$0.26 & 2.07$\pm$0.19 & 6.0\\
\hline
\multicolumn{10}{c}{Quasar}\\
\hline
3C147 & MSO/CSS & 0.545 & 4.454 & 0.231 & 27.92 & 22 & 6.89$\pm$1.51 & 2.69$\pm$0.16 & 47.120\\
\hline
\end{tabular}
\end{table*}

In the leptonic scenario at the basis of the expectation of gamma-ray emission from young radio sources, the relativistic electrons producing the gamma rays via IC scattering are the same ones responsible for the radio emission via synchrotron radiation
\citep{1992ApJ...397L...5M, 1996MNRAS.280...67G, 2010ApJ...720..912A}. Therefore, we investigate the presence of a possible correlation between the radio and gamma-ray emission. Comparing the intrinsic gamma-ray luminosity and the total radio luminosity (5 GHz) of the detected young radio sources a correlation is readily apparent, $\log L_{\gamma} = (0.86\pm 0.18) \times \log L_{\textrm{5\,GHz}} + (7.71 \pm 1.31)$, with Spearman correlation $\rho$ = 0.88. 
%

%
\noindent However, the use of luminosity introduces a redshift bias in samples that have a dynamic range in luminosity distance which is much larger than that in fluxes. For this reason, we investigated a possible correlation between the radio flux density and the \textit{Fermi}-LAT flux, a method that unveiled a significant correlation for the gamma-ray blazar population reported by \citet{2011ApJ...741...30A,2017A&A...606A.138L}. Considering the 1.4-GHz flux density (from the NRAO VLA Sky Survey, NVSS) and the gamma-ray flux for the sub-sample of young radio sources, we do not find any obvious correlation, with Spearman and Pearson correlation coefficient of $\rho=-0.12 $ and $r= 0.05$, respectively. With all the caveats due to the small number of objects, it seems that no direct correlation is present between the radio and gamma-ray emission in our sample, in contrast to that found in the studies previously done using all the AGN detected by \textit{Fermi}-LAT.

\subsection{Undetected young radio sources: stacking analysis results}
\label{sec:stacking_results}
\begin{figure*}
\begin{center}
\rotatebox{0}{\resizebox{!}{65mm}{\includegraphics{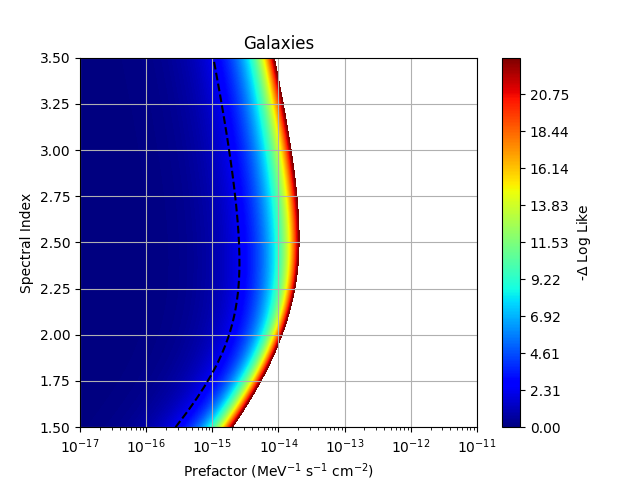}}}
\hspace{0.1cm}
\rotatebox{0}{\resizebox{!}{65mm}{\includegraphics{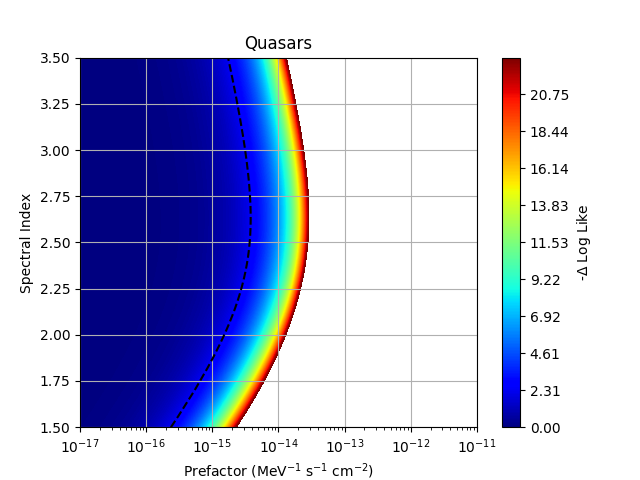}}}
\caption{\small Likelihood profile assuming a simple power-law spectrum for all the undetected galaxies (left panel) and quasars (right panel) separately. The black dashed line represents contour for which a $\Delta \log \mathcal{L}=4.61/2$ is obtained, corresponding to the 95\% confidence level upper limit.}
\label{fig_stacking_all}
\end{center}
\end{figure*}

We searched for a signal from the population of the 151 undetected young radio sources by applying the stacking procedure described in Sect. \ref{sec:stacking_analysis}.
Fig. \ref{fig_stacking_all} shows the 2-dimensional log-likelihood profile obtained from the stacking analysis of the undetected galaxies and quasars. 
The blue region in the plots indicates the most probable parameter values for the considered sample. If a detection was found, a circumscribed blue area should have appeared in the plot, corresponding to the most probable parameter values (see for example Fig. \ref{fig_stacking_simulation}).
Instead, the plots in Fig. \ref{fig_stacking_all} indicate that no significant emission is observed.

Upper limits on the gamma-ray flux for the young radio sources associated with galaxies and quasars are calculated from the black dashed line in Fig. \ref{fig_stacking_all}, corresponding to the 95\% confidence level upper limit, as described in Sect. \ref{sec:stacking_analysis}. Similar results are found both when the stacking analysis is performed on all galaxies and quasars together and when quasars and galaxies are considered separately.
The photon flux upper limits are reported in Table \ref{table_stacking}. The photon indices reported correspond to the values for which the photon flux upper limit is calculated.

\begin{table}
\caption{\small \label{table_stacking} Results of the stacking analysis for all the undetected sources contained in our sample, as well as for the galaxies and quasars only. $F^{*}_{\gamma}$: upper limits on the flux in units of 10$^{-11}$ ph cm$^{-2}$ s$^{-1}$, $\Gamma$: photon index corresponding to the upper limit value.}
\small
\centering
\begin{tabular}{c|cccc}
\hline \hline
Select. & N & TS & $F^{*}_{\gamma}$ & $\Gamma$\\
\hline
All & 151 & 0.3 & 3.29 & 2.53\\
Galaxies & 99 & 0.1 & 4.62 & 2.40\\
Quasars & 52 & 0.2 & 10.09 & 2.64\\
\hline
\end{tabular}
\end{table}



\noindent The upper limits obtained from the stacking analysis are about one order of magnitude below the averaged upper limits of the individual undetected sources ($F^{*}_{av}=7.1 \times 10^{-10}$ ph cm$^{-2}$ s$^{-1}$).
A factor of 10 is compatible with the prediction derived from the analysis of background fluctuations, as discussed in section \ref{verification_stacking_background}.

\begin{figure*}
\begin{center}
\hspace*{-0.5cm}
\rotatebox{0}{\resizebox{!}{65mm}{\includegraphics{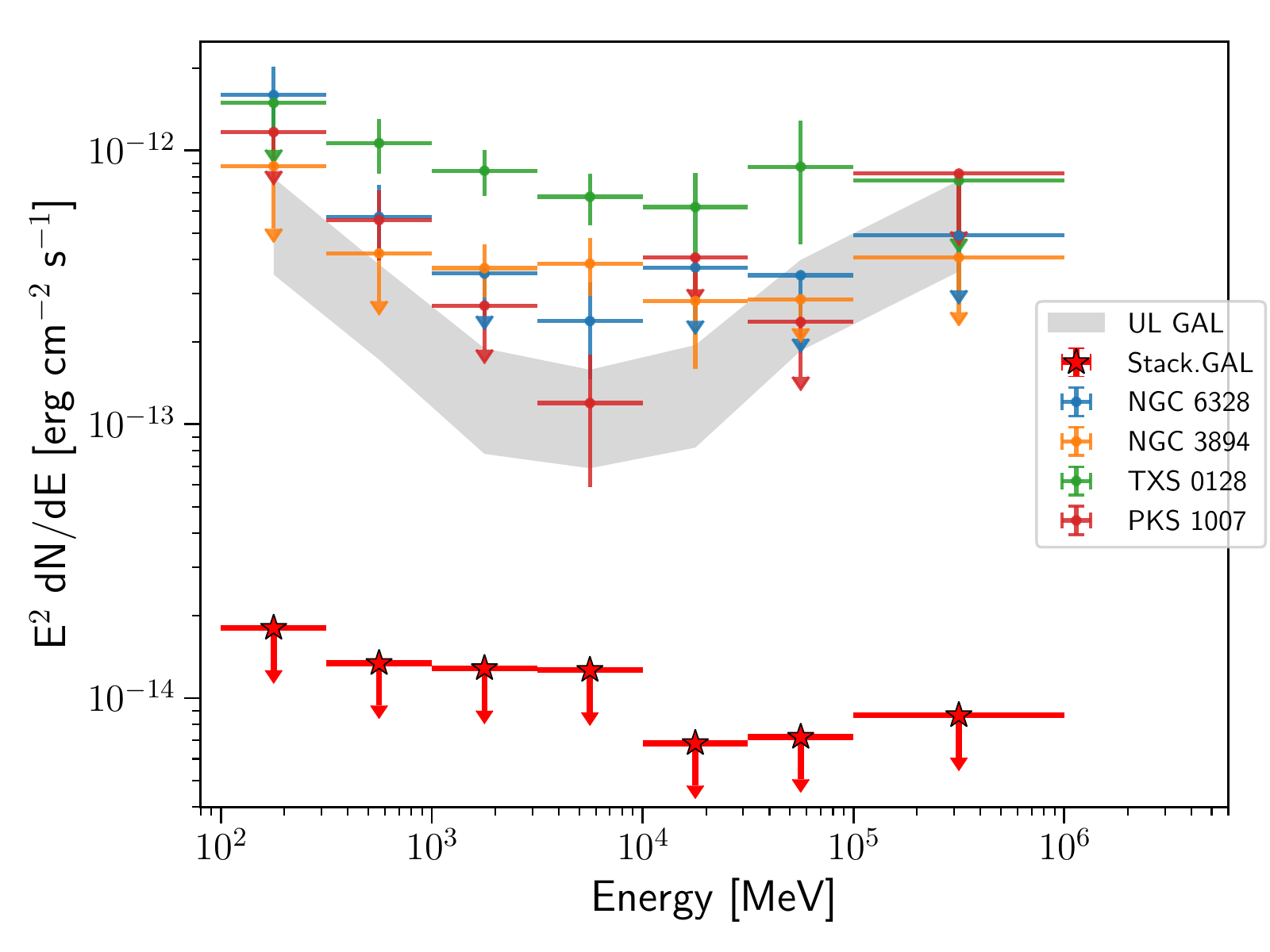}}}
\hspace{0.1cm}
\rotatebox{0}{\resizebox{!}{65mm}{\includegraphics{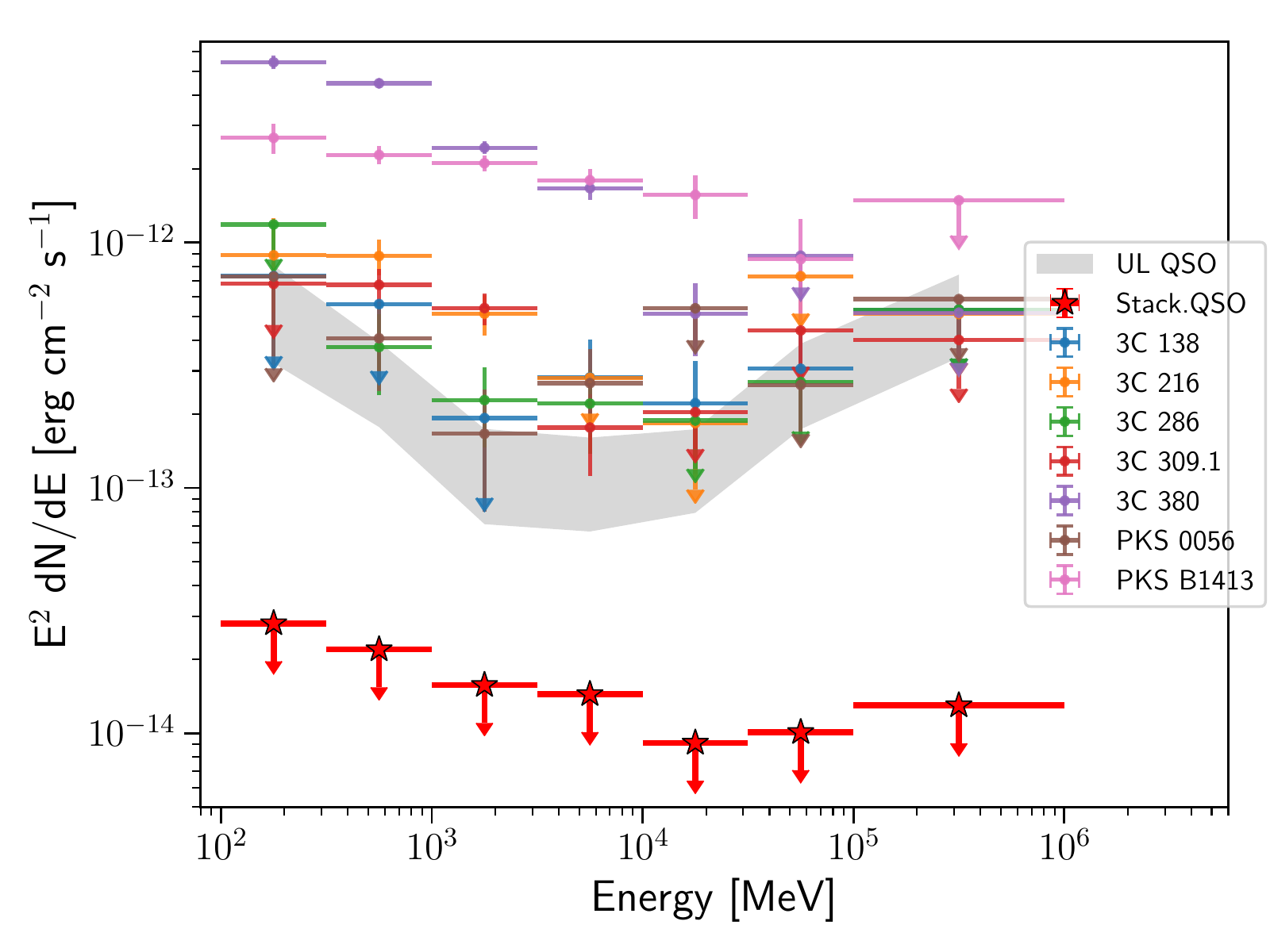}}}
\caption{\small  \label{fig_stacking_sed_gal} {\rm Left panel}: Upper limits for the undetected young radio galaxies as determined from the stacking analysis, compared to the averaged upper limits of the individual undetected radio galaxies (grey band) and the SEDs of the detected ones. {\rm Right panel} panel: the same as for the {\rm left} panel but for quasars.}
\end{center}
\end{figure*}

In order to estimate the upper limits for different energy bins of the undetected sources, we repeat the stacking analysis for seven separate energy ranges, six bands logarithmically spaced between 100\,MeV and 100\,GeV, and a single one between\,100 GeV and 1\,TeV.
The photon index was kept fixed to the value found from the full energy range analysis, i.e. the value reported in Table \ref{table_stacking}.
Fig. \ref{fig_stacking_sed_gal}-left (right) shows the upper limits for the SED of the undetected sources associated with galaxies (quasars), derived with the stacking analysis, compared to the averaged upper limits on the individual undetected galaxies (quasars) and to the SED of the detected ones.

\subsubsection{Stacking analysis of selected sub-samples}
We searched for a possible detection with the stacking analysis using different sub-samples defined by selections of the physical properties. 
We considered the stacking analysis for several sub-samples of nearby ($z$ < 0.07, 0.15, 0.4, and 1) and compact (LS < 0.35, 0.5 and 1 kpc) sources.
Additionally, we also performed the stacking analysis on the sample of \textit{bona-fide} young radio sources from \citet{2014MNRAS.438..463O}, but no significant gamma-ray emission was observed.
Table \ref{table_stacking_subsamples} contains the results of the stacking analysis for each sub-sample based on the different parameters selected.

\begin{table}
\caption{\small \label{table_stacking_subsamples} Results of the stacking analysis for different sub-samples defined by selections on the linear size and redshift ranges. In the last row, labeled as 'O14', are reported the results of the stacking analysis for the sample of \textit{bona-fide} young radio sources created by \citet{2014MNRAS.438..463O}. The results of young radio galaxies are reported on the left while on the right those ones of the quasars. $F^{*}_{\gamma}$: upper limits on the flux in units of 10$^{-11}$ ph cm$^{-2}$ s$^{-1}$.
}
\small
\centering
\hspace*{-0.2cm}
\begin{tabular}{c|cccc|cccc}
\hline \hline
\multicolumn{1}{c}{ } & \multicolumn{4}{c}{Galaxies} &  \multicolumn{4}{c}{Quasars} \\ 
Select. & N & TS & $F^{*}_{\gamma}$ & $\Gamma$ &  N & TS & $F^{*}_{\gamma}$ & $\Gamma$ \\
\hline
LS < 0.35 & 48 & 0.1 & 7.2 & 2.36 & 9 & 0.0 & 58.6 & 2.62\\
LS < 0.5 & 52 & 0.1 & 9.4 & 2.38 & 13 & 0.1 & 46.1 & 2.58\\
LS < 1 & 58 & 0.1 & 11.3 & 2.36 & 17 & 0.1 & 68.7 & 2.68\\
$z$ < 0.07 & 10 & 0.5 & 108 & 2.66 & 0 & - & - & -\\
$z$ < 0.15 & 36 & 0.1 & 58.7 & 2.60 & 0 & - & - & -\\
$z$ < 0.4 & 63 & 0.1 & 10.4 & 2.54 & 4 & 0.0 & 23.7 & 2.48\\
$z$ < 1 & 93 & 0.1 & 5.5 & 2.38 & 25 & 0.1 & 21.3 & 2.54\\
\hline
\begin{tabular}{@{}c@{}}$z$ < 0.07 \\ LS < 0.15 \end{tabular}
 & 8 & 0.8 & 172.9 & 2.78 & 0 & - & - & -\\
\hline
\begin{tabular}{@{}c@{}}$z$ < 0.15 \\ LS < 0.35 \end{tabular}
 & 27 & 0.1 & 31.2 & 2.52 & 0 & - & - & -\\
\hline
\begin{tabular}{@{}c@{}}$z$ < 0.4 \\ LS < 0.5 \end{tabular}
 & 39 & 0.1 & 15.4 & 2.50 & 1 & 0.0 & 297.5 & 2.76\\
\hline
O14 & 37 & 0.1 & 4.6 & 2.50 & 15 & 0.1 & 44.8 & 2.80\\
\hline
\end{tabular}
\end{table}

Despite restricting the stacking analysis to the closest and most compact sources (two among the required criteria for a promising gamma-ray emitter according to \citealt{2008ApJ...680..911S}), no significant gamma-ray emission has been observed in the different sub-samples.

\subsubsection{Stacking analysis: comparison with model predictions}
\label{results_coomparison_predictions}
\begin{figure}
\includegraphics[width=\columnwidth]{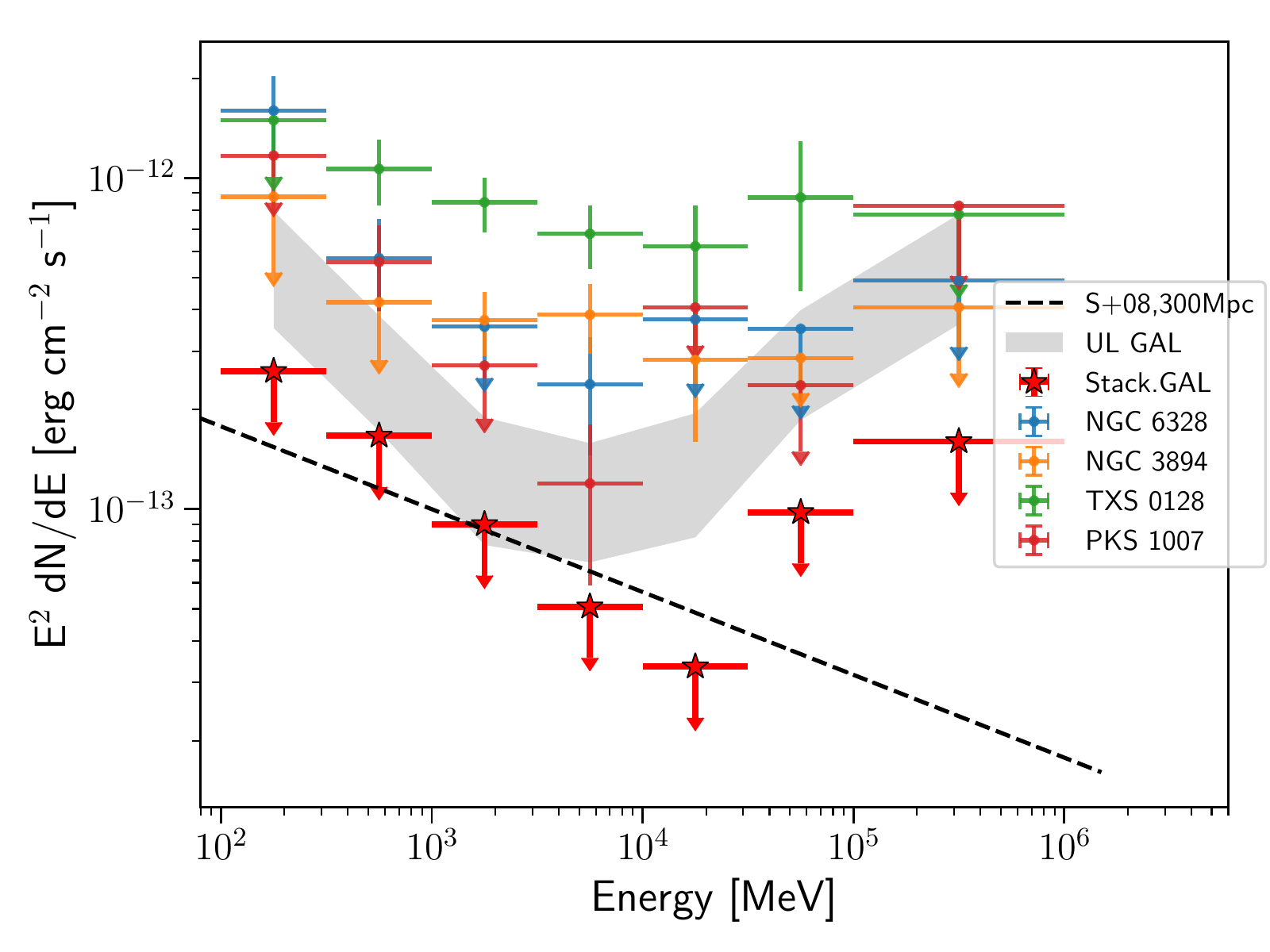}
\caption{\small \label{fig:ul_stacking_near_sources} Upper limits for the undetected young radio galaxies located at z $< 0.07$ ($d_{L} < 300$ Mpc), as determined from the stacking analysis, compared to the expectations for a luminosity distance $d_{L} = 300$ Mpc and projected linear size LS = 100 pc (dashed line) taken from \citet{2008ApJ...680..911S}. The averaged upper limits of the individual undetected radio galaxies (grey band) and the SEDs of the detected young radio galaxies are also plotted.}
\end{figure}

The next test that we made for the analysis was to compare the expectation from the model proposed by \citet{2008ApJ...680..911S} with the results of the stacking analysis. To this aim, we repeated the stacking analysis in energy bins for the sample of sources with $d_{L} \le 300$ Mpc.
In Fig. \ref{fig:ul_stacking_near_sources} we compare the results of this stacking analysis with the model expectation, assuming a source at a luminosity distance $d_{L} = 300$ Mpc. In the model, the gamma-ray emission of the source is due to up-scattering of the UV photons of the disc by the relativistic electrons in the lobes. The predicted luminosity was estimated following \citet{2008ApJ...680..911S}:\\
\begin{align}
\frac{\left(\epsilon {\rm L}_{\epsilon} \right)_{\rm IC/UV}}{10^{42} {\rm erg \ s^{-1}}} \sim  & 2 \frac{\eta_{e}}{\eta_{B}} \left( \frac{{\rm L_{j}}}{{\rm 10^{45} erg \ s^{-1}}}\right)^{1/2} \left( \frac{{\rm LS}}{{\rm 100 pc}} \right)^{-1} \nonumber \\
& \left( \frac{{\rm L}_{\rm UV}}{{\rm 10^{46} erg \ s^{-1}}} \right) \left( \frac{\epsilon}{{\rm 1 GeV}} \right)^{-0.25} 
\label{equation_stawarz}
\end{align}
where $L_{j}$ is the jet power, $L_{UV}$ the UV luminosity and $\eta_{e}/\eta_{B}$ expresses the particle to magnetic field energy density ratio.
The luminosity was calculated for a source with LS = 100 pc, $\eta_{e}/\eta_{B}=1$, $\rm L_{j} = \rm 10^{45} \,erg \ s^{-1}$ and  $\rm L_{UV} = \rm 10^{46}\, erg \ s^{-1}$. We emphasize that the parameters used are very conservative and any larger deviation from the energy equipartition ($\eta_{e}/\eta_{B} >> 1$) will substantially increase the expected IC/UV
flux.\\
For these parameter values, we found that the predicted gamma-ray emission is in tension with the upper limits obtained from the stacking analysis. One possible explanation is that the assumed model parameters are too extreme in terms of e.g. jet power and/or UV luminosity.

Admittedly, the procedure described above is somewhat simplistic as it assumes that all sources have similar fluxes and does not take into account the different values of the physical parameters, such as the linear size, of the sources in the sample. 

\begin{figure}
\includegraphics[width=\columnwidth]{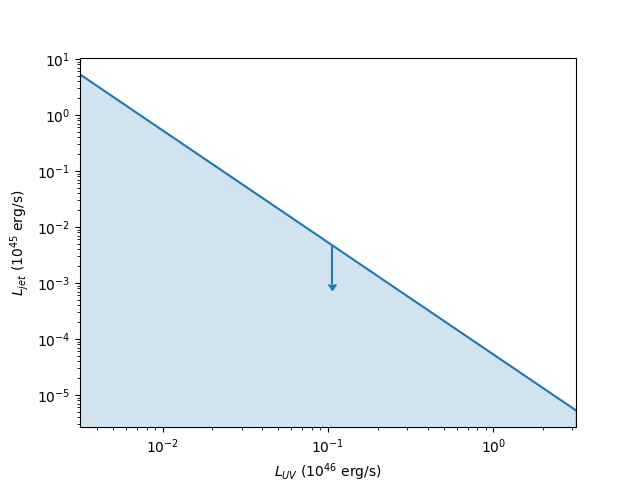}
\caption{\small \label{fig:ul_stacking_rescale} Constraint on the jet power and UV luminosity, estimated with the stacking analysis on undetected galaxies assuming the gamma-ray expectations derived by  \citet{2008ApJ...680..911S}. The blue region defines the allowed values for the two parameters.}
\end{figure}

Following Eq. \ref{equation_stawarz}, we converted the gamma-ray flux upper limits into constraints on the physical parameters of the sources. In particular, we set the known values of $d_{L}$ and LS and derived information on the UV luminosity and jet power from each source.
Since we have two free parameters, we repeated the stacking procedure for several fixed values of the UV luminosity and derived an upper limit on the jet power for each case. This procedure results in an exclusion region in the two-parameter space.
In Fig. \ref{fig:ul_stacking_rescale} the shaded area shows the allowed parameter space in jet power and UV luminosity. For jet powers of 10$^{42}$--10$^{43}$ erg s$^{-1}$ (see Sect. 5), the UV luminosity must be 10$^{45}$ erg s$^{-1}$ or below to be in agreement with the lack of detection in gamma rays.

NGC\,3468 and UGC\,5771\footnote{also known as FIRST\,J105731.1+405646 and 	FIRST\,J103719.3+433515, respectively.} have been excluded from this analysis. In fact, due to their proximity and small size, they would dominate the stacking results.
%

As a further investigation, we checked whether the nearby detected galaxies, NGC\,3894 and NGC\,6328, represent special cases.
We simulated 11 years of Pass 8 data for 100 sources at random positions (see Appendix \ref{verification_stacking_simulation}) assuming the same luminosity and spectral shape as NGC\,3894 ($\Gamma = 2.05$). The simulated sources were located within the nearby Universe at a distance of 300\,Mpc ($z$ = 0.07), six times larger than the redshift of NGC\,3894 ($z$ = 0.0108), and therefore with a simulated flux of $5.5\times 10^{-11}$ ph cm$^{-2}$ s$^{-1}$. 
We performed the stacking analysis on the simulation and found no detection. 
We repeated the procedure for NGC\,6328, simulating sources with $\Gamma = 2.60$, a flux of $2.0\times 10^{-10}$ ph cm$^{-2}$ s$^{-1}$ and located at a distance of 300\,Mpc. Similarly no detection was found with the stacking analysis. 
This indicates that we need a sample of more than a hundred objects with the characteristics of NGC\,3894 and NGC\,6328 up to a distance of 300 Mpc in order to detect significant gamma-ray emission from the stacking analysis.

\begin{figure}
\includegraphics[width=\columnwidth]{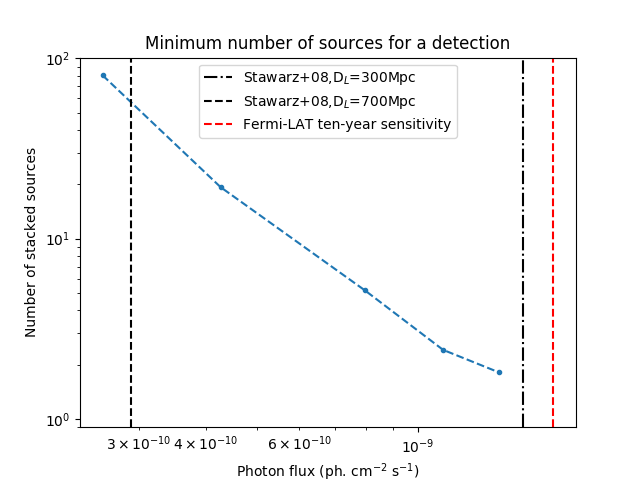}
\caption{\small \label{fig:expectation_stawarz} Number of sources necessary to reach a detection with the stacking procedure as a function of the simulated flux. The vertical black lines show the gamma-ray flux predicted by \citet{2008ApJ...680..911S} for different luminosity distances. See Appendix \ref{verification_stacking_simulation} for more details. The vertical red dashed line represents the averaged \textit{Fermi}-LAT ten-year sensitivity ($F_{E >100\,\textrm{MeV}}=1.8\times 10^{-9}$ MeV cm$^{-2}$ s${-1}$, $\Gamma = 2.0$) for an isolated point source outside the Galactic plane ($\mid b \mid > 25^{\circ}$). }
\end{figure}

Finally, we estimated the number of sources needed to detect (TS$\geq$25) a gamma-ray signal assuming the prediction of \citet{2008ApJ...680..911S}. 
As described in Appendix \ref{verification_stacking_simulation}, we generated five simulated datasets, each one consisting of 100 sources with a fixed input flux in the range ($2 - 20 \times 10^{-10}$ ph cm$^{-2}$ s$^{-1}$) and spectral index of $\Gamma=2.25$ (as in the prediction of \citealt{2008ApJ...680..911S}). For each simulated dataset we applied the stacking procedure (as described in Sect. \ref{sec:analysis_desription}) for an increasing number of sources and estimated the minimum number of sources necessary to reach a detection. 
Fig. \ref{fig:expectation_stawarz} shows these values as a function of the input flux.
We compare these results with the gamma-ray flux expected from galaxies located at luminosity distances $d_L=$ 300 and 700 Mpc. While a young radio galaxy located at $d_L=$ 100 Mpc will be clearly detected ($F_{exp} \sim 1 \times 10^{-8}$ ph cm$^{-2}$ s$^{-1}$), when moving to higher luminosity distances, the expected flux of an individual source is below the LAT sensitivity. 
Therefore, with the stacking procedure, a sample of $\gtrsim$ 2 (60) sources at 300 Mpc (700 Mpc) would be needed to detect a gamma-ray signal.
This estimate is in agreement with the results obtained in our analysis of LAT data, indicating that only the closest sources could be detected by \textit{Fermi}-LAT (see the detection of NGC\,3894, NGC\,6328 and TXS\,0128+554). At higher distances ($\geq$ 300 Mpc), a larger sample of young radio galaxies with characteristics similar to those assumed by \citet{2008ApJ...680..911S}, e.g. LS $\leq$ 100 pc, would be needed to test the lobe scenario.



\section{Discussion}

Before the launch of \textit{Fermi}-LAT, young radio sources were predicted to emerge as a new class of gamma-ray emitting objects. However, after more than ten years of observations, only a handful of sources have been unambiguously detected  \citep{2015ApJ...810...14A,2016ApJ...821L..31M,2020ApJ...892..105A,2020A&A...635A.185P,2020ApJ...899..141L}, with the quasars playing a major role. 

\begin{figure*}
\centering
\includegraphics[width=11cm]{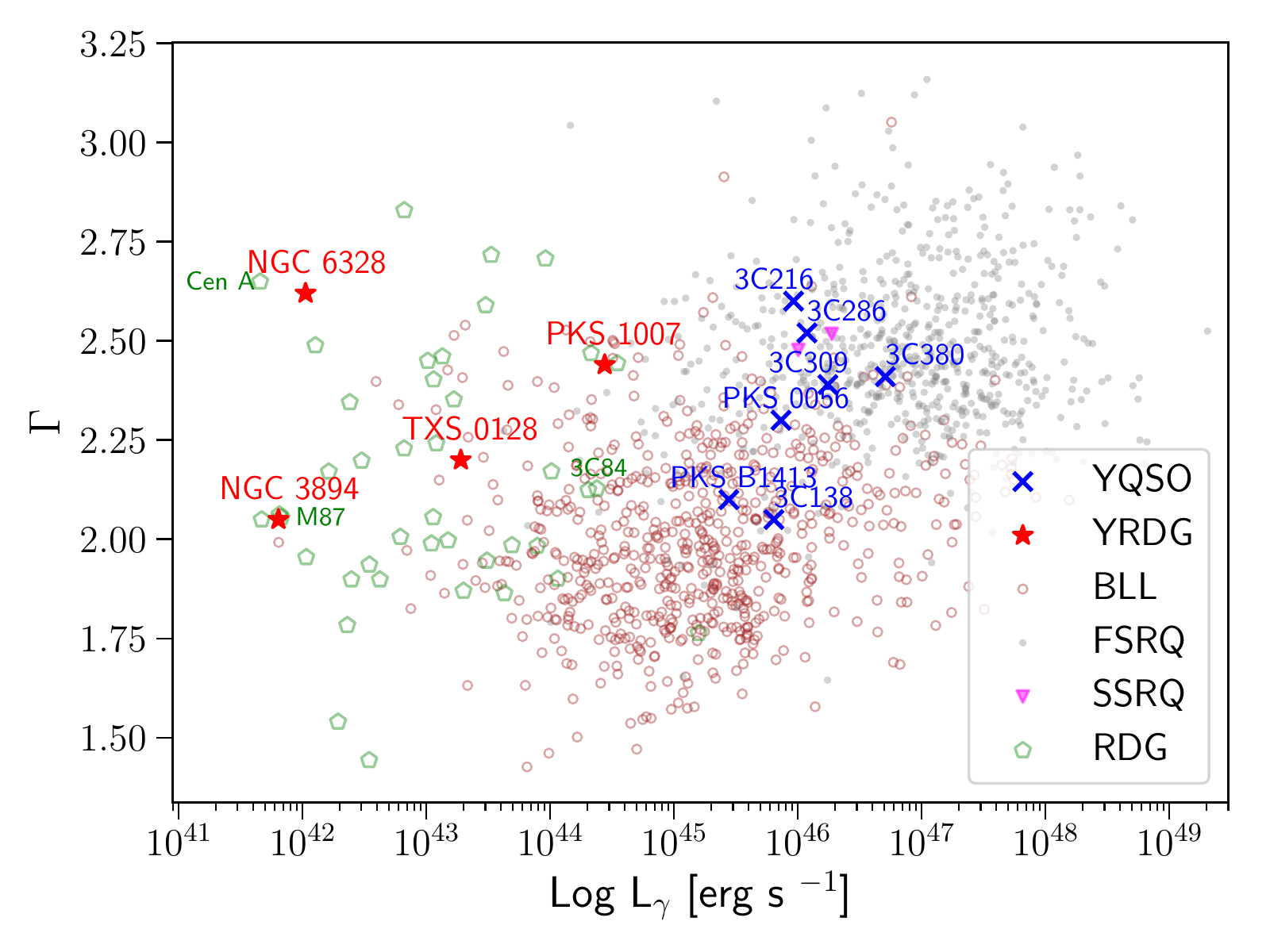}
\caption{\small \label{fig_classification} Diagram of the $\gamma$-ray luminosity vs. photon index for the extragalactic sources with known redshift contained in the 4LAC \citep{2020ApJ...892..105A}. Detected young radio galaxies (YRDG) are represented as red stars, while quasars (YQSO) are represented as blue crosses.} 
\end{figure*}

We analysed 11.3 yr of gamma-ray data collected by {\it Fermi}-LAT for a sample of 162 bona-fide young radio sources. We detect significant gamma-ray emission for 12 per cent (7/59) of quasars, and for 4 per cent (4/103) of galaxies. With a L$_{\gamma}$ between 2.8$\times$10$^{45}$ and 5.1$\times$10$^{46}$ erg s$^{-1}$, quasars are orders of magnitude more luminous than galaxies, whose L$_{\gamma}$ ranges between 6.1$\times$10$^{41}$ and 2.8$\times$10$^{44}$ erg s$^{-1}$. Selection effects likely play a role, in fact quasars are usually at larger redshift with respect to galaxies. However, we note that this difference holds true also for objects at similar redshift, such as the galaxy PKS\,1007$+$142 and the quasar PKS\,B1413$+$135, with the latter being an order of magnitude more luminous in gamma rays than the former\footnote{For this comparison we conservatively assumed a redshift of 0.247.}. One possible explanation is that gamma rays in quasars and radio galaxies have a different origin. In the former, the emission could be produced in the jet, while in the latter in the radio lobes. 
In young radio quasars, effects due to small angles between the jet axis and our line of sight and relativistic jet speed boost the gamma-ray emission (see \citealt{2014ApJ...780..165M}). If we plot the photon index vs L$_{\gamma}$ of the detected sources (Fig. \ref{fig_classification}), we see that young quasars lie in the same region occupied by flat spectrum radio quasars (FSRQs), whereas young galaxies are in the locus of misaligned radio galaxies. Among the misaligned objects there is 3C\,84 \citep[z=0.01756][]{1999PASP..111..438F}, which presents intermittent jet activity and whose last active phase started around 2007 and is still ongoing both in radio and high-energy bands \citep{2010PASJ...62L..11N,2011MNRAS.413.2785B,2018A&A...614A...6S,2015MNRAS.451.4375F,2018ApJ...855...93F,2018MNRAS.475..368H}.

\noindent Further support for boosting effects in quasars comes from the gamma-ray light curves which show significant variability in five quasars (3C\,138, 3C\,216, 3C\,309.1, 3C\,380, and PKS\,B1413$+$135). In particular, for the quasar PKS\,B1413$+$135 we observed a gamma-ray flare at the end of 2019 with a flux increase of a factor of 8.5 with respect to the average value, and up to a factor of 35 if we consider the daily peak flux. These kinds of flux increases have been observed during high activity states in many FSRQs and BL Lacs \citep{2010ApJ...722..520A}.\\
A beamed, jet-origin of gamma-ray emission in quasars is also supported by their radio morphology. The radio emission of the gamma-ray detected quasars mainly comes from the core and the approaching jet. Sometimes the counter-jet is detected at low frequencies only, like in PKS\,B1413$+$135 \citep{1996AJ....111.1839P,2019ApJ...874...43L} and 3C\,138 (Dallacasa et al. in preparation). Superluminal motion of jet features has been observed in 3C\,380, 3C\,309.1 \citep{2019ApJ...874...43L}, and marginally in PKS\,B1413$+$135, confirming the presence of Doppler boosting effects \citep{2005ApJ...622..136G,2019ApJ...874...43L}.
Conversely, gamma-ray detected young radio galaxies show a double lobe-like morphology with minor contribution from the jet \citep{1997AJ....113.2025T,2010MNRAS.408.2261K,2019ApJ...874...43L}, with the possible exception of  NGC\,3894 \citep{Taylor1998}. However, no significant beaming effect seems to take place in this source \citep{2020A&A...635A.185P}. 

Radio lobes may be possible loci for the production of significant gamma-ray emission in unbeamed sources, although a contribution from the jet cannot be ruled out, as suggested for TXS\,0128+554 (Lister et al. 2020). Centaurus A was the first radio source in which gamma-ray emission was unambiguously observed in the lobes \citep{2010Sci...328..725A}, followed some years later by the detection of gamma rays in the lobes of Fornax A \citep{2016ApJ...826....1A}. 
It is difficult to detect high-energy emission from lobes due to their faintness and the challenge of disentangling the lobe contribution from the emission arising from other source regions.

In young radio galaxies, the emission
from highly relativistic regions in the jet is de-beamed due to their
large jet-angles to our lines of sight. Hence, the lobe emission could
dominate in some situations. \citet{2008ApJ...680..911S} predicted different levels of gamma-ray luminosity depending on the radio source size, the jet power and the radiation field of the seed photons which could be Compton-scattered by the relativistic electrons in the lobes. The fact that young radio galaxies are still elusive in gamma rays suggests that the range of values of parameters assumed in the model were too optimistic.

For the radio galaxies in our sample, we estimate the minimum jet power, $P_{\rm j}$ following \citet{2020ApJ...892..116W}:

\begin{align}
P_{\rm j} \sim & 1.5 \times 10^{45} \times \left( \frac{{\rm LS}}{{\rm 100 \,pc}} \right)^{9/7} \left( \frac{\tau_{\rm j}}{{\rm 100 \,yr}} \right)^{-1} \nonumber \\  
& \times \left( \frac{L_{\rm 5 \,GHz}}{10^{42} {\rm erg \,s^{-1}}} \right)^{4/7} {\rm erg\,s^{-1}}
\label{eq_power}
\end{align}

\noindent where LS is the linear size, $\tau_{\rm j}$ is the source age, and L$_{\rm 5 \,GHz}$ is the luminosity at 5 GHz. In Eq. \ref{eq_power} we consider the linear size and the luminosity at 5 GHz reported in Tables \ref{table_sample},\ref{table_sample2} and \ref{table_sample3}. We assume ages between about 100 yr, for the most compact sources, and 10$^5$ yr for sources with LS of several kpc, as derived from radiative and kinematic ages of sources \citep[e.g.][]{1999A&A...345..769M,2002ASPC..258..261F,2003PASA...20...19M,2003PASA...20...69P,2009AN....330..193G}. We end up with minimum jet powers for galaxies between 10$^{40}$ and 10$^{46}$ erg s$^{-1}$. The higher values are obtained for sources at higher redshift, and for 3C\,346, which is among the marginally detected sources from our analysis. However, the majority of the galaxies have L$_{\rm \,5 GHz} < 10^{43}$ 
erg s$^{-1}$ and estimated minimum jet power P$_{\rm j} <10^{44}$ erg s$^{-1}$. As shown in Fig. 11, this requires a UV luminosity above 10$^{45}$ erg s$^{-1}$ in order to detect a cumulative signal by the stacking analysis. This is far from the expectation from \citet{2008ApJ...680..911S}, in which the optimal conditions occurred for sources with jet power $\sim$10$^{46}$ erg s$^{-1}$, LS $<$ 100 pc, and at redshift $<$0.2 ($\sim$1 Gpc). For our estimated jet power, the highest expected gamma-ray luminosity for sources with LS $<$ 100 pc is $<$ 10$^{44}$ erg s$^{-1}$. The fact that young radio galaxies are faint emitters of gamma rays is also suggested by the results of the stacking analysis, which set the upper limit to their emission, as a whole population, an order of magnitude below the \textit{Fermi}-LAT threshold. This indicates that only the closest sources could be detected by \textit{Fermi}-LAT, while if we consider objects at higher and higher redshift, boosting effects are necessary for their detection.

\section{Conclusions}
The goal of our study was to investigate the gamma-ray properties of young-radio sources. To this end we analysed 11.3 years of {\em Fermi}-LAT data for a sample of 162 sources. First, we analysed the gamma-ray data of each source individually to search for a significant detection. Then, we performed a stacking analysis of all undetected sources in order to search for a cumulative signal. For the purpose of our study, having a large sample was the priority, hence we combined different compilations of young radio sources. This resulted in the largest sample of young radio sources used so far for a gamma-ray study (see e.g. \citealt{2014ApJ...780..165M,2016AN....337...59D}). 
The main results can be summarised as follows:
\begin{itemize}
\item{11 sources were significantly (TS $>25$) detected, and we obtained a marginal detection (11 $<$ TS $<$ 25) for nine other sources.
In addition to three radio galaxies and seven quasars already present in the 4FGL, we report the discovery of gamma-ray emission associated with the young radio galaxy PKS\,1007+142;}
\item{young quasars and radio galaxies appear to have distinct gamma-ray properties. On one hand, the former are luminous, typically variable and share the same location as blazar sources in the photon index vs. gamma-ray luminosity plot. Hence, this strongly favors a jet origin for their gamma-ray emission. On the other hand, the latter are compact, gamma-ray faint, not significantly variable and have photon index and L$_{\gamma}$ similar to misaligned sources. Such features are in principle compatible with an origin of the gamma-ray emission (or a fraction of it) in the lobes; }
\item{stacking the LAT data of the remaining below-threshold sources did not result in the detection of a statistically significant signal, neither when the sample were subdivided into bins by redshift nor by linear size. The upper limits obtained with this procedure are, however, tighter than those derived from the individual sources. This enabled a comparison with the model proposed by \citet{2008ApJ...680..911S}, predicting isotropic gamma-ray emission from the compact lobes of young radio galaxies. As a result we can rule out jet powers $\gtrsim$10$^{42}$--10$^{43}$ erg s$^{-1}$ coupled with UV luminosities $>$ 10$^{45}$ erg s$^{-1}$;}
\item{a challenge in the comparison between data and model was presented by the small number of sources at low redshift in our sample. We showed that, already at $\sim$700 Mpc, at least $\gtrsim$60 young radio galaxies would be needed to test the lobe scenario;}
\item{similarly, on the basis of this study we cannot establish whether the undetected sources are gamma-ray analogues of NGC 3894 and NGC 6328, the two closest gamma-ray detected radio galaxies. In fact, at 300 Mpc 100 sources with similar gamma-ray fluxes would still not be sufficient to obtain a stacked signal above the background level.}
\end{itemize}
Our results suggest that young radio sources can be gamma-ray emitters. However, it is likely that
young radio galaxies have typically low gamma-ray luminosities. In the framework of a lobe-origin of such emission, this would rule out sources with powerful jets and bright accretion disks.
For a single source, the detection of gamma-ray emission, either from the jet or the lobes, confirms the presence of a non-thermal component at high energies. This overcomes the limitations that we face in the X-ray band, where the radiation from the accretion and ejection processes cannot be easily disentangled (see e.g. \citealt{2012ApJ...749..107M,2010ApJ...715.1071O,2016ApJ...823...57S} and references therein). In the future, multi-wavelength studies of the gamma-ray detected sources in our sample could be helpful to (i) constrain the source parameters in the first stages of evolution and (ii) to identify the channel through which the bulk of the young source’s energy is released in the ambient environment. The advent of wide-sky surveys will allow a more complete test of the models of high-energy emission from this class of source. A drastic increase in the number of young radio sources will be achieved thanks to the Square Kilometre Array and the next generation VLA, which will expand the samples of young radio sources in redshift and luminosity and improve the characterisation of their radio properties \citep{2015aska.confE.173K,2015aska.confE..71A,2020ApJ...896...18P}.

\section*{Data Availability}
The data \footnote{\textit{Fermi}-LAT data: https://fermi.gsfc.nasa.gov/ssc/data/access/} and tools (\textit{Fermitools}\footnote{https://fermi.gsfc.nasa.gov/ssc/data/analysis/software/} and \textit{fermipy} \footnote{https://fermipy.readthedocs.io/en/latest/}) underlying this article are publicly available.

\subsection*{ACKNOWLEDGMENTS}
The effort of the LAT-team Calibration \& Analysis Working Group to develop Pass 8, and also the LAT AGN Working Group for the support given to this project, are gratefully acknowledged.

The Fermi LAT Collaboration acknowledges generous ongoing support from a number of agencies and institutes that have supported both the development and the operation of the LAT as well as scientific data analysis. These include the National Aeronautics and Space Administration and the Department of Energy in the United States, the Commissariat à l'Energie Atomique and the Centre National de la Recherche Scientifique / Institut National de Physique Nucléaire et de Physique des Particules in France, the Agenzia Spaziale Italiana and the Istituto Nazionale di Fisica Nucleare in Italy, the Ministry of Education, Culture, Sports, Science and Technology (MEXT), High Energy Accelerator Research Organization (KEK) and Japan Aerospace Exploration Agency (JAXA) in Japan, and the K. A. Wallenberg Foundation, the Swedish Research Council and the Swedish National Space Board in Sweden.

Additional support for science analysis during the operations phase is gratefully acknowledged from the Istituto Nazionale di Astrofisica in Italy and the Centre National d'Etudes Spatiales in France. This work is performed in part under DOE Contract DE-AC02-76SF00515.

\bibliographystyle{mnras}
\bibliography{bibliography} 

\begin{thebibliography}{}
\makeatletter
\relax
\def\mn@urlcharsother{\let\do\@makeother \do\$\do\&\do\#\do\^\do\_\do\%\do\~}
\def\mn@doi{\begingroup\mn@urlcharsother \@ifnextchar [ {\mn@doi@}
  {\mn@doi@[]}}
\def\mn@doi@[#1]#2{\def\@tempa{#1}\ifx\@tempa\@empty \href
  {http://dx.doi.org/#2} {doi:#2}\else \href {http://dx.doi.org/#2} {#1}\fi
  \endgroup}
\def\mn@eprint#1#2{\mn@eprint@#1:#2::\@nil}
\def\mn@eprint@arXiv#1{\href {http://arxiv.org/abs/#1} {{\tt arXiv:#1}}}
\def\mn@eprint@dblp#1{\href {http://dblp.uni-trier.de/rec/bibtex/#1.xml}
  {dblp:#1}}
\def\mn@eprint@#1:#2:#3:#4\@nil{\def\@tempa {#1}\def\@tempb {#2}\def\@tempc
  {#3}\ifx \@tempc \@empty \let \@tempc \@tempb \let \@tempb \@tempa \fi \ifx
  \@tempb \@empty \def\@tempb {arXiv}\fi \@ifundefined
  {mn@eprint@\@tempb}{\@tempb:\@tempc}{\expandafter \expandafter \csname
  mn@eprint@\@tempb\endcsname \expandafter{\@tempc}}}

\bibitem[\protect\citeauthoryear{{Abdo} et~al.,}{{Abdo}
  et~al.}{2009}]{2009PhRvD..80l2004A}
{Abdo} A.~A.,  et~al., 2009, \mn@doi [\prd] {10.1103/PhysRevD.80.122004}, \href
  {http://adsabs.harvard.edu/abs/2009PhRvD..80l2004A} {80, 122004}

\bibitem[\protect\citeauthoryear{{Abdo} et~al.,}{{Abdo}
  et~al.}{2010a}]{2010Sci...328..725A}
{Abdo} A.~A.,  et~al., 2010a, \mn@doi [Science] {10.1126/science.1184656},
  \href {https://ui.adsabs.harvard.edu/abs/2010Sci...328..725A} {328, 725}

\bibitem[\protect\citeauthoryear{{Abdo} et~al.,}{{Abdo}
  et~al.}{2010b}]{2010ApJ...720..912A}
{Abdo} A.~A.,  et~al., 2010b, \mn@doi [\apj] {10.1088/0004-637X/720/1/912},
  \href {https://ui.adsabs.harvard.edu/abs/2010ApJ...720..912A} {720, 912}

\bibitem[\protect\citeauthoryear{{Abdo} et~al.,}{{Abdo}
  et~al.}{2010c}]{2010ApJ...722..520A}
{Abdo} A.~A.,  et~al., 2010c, \mn@doi [\apj] {10.1088/0004-637X/722/1/520},
  \href {https://ui.adsabs.harvard.edu/abs/2010ApJ...722..520A} {722, 520}

\bibitem[\protect\citeauthoryear{{Abdollahi} et~al.,}{{Abdollahi}
  et~al.}{2020}]{2020ApJS..247...33A}
{Abdollahi} S.,  et~al., 2020, \mn@doi [\apjs] {10.3847/1538-4365/ab6bcb},
  \href {https://ui.adsabs.harvard.edu/abs/2020ApJS..247...33A} {247, 33}

\bibitem[\protect\citeauthoryear{{Ackermann} et~al.,}{{Ackermann}
  et~al.}{2011}]{2011ApJ...741...30A}
{Ackermann} M.,  et~al., 2011, \mn@doi [\apj] {10.1088/0004-637X/741/1/30},
  \href {https://ui.adsabs.harvard.edu/abs/2011ApJ...741...30A} {741, 30}

\bibitem[\protect\citeauthoryear{{Ackermann} et~al.,}{{Ackermann}
  et~al.}{2015}]{2015ApJ...810...14A}
{Ackermann} M.,  et~al., 2015, \mn@doi [\apj] {10.1088/0004-637X/810/1/14},
  \href {https://ui.adsabs.harvard.edu/abs/2015ApJ...810...14A} {810, 14}

\bibitem[\protect\citeauthoryear{{Ackermann} et~al.,}{{Ackermann}
  et~al.}{2016}]{2016ApJ...826....1A}
{Ackermann} M.,  et~al., 2016, \mn@doi [\apj] {10.3847/0004-637X/826/1/1},
  \href {https://ui.adsabs.harvard.edu/abs/2016ApJ...826....1A} {826, 1}

\bibitem[\protect\citeauthoryear{{Afonso}, {Casanellas}, {Prandoni}, {Jarvis},
  {Lorenzoni}, {Magliocchetti}  \& {Seymour}}{{Afonso}
  et~al.}{2015}]{2015aska.confE..71A}
{Afonso} J.,  {Casanellas} J.,  {Prandoni} I.,  {Jarvis} M.,  {Lorenzoni} S.,
  {Magliocchetti} M.,   {Seymour} N.,  2015, in Advancing Astrophysics with the
  Square Kilometre Array (AASKA14). p.~71 (\mn@eprint {arXiv} {1412.6040})

\bibitem[\protect\citeauthoryear{{Ajello} et~al.,}{{Ajello}
  et~al.}{2020a}]{2020ApJ...892..105A}
{Ajello} M.,  et~al., 2020a, \mn@doi [\apj] {10.3847/1538-4357/ab791e}, \href
  {https://ui.adsabs.harvard.edu/abs/2020ApJ...892..105A} {892, 105}

\bibitem[\protect\citeauthoryear{{Ajello}, {Di Mauro}, {Paliya}  \&
  {Garrappa}}{{Ajello} et~al.}{2020b}]{2020ApJ...894...88A}
{Ajello} M.,  {Di Mauro} M.,  {Paliya} V.~S.,   {Garrappa} S.,  2020b, \mn@doi
  [\apj] {10.3847/1538-4357/ab86a6}, \href
  {https://ui.adsabs.harvard.edu/abs/2020ApJ...894...88A} {894, 88}

\bibitem[\protect\citeauthoryear{{Angioni}, {Cheung}  \& {Buson}}{{Angioni}
  et~al.}{2019}]{2019ATel13049....1A}
{Angioni} R.,  {Cheung} C.~C.,   {Buson} S.,  2019, The Astronomer's Telegram,
  \href {https://ui.adsabs.harvard.edu/abs/2019ATel13049....1A} {13049, 1}

\bibitem[\protect\citeauthoryear{{Atwood} et~al.,}{{Atwood}
  et~al.}{2009}]{2009ApJ...697.1071A}
{Atwood} W.~B.,  et~al., 2009, \mn@doi [\apj] {10.1088/0004-637X/697/2/1071},
  \href {http://adsabs.harvard.edu/abs/2009ApJ...697.1071A} {697, 1071}

\bibitem[\protect\citeauthoryear{{Brown} \& {Adams}}{{Brown} \&
  {Adams}}{2011}]{2011MNRAS.413.2785B}
{Brown} A.~M.,  {Adams} J.,  2011, \mn@doi [\mnras]
  {10.1111/j.1365-2966.2011.18351.x}, \href
  {https://ui.adsabs.harvard.edu/abs/2011MNRAS.413.2785B} {413, 2785}

\bibitem[\protect\citeauthoryear{{Bruel}, {Burnett}, {Digel}, {Johannesson},
  {Omodei}  \& {Wood}}{{Bruel} et~al.}{2018}]{2018arXiv181011394B}
{Bruel} P.,  {Burnett} T.~H.,  {Digel} S.~W.,  {Johannesson} G.,  {Omodei} N.,
   {Wood} M.,  2018, arXiv e-prints, \href
  {https://ui.adsabs.harvard.edu/abs/2018arXiv181011394B} {p. arXiv:1810.11394}

\bibitem[\protect\citeauthoryear{{Carilli}, {Perlman}  \& {Stocke}}{{Carilli}
  et~al.}{1992}]{1992ApJ...400L..13C}
{Carilli} C.~L.,  {Perlman} E.~S.,   {Stocke} J.~T.,  1992, \mn@doi [\apjl]
  {10.1086/186637}, \href
  {https://ui.adsabs.harvard.edu/abs/1992ApJ...400L..13C} {400, L13}

\bibitem[\protect\citeauthoryear{{D'Ammando} et~al.,}{{D'Ammando}
  et~al.}{2015}]{2015MNRAS.450.3975D}
{D'Ammando} F.,  et~al., 2015, \mn@doi [\mnras] {10.1093/mnras/stv909}, \href
  {https://ui.adsabs.harvard.edu/abs/2015MNRAS.450.3975D} {450, 3975}

\bibitem[\protect\citeauthoryear{{D'Ammando}, {Orienti}, {Giroletti}  \& {Fermi
  Large Area Telescope Collaboration}}{{D'Ammando}
  et~al.}{2016}]{2016AN....337...59D}
{D'Ammando} F.,  {Orienti} M.,  {Giroletti} M.,   {Fermi Large Area Telescope
  Collaboration} 2016, \mn@doi [Astronomische Nachrichten]
  {10.1002/asna.201512265}, \href
  {https://ui.adsabs.harvard.edu/abs/2016AN....337...59D} {337, 59}

\bibitem[\protect\citeauthoryear{{Fabian}, {Lohfink}, {Kara}, {Parker},
  {Vasudevan}  \& {Reynolds}}{{Fabian} et~al.}{2015}]{2015MNRAS.451.4375F}
{Fabian} A.~C.,  {Lohfink} A.,  {Kara} E.,  {Parker} M.~L.,  {Vasudevan} R.,
  {Reynolds} C.~S.,  2015, \mn@doi [\mnras] {10.1093/mnras/stv1218}, \href
  {https://ui.adsabs.harvard.edu/abs/2015MNRAS.451.4375F} {451, 4375}

\bibitem[\protect\citeauthoryear{{Falco} et~al.,}{{Falco}
  et~al.}{1999}]{1999PASP..111..438F}
{Falco} E.~E.,  et~al., 1999, \mn@doi [\pasp] {10.1086/316343}, \href
  {https://ui.adsabs.harvard.edu/abs/1999PASP..111..438F} {111, 438}

\bibitem[\protect\citeauthoryear{{Fanti} \& {Fanti}}{{Fanti} \&
  {Fanti}}{2002}]{2002ASPC..258..261F}
{Fanti} C.,  {Fanti} R.,  2002, in {Maiolino} R.,  {Marconi} A.,   {Nagar} N.,
  eds,  Astronomical Society of the Pacific Conference Series Vol. 258, Issues
  in Unification of Active Galactic Nuclei. p.~261

\bibitem[\protect\citeauthoryear{{Fanti}, {Fanti}, {Schilizzi}, {Spencer}, {Nan
  Rendong}, {Parma}, {van Breugel}  \& {Venturi}}{{Fanti}
  et~al.}{1990}]{1990A&A...231..333F}
{Fanti} R.,  {Fanti} C.,  {Schilizzi} R.~T.,  {Spencer} R.~E.,  {Nan Rendong}
  {Parma} P.,  {van Breugel} W.~J.~M.,   {Venturi} T.,  1990, \aap, \href
  {https://ui.adsabs.harvard.edu/abs/1990A&A...231..333F} {231, 333}

\bibitem[\protect\citeauthoryear{{Fanti}, {Fanti}, {Dallacasa}, {Schilizzi},
  {Spencer}  \& {Stanghellini}}{{Fanti} et~al.}{1995}]{1995A&A...302..317F}
{Fanti} C.,  {Fanti} R.,  {Dallacasa} D.,  {Schilizzi} R.~T.,  {Spencer} R.~E.,
    {Stanghellini} C.,  1995, \aap, \href
  {https://ui.adsabs.harvard.edu/abs/1995A&A...302..317F} {302, 317}

\bibitem[\protect\citeauthoryear{{Fukazawa} et~al.,}{{Fukazawa}
  et~al.}{2018}]{2018ApJ...855...93F}
{Fukazawa} Y.,  et~al., 2018, \mn@doi [\apj] {10.3847/1538-4357/aaabc0}, \href
  {https://ui.adsabs.harvard.edu/abs/2018ApJ...855...93F} {855, 93}

\bibitem[\protect\citeauthoryear{{Ghisellini} \& {Madau}}{{Ghisellini} \&
  {Madau}}{1996}]{1996MNRAS.280...67G}
{Ghisellini} G.,  {Madau} P.,  1996, \mn@doi [\mnras] {10.1093/mnras/280.1.67},
  \href {https://ui.adsabs.harvard.edu/abs/1996MNRAS.280...67G} {280, 67}

\bibitem[\protect\citeauthoryear{{Giroletti} \& {Polatidis}}{{Giroletti} \&
  {Polatidis}}{2009}]{2009AN....330..193G}
{Giroletti} M.,  {Polatidis} A.,  2009, \mn@doi [Astronomische Nachrichten]
  {10.1002/asna.200811154}, \href
  {https://ui.adsabs.harvard.edu/abs/2009AN....330..193G} {330, 193}

\bibitem[\protect\citeauthoryear{{Gugliucci}, {Taylor}, {Peck}  \&
  {Giroletti}}{{Gugliucci} et~al.}{2005}]{2005ApJ...622..136G}
{Gugliucci} N.~E.,  {Taylor} G.~B.,  {Peck} A.~B.,   {Giroletti} M.,  2005,
  \mn@doi [\apj] {10.1086/427934}, \href
  {https://ui.adsabs.harvard.edu/abs/2005ApJ...622..136G} {622, 136}

\bibitem[\protect\citeauthoryear{{Harwood} et~al.,}{{Harwood}
  et~al.}{2017}]{2017MNRAS.469..639H}
{Harwood} J.~J.,  et~al., 2017, \mn@doi [\mnras] {10.1093/mnras/stx820}, \href
  {https://ui.adsabs.harvard.edu/abs/2017MNRAS.469..639H} {469, 639}

\bibitem[\protect\citeauthoryear{{Hodgson} et~al.,}{{Hodgson}
  et~al.}{2018}]{2018MNRAS.475..368H}
{Hodgson} J.~A.,  et~al., 2018, \mn@doi [\mnras] {10.1093/mnras/stx3041}, \href
  {https://ui.adsabs.harvard.edu/abs/2018MNRAS.475..368H} {475, 368}

\bibitem[\protect\citeauthoryear{{Jamrozy}, {Machalski}, {Mack}  \&
  {Klein}}{{Jamrozy} et~al.}{2005}]{2005A&A...433..467J}
{Jamrozy} M.,  {Machalski} J.,  {Mack} K.~H.,   {Klein} U.,  2005, \mn@doi
  [\aap] {10.1051/0004-6361:20042007}, \href
  {https://ui.adsabs.harvard.edu/abs/2005A&A...433..467J} {433, 467}

\bibitem[\protect\citeauthoryear{{Kapinska}, {Hardcastle}, {Jackson}, {An},
  {Baan}  \& {Jarvis}}{{Kapinska} et~al.}{2015}]{2015aska.confE.173K}
{Kapinska} A.~D.,  {Hardcastle} M.,  {Jackson} C.,  {An} T.,  {Baan} W.,
  {Jarvis} M.,  2015, in Advancing Astrophysics with the Square Kilometre Array
  (AASKA14). p.~173 (\mn@eprint {arXiv} {1412.5884})

\bibitem[\protect\citeauthoryear{{Kino} \& {Asano}}{{Kino} \&
  {Asano}}{2011}]{2011MNRAS.412L..20K}
{Kino} M.,  {Asano} K.,  2011, \mn@doi [\mnras]
  {10.1111/j.1745-3933.2010.00996.x}, \href
  {https://ui.adsabs.harvard.edu/abs/2011MNRAS.412L..20K} {412, L20}

\bibitem[\protect\citeauthoryear{{Kosmaczewski} et~al.,}{{Kosmaczewski}
  et~al.}{2020}]{2020ApJ...897..164K}
{Kosmaczewski} E.,  et~al., 2020, \mn@doi [\apj] {10.3847/1538-4357/ab9b1f},
  \href {https://ui.adsabs.harvard.edu/abs/2020ApJ...897..164K} {897, 164}

\bibitem[\protect\citeauthoryear{{Kunert-Bajraszewska}, {Gawro{\'n}ski},
  {Labiano}  \& {Siemiginowska}}{{Kunert-Bajraszewska}
  et~al.}{2010}]{2010MNRAS.408.2261K}
{Kunert-Bajraszewska} M.,  {Gawro{\'n}ski} M.~P.,  {Labiano} A.,
  {Siemiginowska} A.,  2010, \mn@doi [\mnras]
  {10.1111/j.1365-2966.2010.17271.x}, \href
  {https://ui.adsabs.harvard.edu/abs/2010MNRAS.408.2261K} {408, 2261}

\bibitem[\protect\citeauthoryear{{Liao} \& {Gu}}{{Liao} \&
  {Gu}}{2020}]{2020MNRAS.491...92L}
{Liao} M.,  {Gu} M.,  2020, \mn@doi [\mnras] {10.1093/mnras/stz2981}, \href
  {https://ui.adsabs.harvard.edu/abs/2020MNRAS.491...92L} {491, 92}

\bibitem[\protect\citeauthoryear{{Lico}, {Giroletti}, {Orienti}, {Costamante},
  {Pavlidou}, {D'Ammando}  \& {Tavecchio}}{{Lico}
  et~al.}{2017}]{2017A&A...606A.138L}
{Lico} R.,  {Giroletti} M.,  {Orienti} M.,  {Costamante} L.,  {Pavlidou} V.,
  {D'Ammando} F.,   {Tavecchio} F.,  2017, \mn@doi [\aap]
  {10.1051/0004-6361/201731116}, \href
  {https://ui.adsabs.harvard.edu/abs/2017A&A...606A.138L} {606, A138}

\bibitem[\protect\citeauthoryear{{Lister} et~al.,}{{Lister}
  et~al.}{2019}]{2019ApJ...874...43L}
{Lister} M.~L.,  et~al., 2019, \mn@doi [\apj] {10.3847/1538-4357/ab08ee}, \href
  {https://ui.adsabs.harvard.edu/abs/2019ApJ...874...43L} {874, 43}

\bibitem[\protect\citeauthoryear{{Lister}, {Homan}, {Kovalev}, {Mand al},
  {Pushkarev}  \& {Siemiginowska}}{{Lister} et~al.}{2020}]{2020ApJ...899..141L}
{Lister} M.~L.,  {Homan} D.~C.,  {Kovalev} Y.~Y.,  {Mand al} S.,  {Pushkarev}
  A.~B.,   {Siemiginowska} A.,  2020, \mn@doi [\apj]
  {10.3847/1538-4357/aba18d}, \href
  {https://ui.adsabs.harvard.edu/abs/2020ApJ...899..141L} {899, 141}

\bibitem[\protect\citeauthoryear{{Maraschi}, {Ghisellini}  \&
  {Celotti}}{{Maraschi} et~al.}{1992}]{1992ApJ...397L...5M}
{Maraschi} L.,  {Ghisellini} G.,   {Celotti} A.,  1992, \mn@doi [\apjl]
  {10.1086/186531}, \href
  {https://ui.adsabs.harvard.edu/abs/1992ApJ...397L...5M} {397, L5}

\bibitem[\protect\citeauthoryear{{Massaro}, {Thompson}  \& {Ferrara}}{{Massaro}
  et~al.}{2015}]{2015A&ARv..24....2M}
{Massaro} F.,  {Thompson} D.~J.,   {Ferrara} E.~C.,  2015, \mn@doi [\aapr]
  {10.1007/s00159-015-0090-6}, \href
  {https://ui.adsabs.harvard.edu/abs/2015A&ARv..24....2M} {24, 2}

\bibitem[\protect\citeauthoryear{{Mattox} et~al.,}{{Mattox}
  et~al.}{1996}]{1996ApJ...461..396M}
{Mattox} J.~R.,  et~al., 1996, \mn@doi [\apj] {10.1086/177068}, \href
  {https://ui.adsabs.harvard.edu/abs/1996ApJ...461..396M} {461, 396}

\bibitem[\protect\citeauthoryear{{Migliori}, {Siemiginowska}  \&
  {Celotti}}{{Migliori} et~al.}{2012}]{2012ApJ...749..107M}
{Migliori} G.,  {Siemiginowska} A.,   {Celotti} A.,  2012, \mn@doi [\apj]
  {10.1088/0004-637X/749/2/107}, \href
  {https://ui.adsabs.harvard.edu/abs/2012ApJ...749..107M} {749, 107}

\bibitem[\protect\citeauthoryear{{Migliori}, {Siemiginowska}, {Kelly},
  {Stawarz}, {Celotti}  \& {Begelman}}{{Migliori}
  et~al.}{2014}]{2014ApJ...780..165M}
{Migliori} G.,  {Siemiginowska} A.,  {Kelly} B.~C.,  {Stawarz} {\L}.,
  {Celotti} A.,   {Begelman} M.~C.,  2014, \mn@doi [\apj]
  {10.1088/0004-637X/780/2/165}, \href
  {https://ui.adsabs.harvard.edu/abs/2014ApJ...780..165M} {780, 165}

\bibitem[\protect\citeauthoryear{{Migliori}, {Siemiginowska}, {Sobolewska},
  {Loh}, {Corbel}, {Ostorero}  \& {Stawarz}}{{Migliori}
  et~al.}{2016}]{2016ApJ...821L..31M}
{Migliori} G.,  {Siemiginowska} A.,  {Sobolewska} M.,  {Loh} A.,  {Corbel} S.,
  {Ostorero} L.,   {Stawarz} {\L}.,  2016, \mn@doi [\apj]
  {10.3847/2041-8205/821/2/L31}, \href
  {https://ui.adsabs.harvard.edu/abs/2016ApJ...821L..31M} {821, L31}

\bibitem[\protect\citeauthoryear{{Murgia}}{{Murgia}}{2003}]{2003PASA...20...19M}
{Murgia} M.,  2003, \mn@doi [\pasa] {10.1071/AS02033}, \href
  {https://ui.adsabs.harvard.edu/abs/2003PASA...20...19M} {20, 19}

\bibitem[\protect\citeauthoryear{{Murgia}, {Fanti}, {Fanti}, {Gregorini},
  {Klein}, {Mack}  \& {Vigotti}}{{Murgia} et~al.}{1999}]{1999A&A...345..769M}
{Murgia} M.,  {Fanti} C.,  {Fanti} R.,  {Gregorini} L.,  {Klein} U.,  {Mack}
  K.~H.,   {Vigotti} M.,  1999, \aap, \href
  {https://ui.adsabs.harvard.edu/abs/1999A&A...345..769M} {345, 769}

\bibitem[\protect\citeauthoryear{{Nagai} et~al.,}{{Nagai}
  et~al.}{2010}]{2010PASJ...62L..11N}
{Nagai} H.,  et~al., 2010, \mn@doi [\pasj] {10.1093/pasj/62.2.L11}, \href
  {https://ui.adsabs.harvard.edu/abs/2010PASJ...62L..11N} {62, L11}

\bibitem[\protect\citeauthoryear{{Nolan} et~al.,}{{Nolan}
  et~al.}{2012}]{2012ApJS..199...31N}
{Nolan} P.~L.,  et~al., 2012, \mn@doi [\apjs] {10.1088/0067-0049/199/2/31},
  \href {https://ui.adsabs.harvard.edu/abs/2012ApJS..199...31N} {199, 31}

\bibitem[\protect\citeauthoryear{{O'Dea}}{{O'Dea}}{1998}]{1998PASP..110..493O}
{O'Dea} C.~P.,  1998, \mn@doi [\pasp] {10.1086/316162}, \href
  {https://ui.adsabs.harvard.edu/abs/1998PASP..110..493O} {110, 493}

\bibitem[\protect\citeauthoryear{{O'Dea} \& {Baum}}{{O'Dea} \&
  {Baum}}{1997}]{1997AJ....113..148O}
{O'Dea} C.~P.,  {Baum} S.~A.,  1997, \mn@doi [\aj] {10.1086/118241}, \href
  {https://ui.adsabs.harvard.edu/abs/1997AJ....113..148O} {113, 148}

\bibitem[\protect\citeauthoryear{{O'Dea} \& {Saikia}}{{O'Dea} \&
  {Saikia}}{2021}]{2021A&ARv..29....3O}
{O'Dea} C.~P.,  {Saikia} D.~J.,  2021, \mn@doi [\aapr]
  {10.1007/s00159-021-00131-w}, \href
  {https://ui.adsabs.harvard.edu/abs/2021A&ARv..29....3O} {29, 3}

\bibitem[\protect\citeauthoryear{{Orienti} \& {Dallacasa}}{{Orienti} \&
  {Dallacasa}}{2008}]{2008A&A...487..885O}
{Orienti} M.,  {Dallacasa} D.,  2008, \mn@doi [\aap]
  {10.1051/0004-6361:200809948}, \href
  {https://ui.adsabs.harvard.edu/abs/2008A&A...487..885O} {487, 885}

\bibitem[\protect\citeauthoryear{{Orienti} \& {Dallacasa}}{{Orienti} \&
  {Dallacasa}}{2014}]{2014MNRAS.438..463O}
{Orienti} M.,  {Dallacasa} D.,  2014, \mn@doi [\mnras] {10.1093/mnras/stt2217},
  \href {https://ui.adsabs.harvard.edu/abs/2014MNRAS.438..463O} {438, 463}

\bibitem[\protect\citeauthoryear{{Orienti}, {Dallacasa}, {Fanti}, {Fanti},
  {Tinti}  \& {Stanghellini}}{{Orienti} et~al.}{2004}]{2004A&A...426..463O}
{Orienti} M.,  {Dallacasa} D.,  {Fanti} C.,  {Fanti} R.,  {Tinti} S.,
  {Stanghellini} C.,  2004, \mn@doi [\aap] {10.1051/0004-6361:20041204}, \href
  {https://ui.adsabs.harvard.edu/abs/2004A&A...426..463O} {426, 463}

\bibitem[\protect\citeauthoryear{{Ostorero} et~al.,}{{Ostorero}
  et~al.}{2010}]{2010ApJ...715.1071O}
{Ostorero} L.,  et~al., 2010, \mn@doi [\apj] {10.1088/0004-637X/715/2/1071},
  \href {https://ui.adsabs.harvard.edu/abs/2010ApJ...715.1071O} {715, 1071}

\bibitem[\protect\citeauthoryear{{Patil} et~al.,}{{Patil}
  et~al.}{2020}]{2020ApJ...896...18P}
{Patil} P.,  et~al., 2020, \mn@doi [\apj] {10.3847/1538-4357/ab9011}, \href
  {https://ui.adsabs.harvard.edu/abs/2020ApJ...896...18P} {896, 18}

\bibitem[\protect\citeauthoryear{{Perlman}, {Carilli}, {Stocke}  \&
  {Conway}}{{Perlman} et~al.}{1996}]{1996AJ....111.1839P}
{Perlman} E.~S.,  {Carilli} C.~L.,  {Stocke} J.~T.,   {Conway} J.,  1996,
  \mn@doi [\aj] {10.1086/117922}, \href
  {https://ui.adsabs.harvard.edu/abs/1996AJ....111.1839P} {111, 1839}

\bibitem[\protect\citeauthoryear{{Phillips} \& {Mutel}}{{Phillips} \&
  {Mutel}}{1982}]{1982A&A...106...21P}
{Phillips} R.~B.,  {Mutel} R.~L.,  1982, \aap, \href
  {https://ui.adsabs.harvard.edu/abs/1982A&A...106...21P} {106, 21}

\bibitem[\protect\citeauthoryear{{Polatidis} \& {Conway}}{{Polatidis} \&
  {Conway}}{2003}]{2003PASA...20...69P}
{Polatidis} A.~G.,  {Conway} J.~E.,  2003, \mn@doi [\pasa] {10.1071/AS02053},
  \href {https://ui.adsabs.harvard.edu/abs/2003PASA...20...69P} {20, 69}

\bibitem[\protect\citeauthoryear{{Principe} \& {Malyshev}}{{Principe} \&
  {Malyshev}}{2017}]{2017AIPC.1792g0016P}
{Principe} G.,  {Malyshev} D.,  2017, in 6th International Symposium on High
  Energy Gamma-Ray Astronomy. p. 070016 (\mn@eprint {arXiv} {1610.01351}),
  \mn@doi{10.1063/1.4969013}

\bibitem[\protect\citeauthoryear{{Principe}, {Malyshev}, {Ballet}  \&
  {Funk}}{{Principe} et~al.}{2018}]{2018A&A...618A..22P}
{Principe} G.,  {Malyshev} D.,  {Ballet} J.,   {Funk} S.,  2018, \mn@doi [\aap]
  {10.1051/0004-6361/201833116}, \href
  {https://ui.adsabs.harvard.edu/abs/2018A&A...618A..22P} {618, A22}

\bibitem[\protect\citeauthoryear{{Principe}, {Malyshev}, {Ballet}  \&
  {Funk}}{{Principe} et~al.}{2019}]{2019RLSFN.tmp....7P}
{Principe} G.,  {Malyshev} D.,  {Ballet} J.,   {Funk} S.,  2019, \mn@doi
  [Rendiconti Lincei. Scienze Fisiche e Naturali] {10.1007/s12210-019-00771-2},
  \href {https://ui.adsabs.harvard.edu/abs/2019RLSFN.tmp....7P} {p.~7}

\bibitem[\protect\citeauthoryear{{Principe} et~al.,}{{Principe}
  et~al.}{2020}]{2020A&A...635A.185P}
{Principe} G.,  et~al., 2020, \mn@doi [\aap] {10.1051/0004-6361/201937049},
  \href {https://ui.adsabs.harvard.edu/abs/2020A&A...635A.185P} {635, A185}

\bibitem[\protect\citeauthoryear{{Readhead}, {Taylor}, {Xu}, {Pearson},
  {Wilkinson}  \& {Polatidis}}{{Readhead} et~al.}{1996}]{1996ApJ...460..612R}
{Readhead} A.~C.~S.,  {Taylor} G.~B.,  {Xu} W.,  {Pearson} T.~J.,  {Wilkinson}
  P.~N.,   {Polatidis} A.~G.,  1996, \mn@doi [\apj] {10.1086/176996}, \href
  {https://ui.adsabs.harvard.edu/abs/1996ApJ...460..612R} {460, 612}

\bibitem[\protect\citeauthoryear{{Readhead} et~al.,}{{Readhead}
  et~al.}{2021}]{2021ApJ...907...61R}
{Readhead} A.~C.~S.,  et~al., 2021, \mn@doi [\apj] {10.3847/1538-4357/abd08c},
  \href {https://ui.adsabs.harvard.edu/abs/2021ApJ...907...61R} {907, 61}

\bibitem[\protect\citeauthoryear{{Rossetti}, {Fanti}, {Fanti}, {Dallacasa}  \&
  {Stanghellini}}{{Rossetti} et~al.}{2006}]{2006A&A...449...49R}
{Rossetti} A.,  {Fanti} C.,  {Fanti} R.,  {Dallacasa} D.,   {Stanghellini} C.,
  2006, \mn@doi [\aap] {10.1051/0004-6361:20053945}, \href
  {https://ui.adsabs.harvard.edu/abs/2006A&A...449...49R} {449, 49}

\bibitem[\protect\citeauthoryear{{Sahakyan}, {Baghmanyan}  \&
  {Zargaryan}}{{Sahakyan} et~al.}{2018}]{2018A&A...614A...6S}
{Sahakyan} N.,  {Baghmanyan} V.,   {Zargaryan} D.,  2018, \mn@doi [\aap]
  {10.1051/0004-6361/201732304}, \href
  {https://ui.adsabs.harvard.edu/abs/2018A&A...614A...6S} {614, A6}

\bibitem[\protect\citeauthoryear{{Siemiginowska}, {Sobolewska}, {Migliori},
  {Guainazzi}, {Hardcastle}, {Ostorero}  \& {Stawarz}}{{Siemiginowska}
  et~al.}{2016}]{2016ApJ...823...57S}
{Siemiginowska} A.,  {Sobolewska} M.,  {Migliori} G.,  {Guainazzi} M.,
  {Hardcastle} M.,  {Ostorero} L.,   {Stawarz} {\L}.,  2016, \mn@doi [\apj]
  {10.3847/0004-637X/823/1/57}, \href
  {https://ui.adsabs.harvard.edu/abs/2016ApJ...823...57S} {823, 57}

\bibitem[\protect\citeauthoryear{{Snellen}, {Mack}, {Schilizzi}  \&
  {Tschager}}{{Snellen} et~al.}{2004}]{2004MNRAS.348..227S}
{Snellen} I.~A.~G.,  {Mack} K.~H.,  {Schilizzi} R.~T.,   {Tschager} W.,  2004,
  \mn@doi [\mnras] {10.1111/j.1365-2966.2004.07337.x}, \href
  {https://ui.adsabs.harvard.edu/abs/2004MNRAS.348..227S} {348, 227}

\bibitem[\protect\citeauthoryear{{Stawarz}, {Ostorero}, {Begelman}, {Moderski},
  {Kataoka}  \& {Wagner}}{{Stawarz} et~al.}{2008}]{2008ApJ...680..911S}
{Stawarz} {\L}.,  {Ostorero} L.,  {Begelman} M.~C.,  {Moderski} R.,  {Kataoka}
  J.,   {Wagner} S.,  2008, \mn@doi [\apj] {10.1086/587781}, \href
  {https://ui.adsabs.harvard.edu/abs/2008ApJ...680..911S} {680, 911}

\bibitem[\protect\citeauthoryear{{Taylor}, {Wrobel}  \& {Vermeulen}}{{Taylor}
  et~al.}{1998}]{Taylor1998}
{Taylor} G.~B.,  {Wrobel} J.~M.,   {Vermeulen} R.~C.,  1998, \mn@doi [\apj]
  {10.1086/305586}, \href
  {https://ui.adsabs.harvard.edu/abs/1998ApJ...498..619T} {498, 619}

\bibitem[\protect\citeauthoryear{{Tingay} et~al.,}{{Tingay}
  et~al.}{1997}]{1997AJ....113.2025T}
{Tingay} S.~J.,  et~al., 1997, \mn@doi [\aj] {10.1086/118414}, \href
  {https://ui.adsabs.harvard.edu/abs/1997AJ....113.2025T} {113, 2025}

\bibitem[\protect\citeauthoryear{{Vedantham} et~al.,}{{Vedantham}
  et~al.}{2017}]{2017ApJ...845...90V}
{Vedantham} H.~K.,  et~al., 2017, \mn@doi [\apj] {10.3847/1538-4357/aa7741},
  \href {https://ui.adsabs.harvard.edu/abs/2017ApJ...845...90V} {845, 90}

\bibitem[\protect\citeauthoryear{{W{\'o}jtowicz}, {Stawarz}, {Cheung},
  {Ostorero}, {Kosmaczewski}  \& {Siemiginowska}}{{W{\'o}jtowicz}
  et~al.}{2020}]{2020ApJ...892..116W}
{W{\'o}jtowicz} A.,  {Stawarz} {\l}.,  {Cheung} C.~C.,  {Ostorero} L.,
  {Kosmaczewski} E.,   {Siemiginowska} A.,  2020, \mn@doi [\apj]
  {10.3847/1538-4357/ab7930}, \href
  {https://ui.adsabs.harvard.edu/abs/2020ApJ...892..116W} {892, 116}

\bibitem[\protect\citeauthoryear{{Wood}, {Caputo}, {Charles}, {Di Mauro},
  {Magill}  \& {Jeremy Perkins for the Fermi-LAT Collaboration}}{{Wood}
  et~al.}{2017}]{2017arXiv170709551W}
{Wood} M.,  {Caputo} R.,  {Charles} E.,  {Di Mauro} M.,  {Magill} J.,   {Jeremy
  Perkins for the Fermi-LAT Collaboration} 2017, preprint, \href
  {http://adsabs.harvard.edu/abs/2017arXiv170709551W} {} (\mn@eprint {arXiv}
  {1707.09551})

\bibitem[\protect\citeauthoryear{{Zhang}, {Zhang}, {Gan}, {Yi}, {Wang}  \&
  {Liang}}{{Zhang} et~al.}{2020}]{2020ApJ...899....2Z}
{Zhang} J.,  {Zhang} H.-M.,  {Gan} Y.-Y.,  {Yi} T.-F.,  {Wang} J.-F.,   {Liang}
  E.-W.,  2020, \mn@doi [\apj] {10.3847/1538-4357/aba2cd}, \href
  {https://ui.adsabs.harvard.edu/abs/2020ApJ...899....2Z} {899, 2}

\bibitem[\protect\citeauthoryear{{de Vries}, {Snellen}, {Schilizzi}, {Mack}  \&
  {Kaiser}}{{de Vries} et~al.}{2009}]{2009A&A...498..641D}
{de Vries} N.,  {Snellen} I.~A.~G.,  {Schilizzi} R.~T.,  {Mack} K.~H.,
  {Kaiser} C.~R.,  2009, \mn@doi [\aap] {10.1051/0004-6361/200811145}, \href
  {https://ui.adsabs.harvard.edu/abs/2009A&A...498..641D} {498, 641}

\makeatother
\end{thebibliography}

\appendix
\section{Appendix}
\label{sec:appendix}

\subsection{Validation of the stacking analysis method}
\subsubsection{Validation on simulated sources}
\label{verification_stacking_simulation}
\begin{figure}[h!]
\begin{center}
\rotatebox{0}{\resizebox{!}{55mm}{\includegraphics{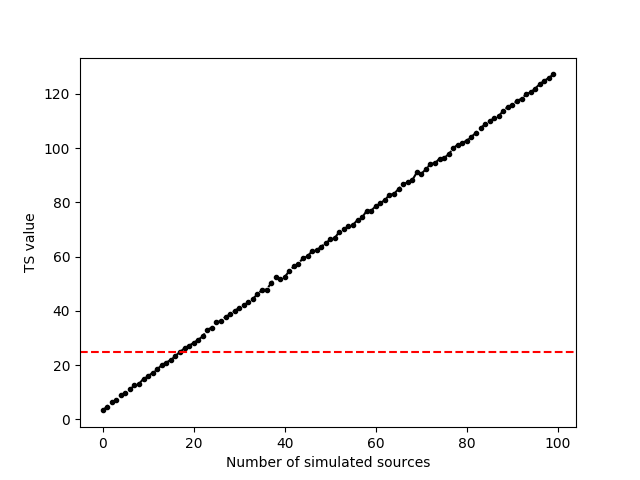}}}
\hspace{0.1cm}
\rotatebox{0}{\resizebox{!}{55mm}{\includegraphics{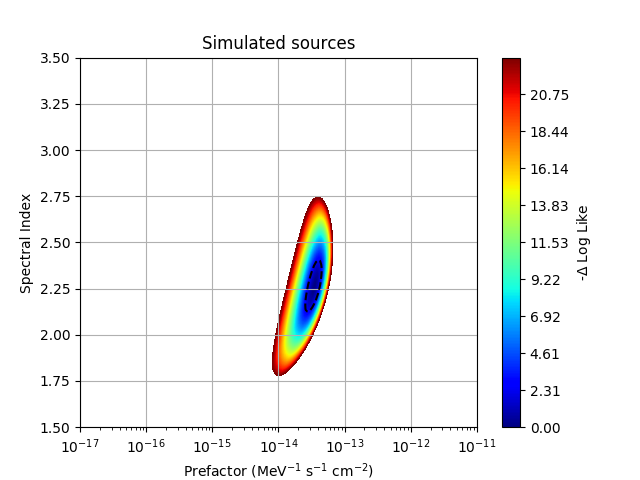}}}
\caption{\small Left: TS value obtained when stacking an increasing number of simulated sources.
Right: Likelihood profile assuming a simple PL
spectrum for the 100 simulated sources using a PL with $\Gamma=2.25$, normalisation $N_{0} = 3 \times 10^{-14}$ (MeV$^{-1}$ s$^{-1}$ cm$^{-2}$) and an energy scale of $E_0=1$ GeV.}
\label{fig_stacking_simulation} 
\end{center}
\end{figure}

To verify the robustness of the stacking method, we simulated 11 years of Pass 8 LAT data for 100 sources at $\mid b \mid > 10 ^{\circ}$, in random positions which lie at a minimum distance of 1$^{\circ}$ from any other source in the 4FGL catalog \citep{2020ApJS..247...33A}. The spectra of the simulated sources are modeled using a power law with spectral index $\Gamma=2.25$, normalisation 
$N_{0} = 3 \times 10^{-14}$ (MeV$^{-1}$ s$^{-1}$ cm$^{-2}$)
and energy scale $E_0=1$ GeV, corresponding to an integrated photon flux of $4.3 \times 10^{-10}$ (ph cm$^{-2}$ s$^{-1}$).
The simulations include the diffuse (Galactic
and isotropic) emission as well as all the point sources from the 4FGL catalog. 
The normalisation of the simulated sources was chosen such that their resulting significance after the analysis lies below 5 $\sigma$. In fact, none of the simulated sources are detected. A stacking analysis was performed for the simulated dataset as described in Sect. \ref{sec:stacking_analysis}. We repeated this analysis for an increasing number of simulated sources and calculated the TS as defined in Sect. \ref{sec:analysis_desription}. The results are presented in Fig. \ref{fig_stacking_simulation} (left),
showing that the TS increases linearly with the number of sources included in the stacking procedure.
Fig. \ref{fig_stacking_simulation} (right) shows the likelihood profile obtained with the full sample of simulated sources, showing a detection with a TS $\sim$ 116. 

[h!]

\noindent The best-fit photon index and normalisation are $\Gamma = 2.26 \pm 0.13$ and $N_0 = (3.5\pm1.0) \times 10^{-14}$ MeV$^{-1}$ s$^{-1}$ cm$^{-2}$, respectively, which are in agreement with the input values of the simulated sources. 
From this test we conclude that the stacking analysis provides a powerful method to detect sources 
of the same population, assuming they have similar spectral properties.

\noindent Similarly, we repeated the simulation and stacking procedure for five flux values, varying $N_0$ in the range ($(1-10) \times 10^{-14}$ MeV$^{-1}$ s$^{-1}$ cm$^{-2}$) and using a PL with index $\Gamma=2.25$. For each simulation we calculated the number of sources necessary to have a detection (TS=25).


\subsubsection{Validation on the background}
\label{verification_stacking_background}
\begin{figure}
\centering
\includegraphics[width=.8\columnwidth]{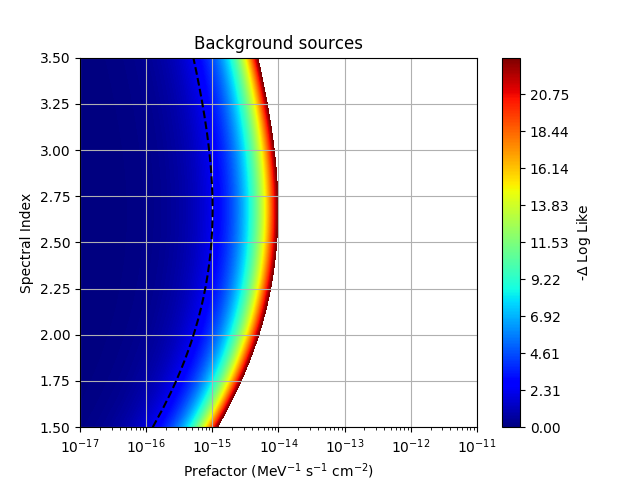}
\caption{\small \label{fig_stacking_background} Likelihood profile assuming a simple power-law spectrum for 100 random positions where no gamma-ray sources are present.}
\end{figure}

In order to verify the quality of our results from applying the stacking analysis and ensure that they could be distinguished from simple background fluctuations, we performed the same analysis for the background.
The analysis was performed using 100 random positions that fulfill the following criteria: at least at 1$^\circ$ away from all 4FGL sources and from the sources in our sample and outside of the Galactic plane, i.e. with Galactic latitude $|b|>10$.
The result of the stacking on the background sources is presented in Fig. \ref{fig_stacking_background} and it shows no significant detection with a TS $\sim$ 0.3. The flux upper limit is calculated from the contour corresponding to the 95\% confidence level, as described in Sect. \ref{sec:stacking_analysis}, resulting in $2.82 \times 10^{-11}$ ph cm$^{-2}$ s$^{-1}$. 

\noindent In order to check the consistency of the results obtained from real data with background fluctuations, we calculated the photon flux upper limit for a smaller number of sources. Indeed, the upper limit value decreases with the increasing number of sources included in the stacking analysis. 
In particular, we performed the stacking analysis on a selection of sources with the same number of below-threshold galaxies (99) and quasars (52) as in our sample.
The resulting photon flux upper limits are $2.85 \times 10^{-11}$ ph  s$^{-1}$ and $8.25 \times 10^{-11}$ ph cm$^{-2}$ s$^{-1}$, respectively.




\subsection{Results on individual young radio sources}
\label{appendix_detected_sources}
\label{appendix_tables}
In this section we present the gamma-ray SED and lightcurves of the detected galaxies and quasars. Fig. \ref{fig:sed_lightcurve1}, \ref{fig:sed_lightcurve2},\ref{fig:sed_lightcurve3} show the SED and light-curve plots of the detected sources.
In Tables \ref{table_sample}, \ref{table_sample2}, and \ref{table_sample3}, we report the physical and radio parameters as well as the resulting gamma-ray characteristics obtained in this work for all sources selected for this study. 

\begin{figure}
\centering
\includegraphics[width=60mm]{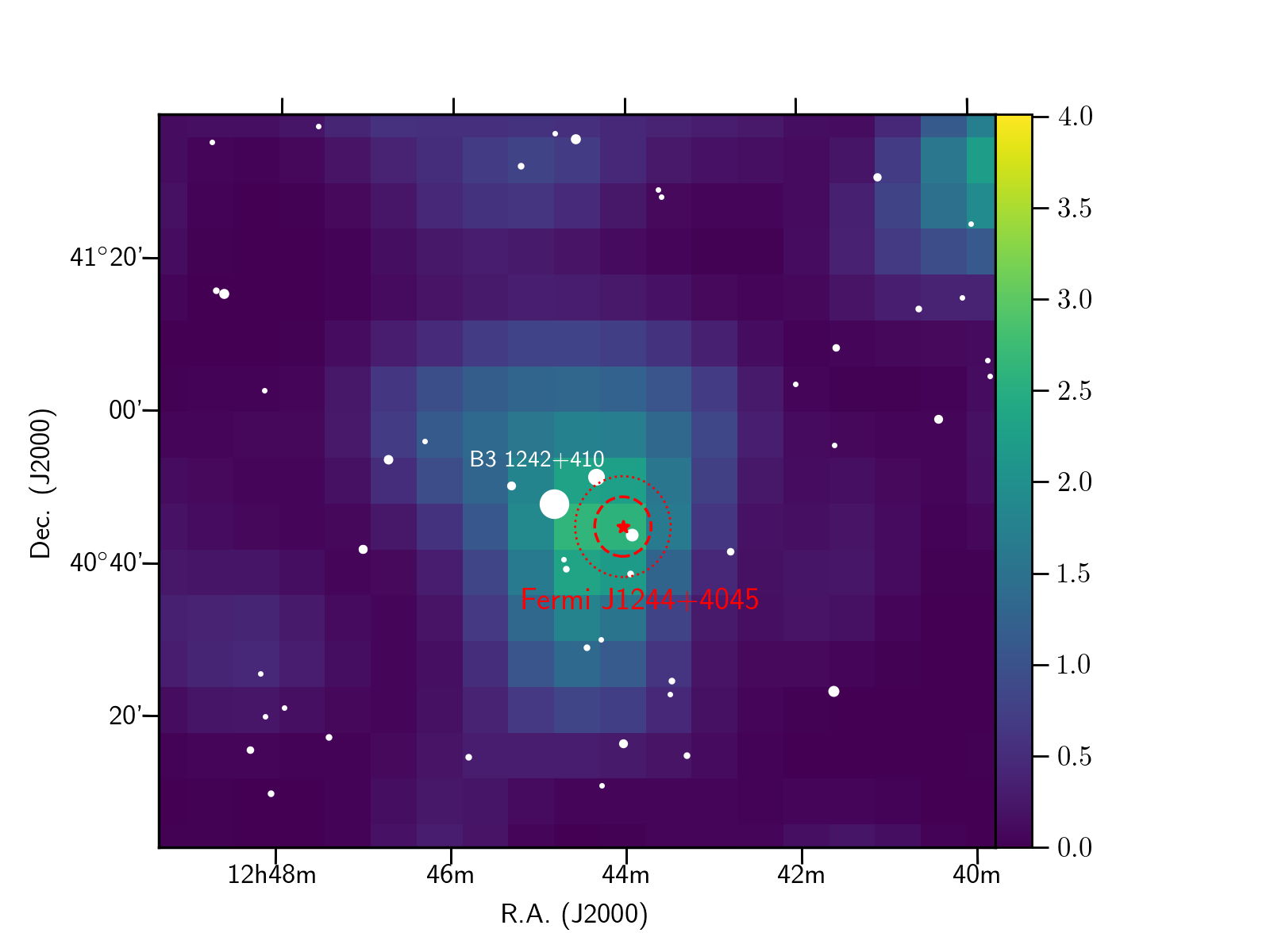}
\caption{\small \textit{Fermi}-LAT TS map (in sigma units) above 1 GeV of the region around B3\,1242$+$410, the red star, dashed and dotted circles represent the central position and the 68\% and 95\% confidence-level uncertainty (R$_{68}=0.07^{\circ}$, R$_{95}=0.12^{\circ}$) of the gamma-ray source, respectively. White dots show radio sources from the NVSS survey arbitrarily scaled depending on their radio flux.}
\label{fig:map_B3_1242}
\end{figure}

\begin{figure*}
\begin{center}
\rotatebox{0}{\resizebox{!}{50mm}{\includegraphics{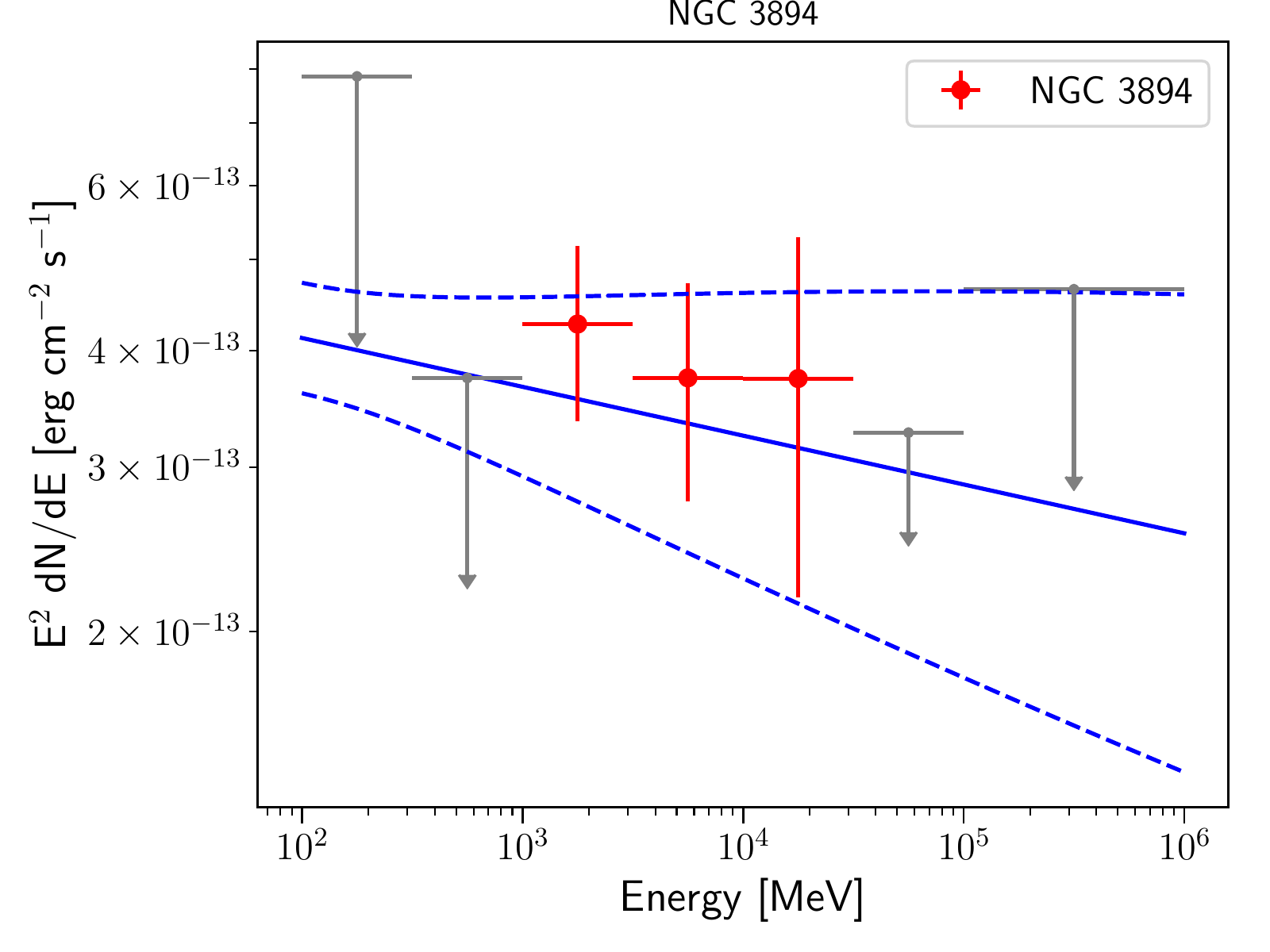}}}
\hspace{0.1cm}
\rotatebox{0}{\resizebox{!}{50mm}{\includegraphics{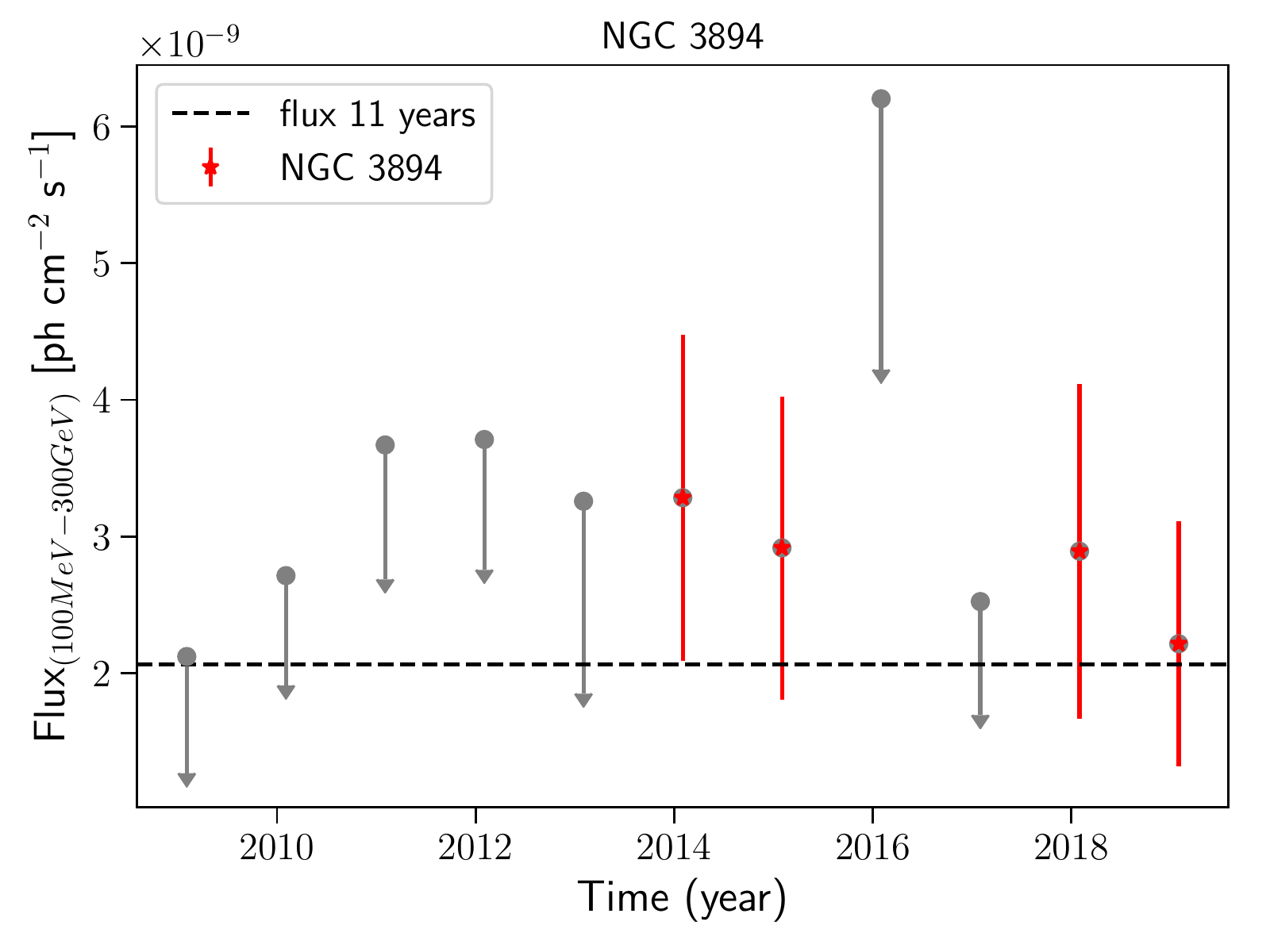}}}
\\
\rotatebox{0}{\resizebox{!}{50mm}{\includegraphics{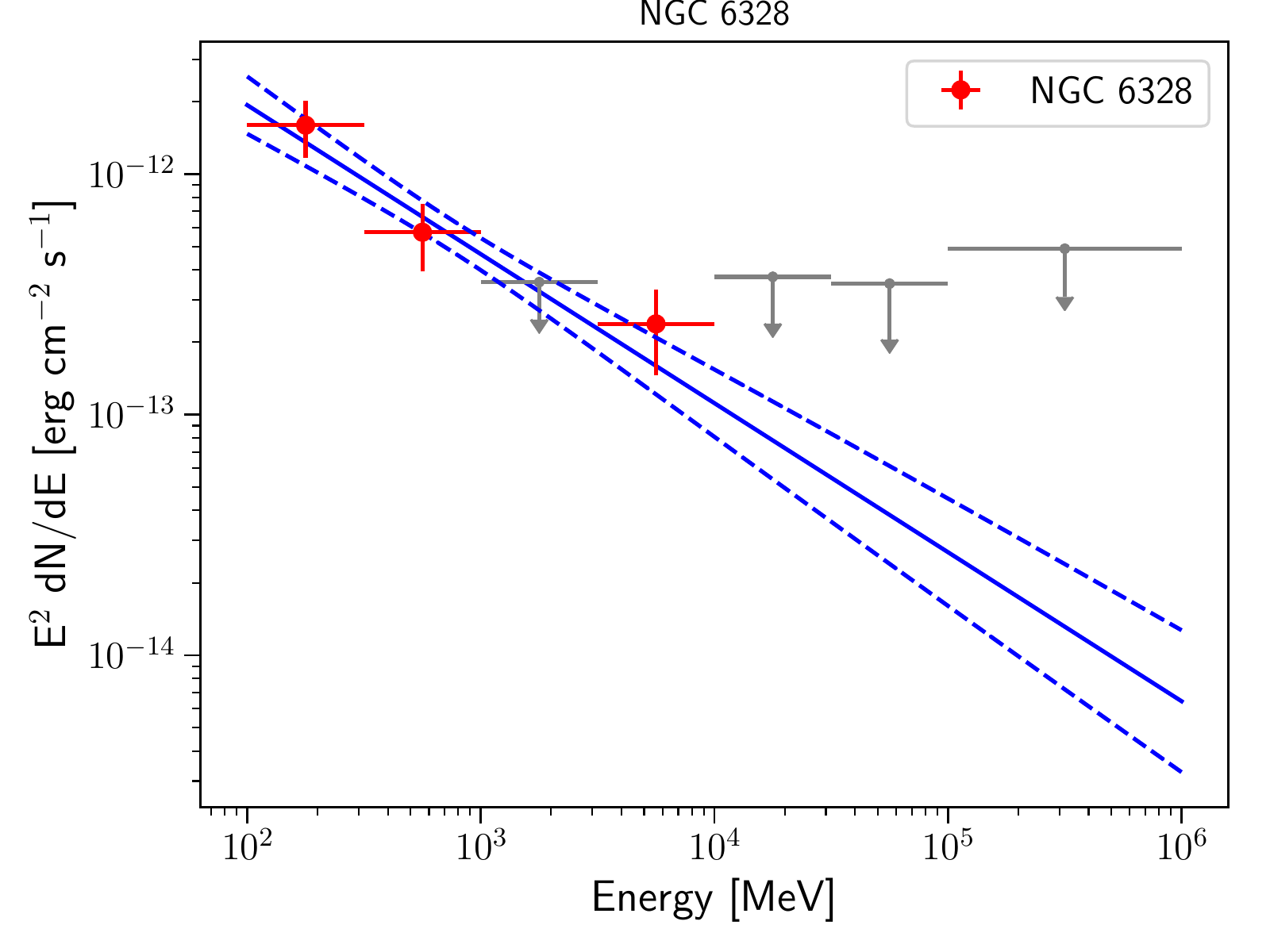}}}
\hspace{0.1cm}
\rotatebox{0}{\resizebox{!}{50mm}{\includegraphics{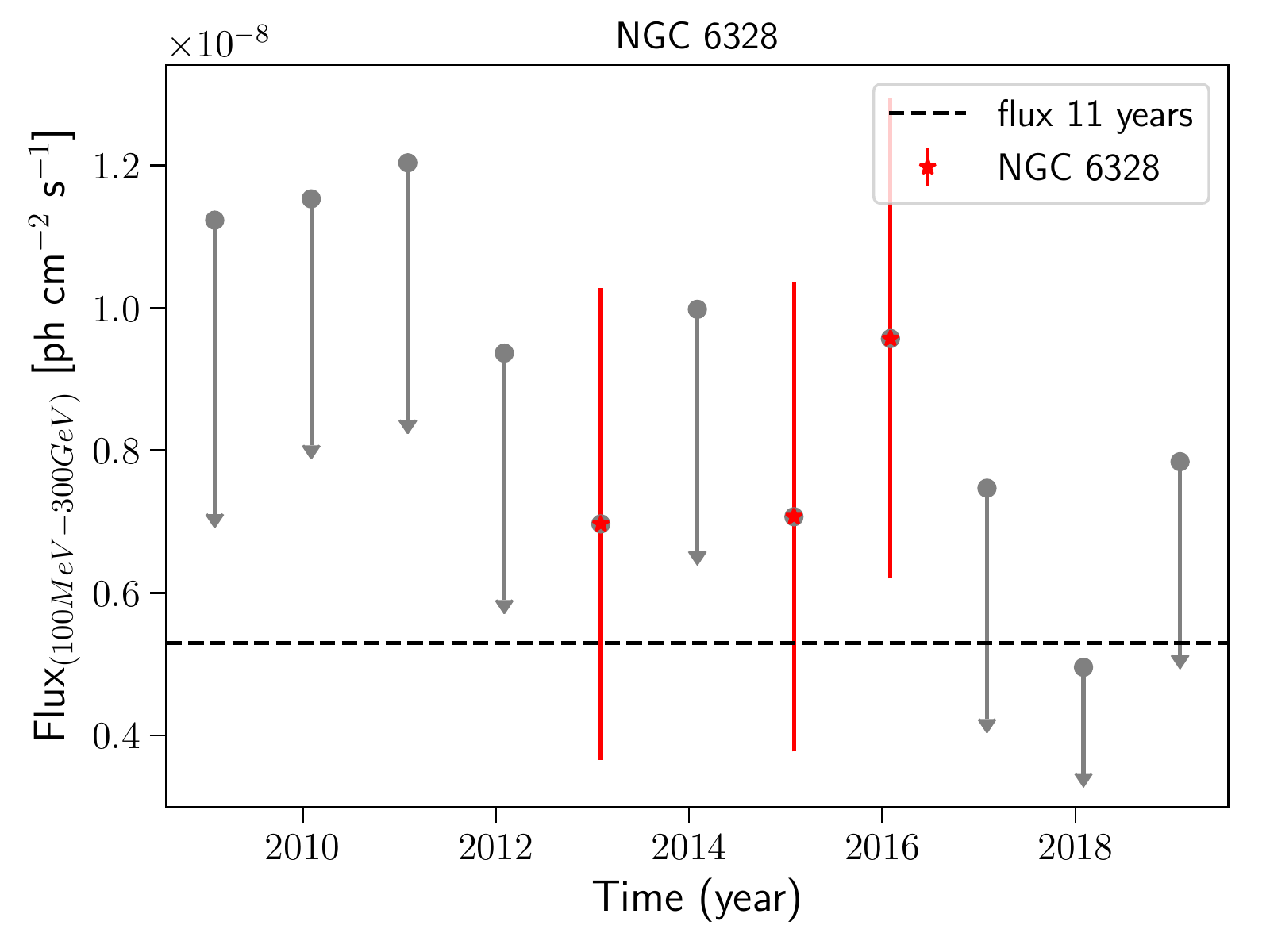}}}
\\
\rotatebox{0}{\resizebox{!}{50mm}{\includegraphics{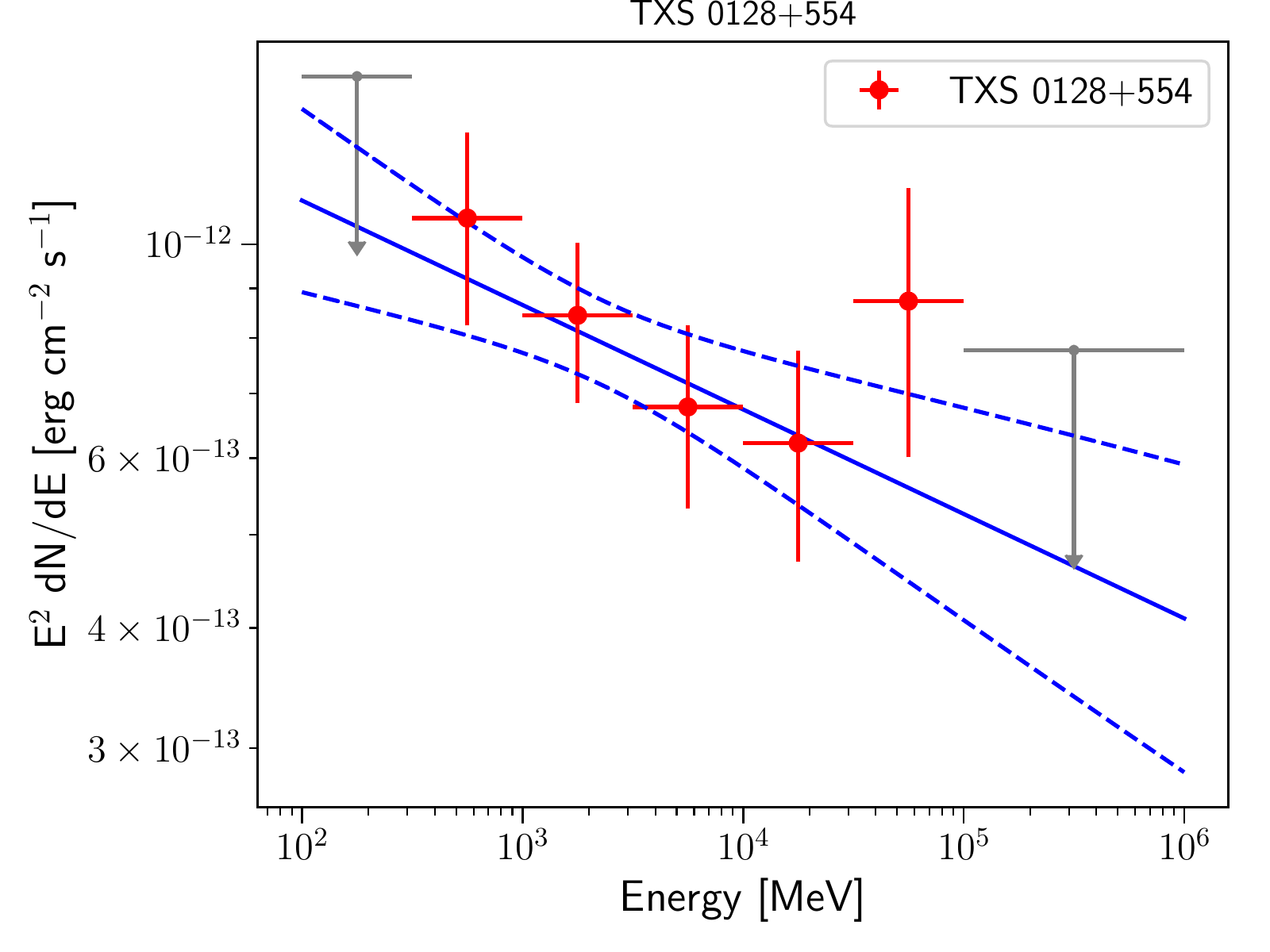}}}
\hspace{0.1cm}
\rotatebox{0}{\resizebox{!}{51mm}{\includegraphics{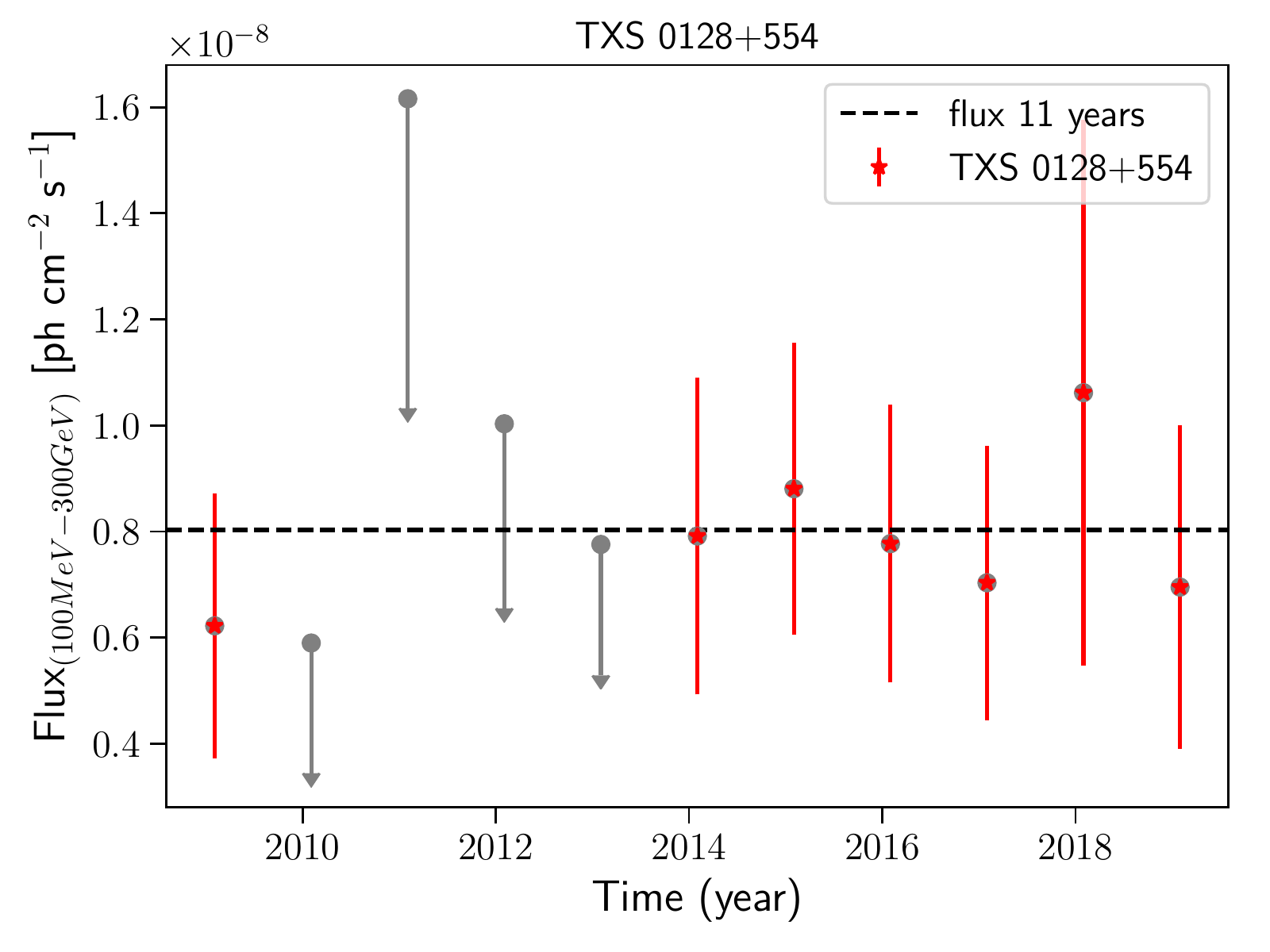}}}
\caption{\small Left panels: \textit{Fermi}-LAT flux points obtained between 100\,MeV and 1\,TeV. The arrows refer to the $1\sigma$ upper limit on the source flux. The SED has been fitted with PL (blue line). 
Right panels: \textit{Fermi}-LAT one-year binned light curve. The arrows refer to the $1\sigma$ upper limit on the source flux. The dashed line represents the averaged flux for the entire period. The flux values have been estimated for the energy range 100 MeV--300 GeV. Upper limits are computed when TS $<$ 4 for both the SED and light curve plots. The plots are for the sources: NGC\,3894 (top), NGC\,6328 (middle) and TXS\,0128+445 (bottom).}
\label{fig:sed_lightcurve1}
\end{center}
\end{figure*}

\begin{figure*}
\begin{center}
\rotatebox{0}{\resizebox{!}{52mm}{\includegraphics{./plots/PKS_1007plus142_sed_fermi.pdf}}}
\hspace{0.1cm}
\rotatebox{0}{\resizebox{!}{52mm}{\includegraphics{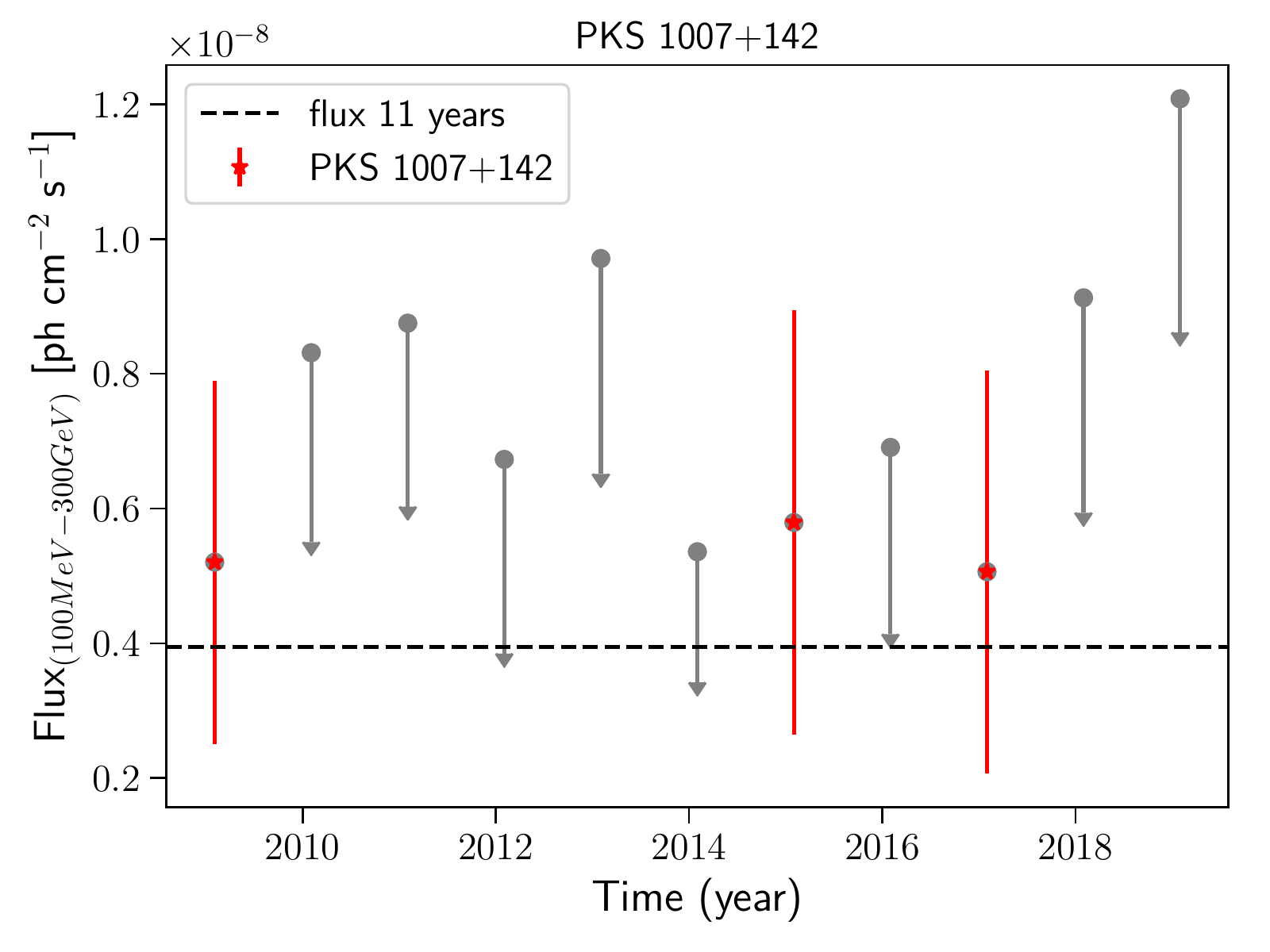}}}
\\
\rotatebox{0}{\resizebox{!}{52mm}{\includegraphics{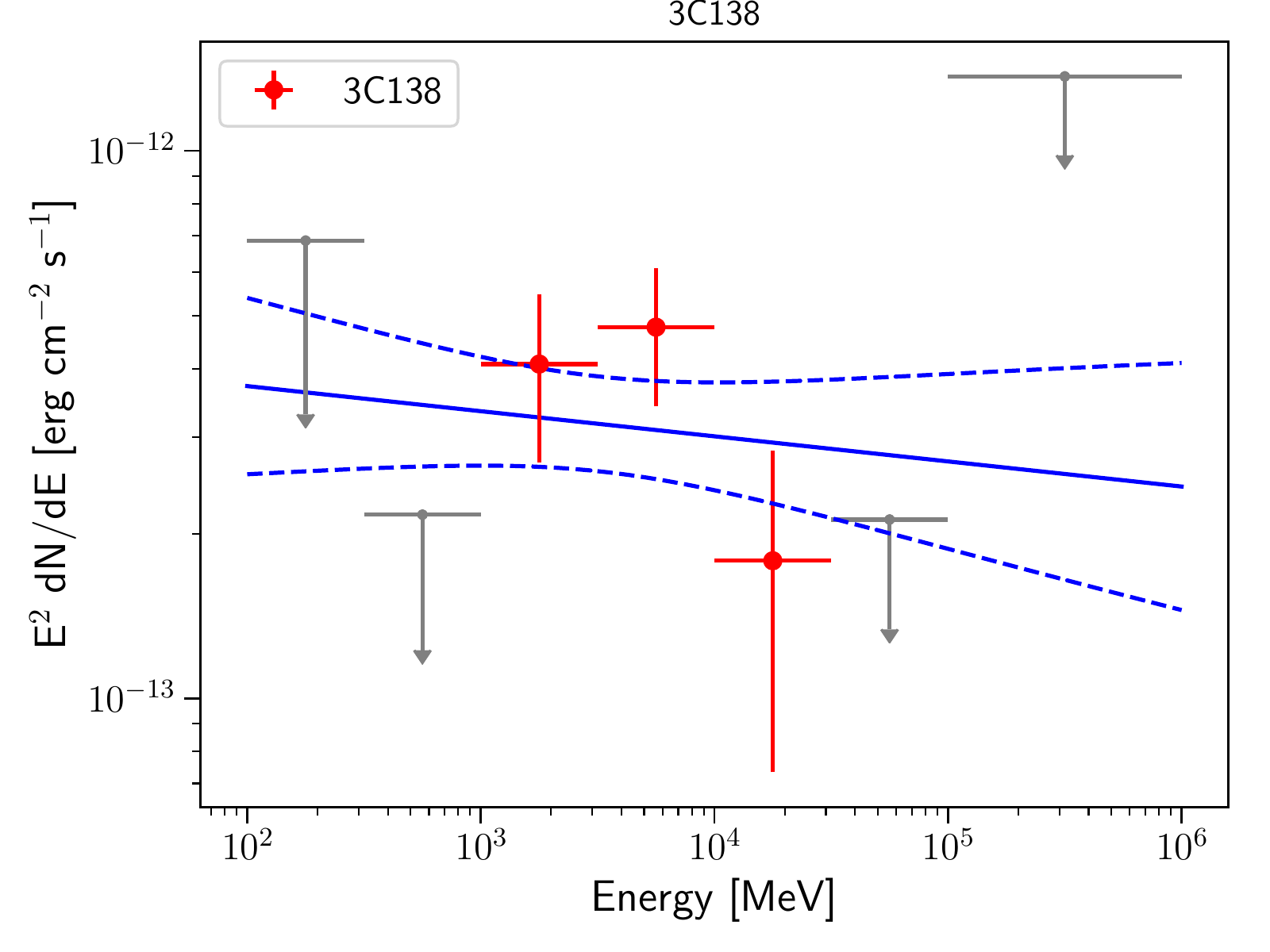}}}
\hspace{0.1cm}
\rotatebox{0}{\resizebox{!}{52mm}{\includegraphics{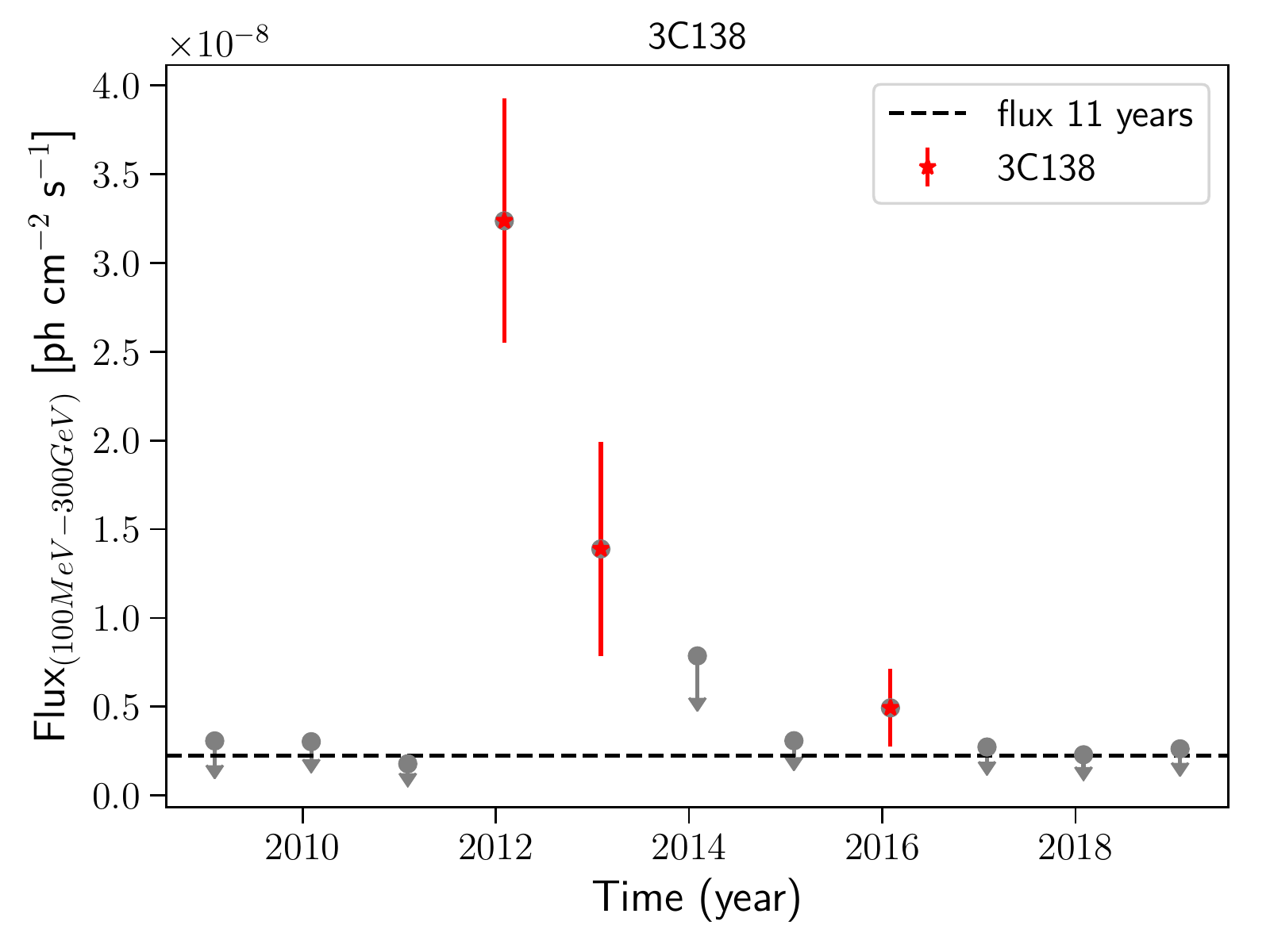}}}
\\
\rotatebox{0}{\resizebox{!}{52mm}{\includegraphics{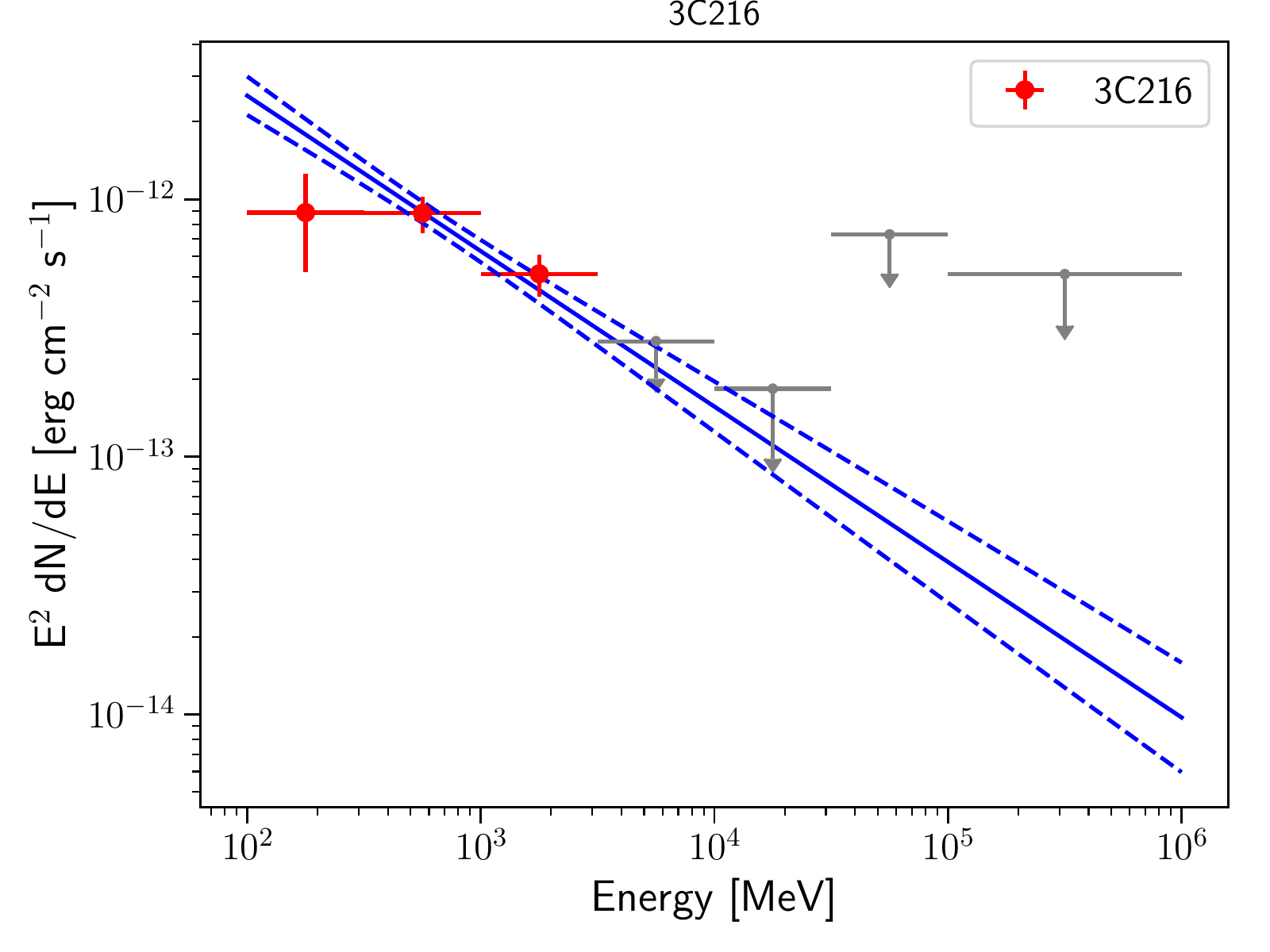}}}
\hspace{0.1cm}
\rotatebox{0}{\resizebox{!}{52mm}{\includegraphics{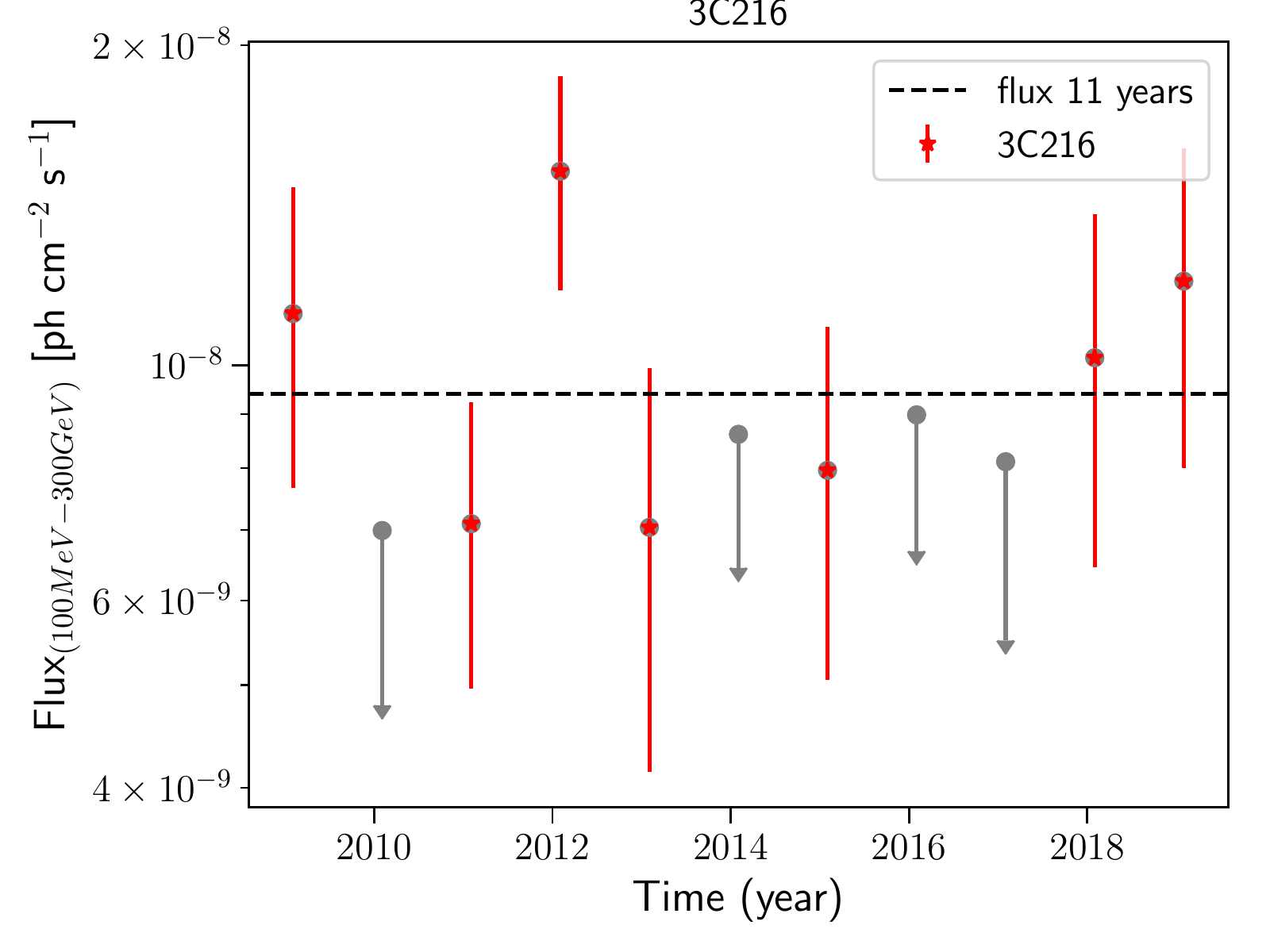}}}
\\
\rotatebox{0}{\resizebox{!}{52mm}{\includegraphics{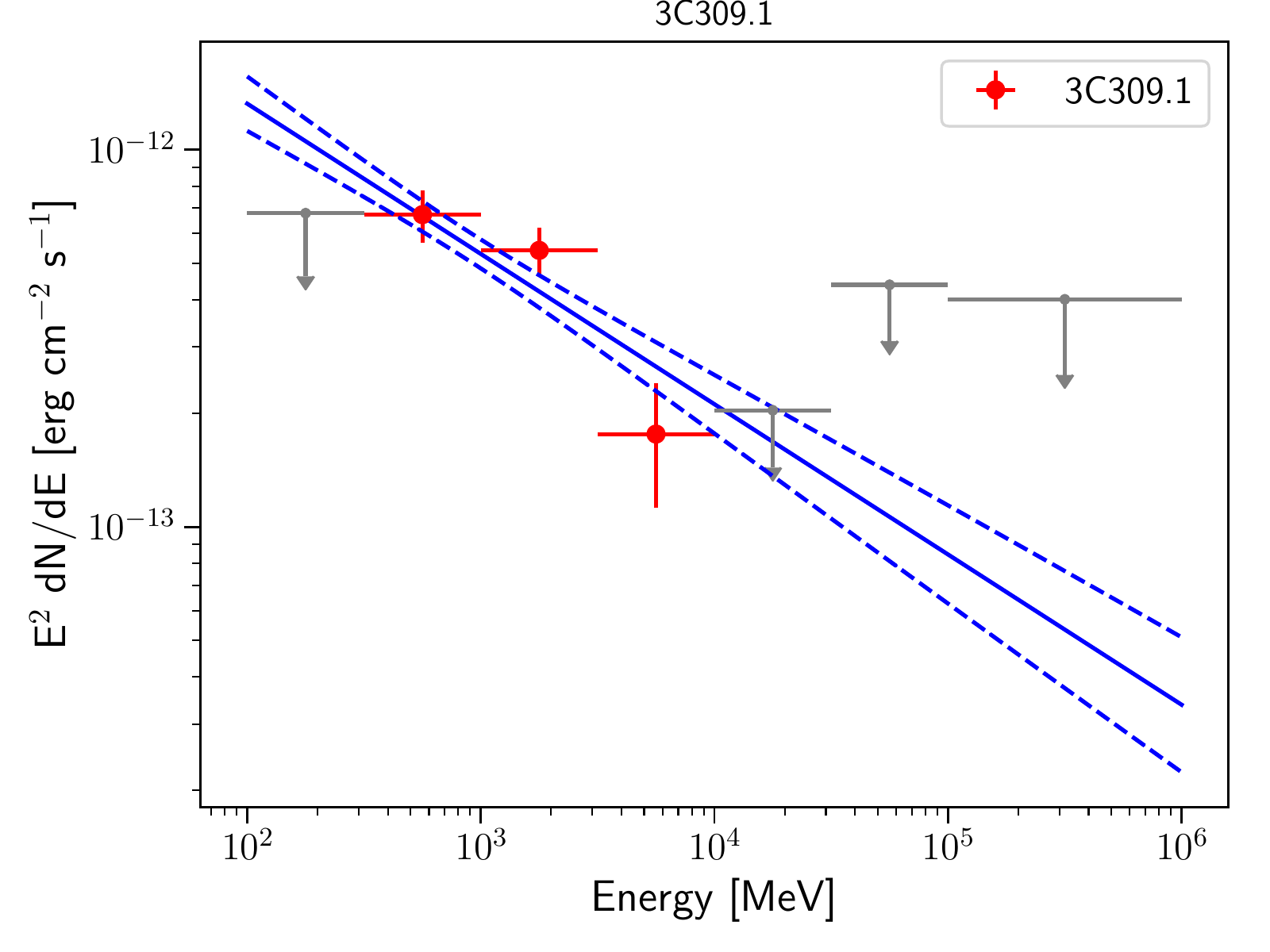}}}
\hspace{0.1cm}
\rotatebox{0}{\resizebox{!}{52mm}{\includegraphics{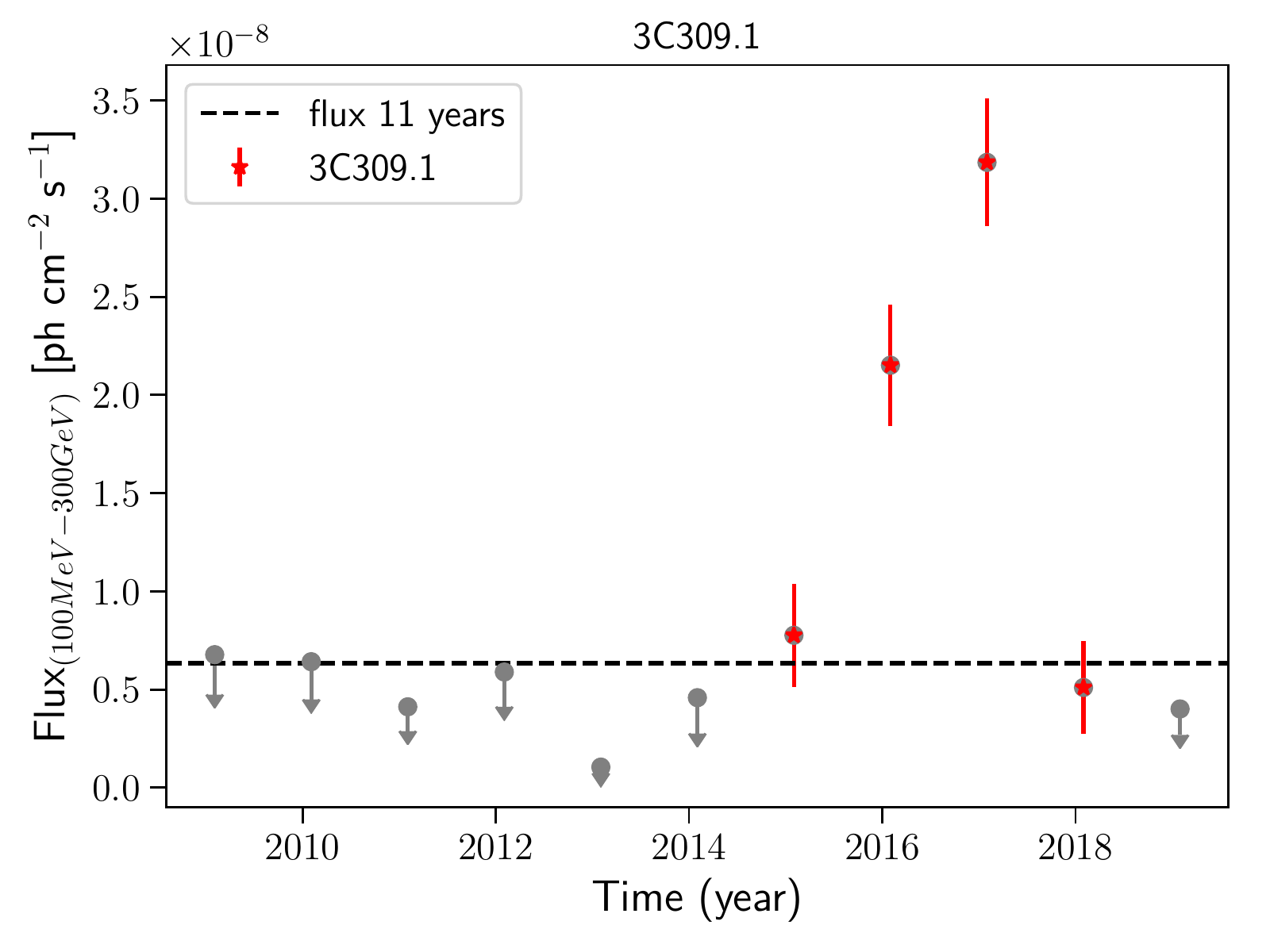}}}
\caption{\small The labels are the same
as in Fig.\ref{fig:sed_lightcurve1}.
The plots are for and the newly detected galaxy PKS\,1007+142 (first row) and the quasars: 3C\,138 (second row), 3C\,216 (third row) and 3C\,309.1 (fourth row).}
\label{fig:sed_lightcurve2}
\end{center}
\end{figure*}

\begin{figure*}
\begin{center}
\rotatebox{0}{\resizebox{!}{52mm}{\includegraphics{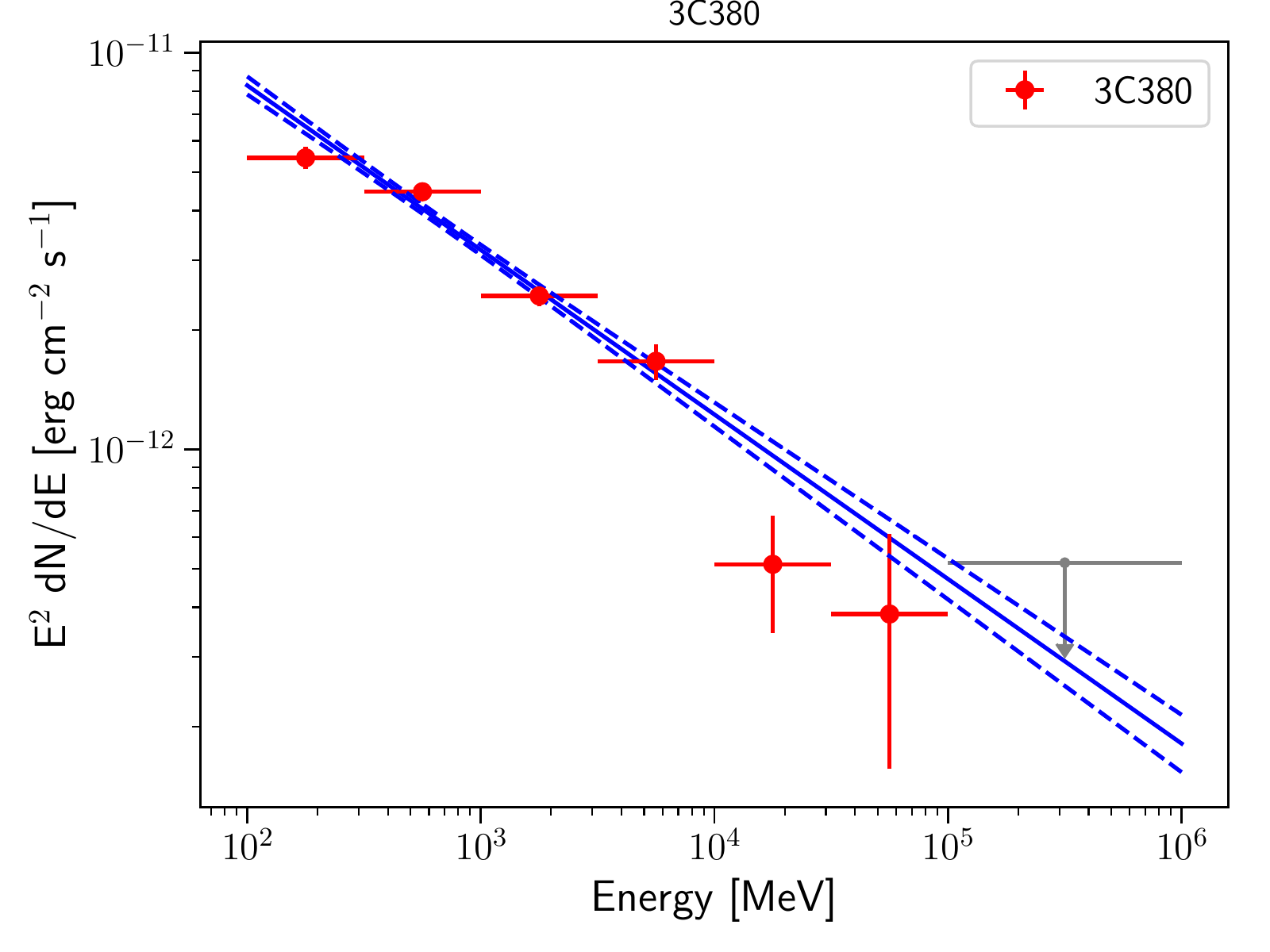}}}
\hspace{0.1cm}
\rotatebox{0}{\resizebox{!}{52mm}{\includegraphics{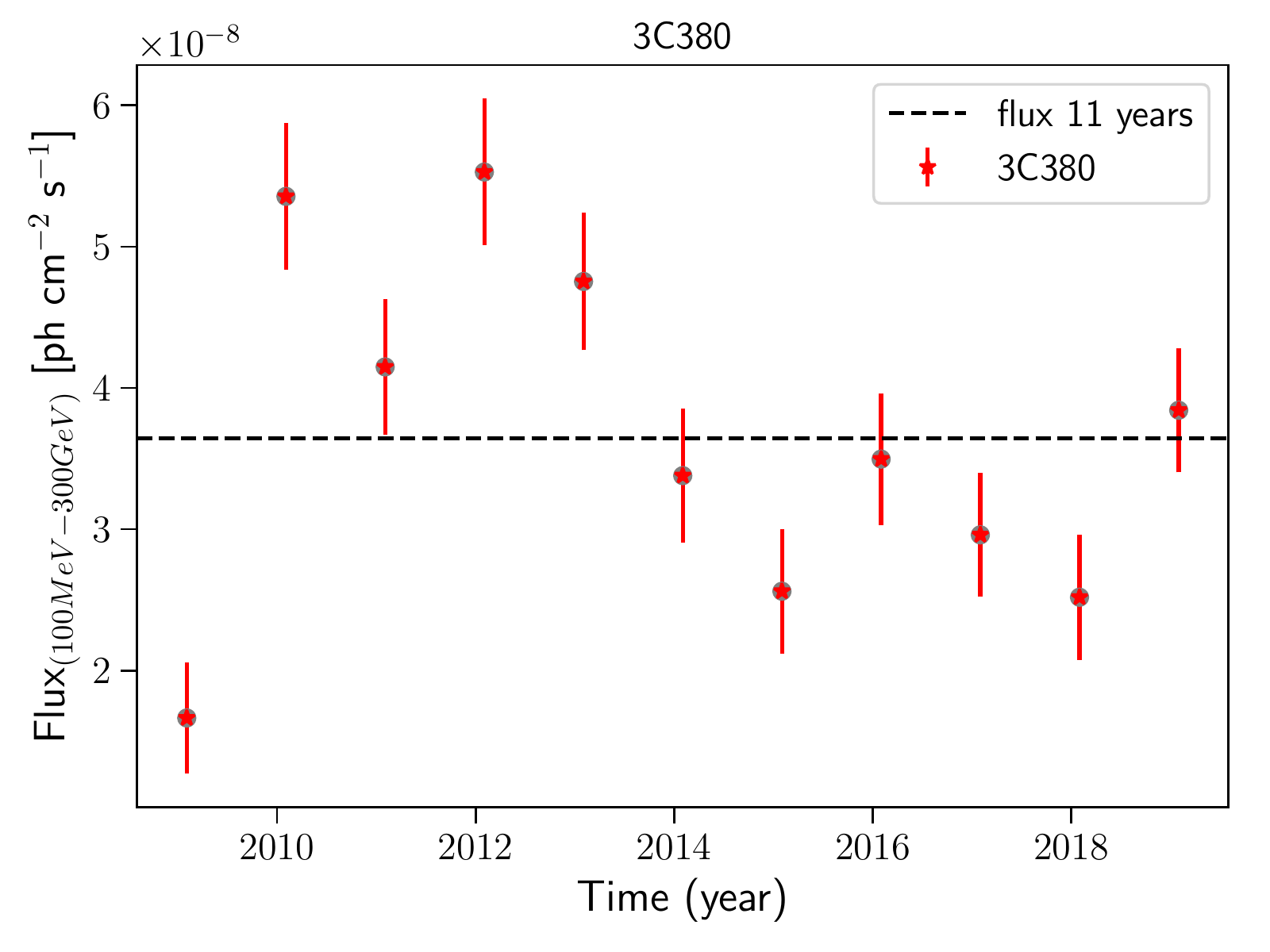}}}
\\
\rotatebox{0}{\resizebox{!}{52mm}{\includegraphics{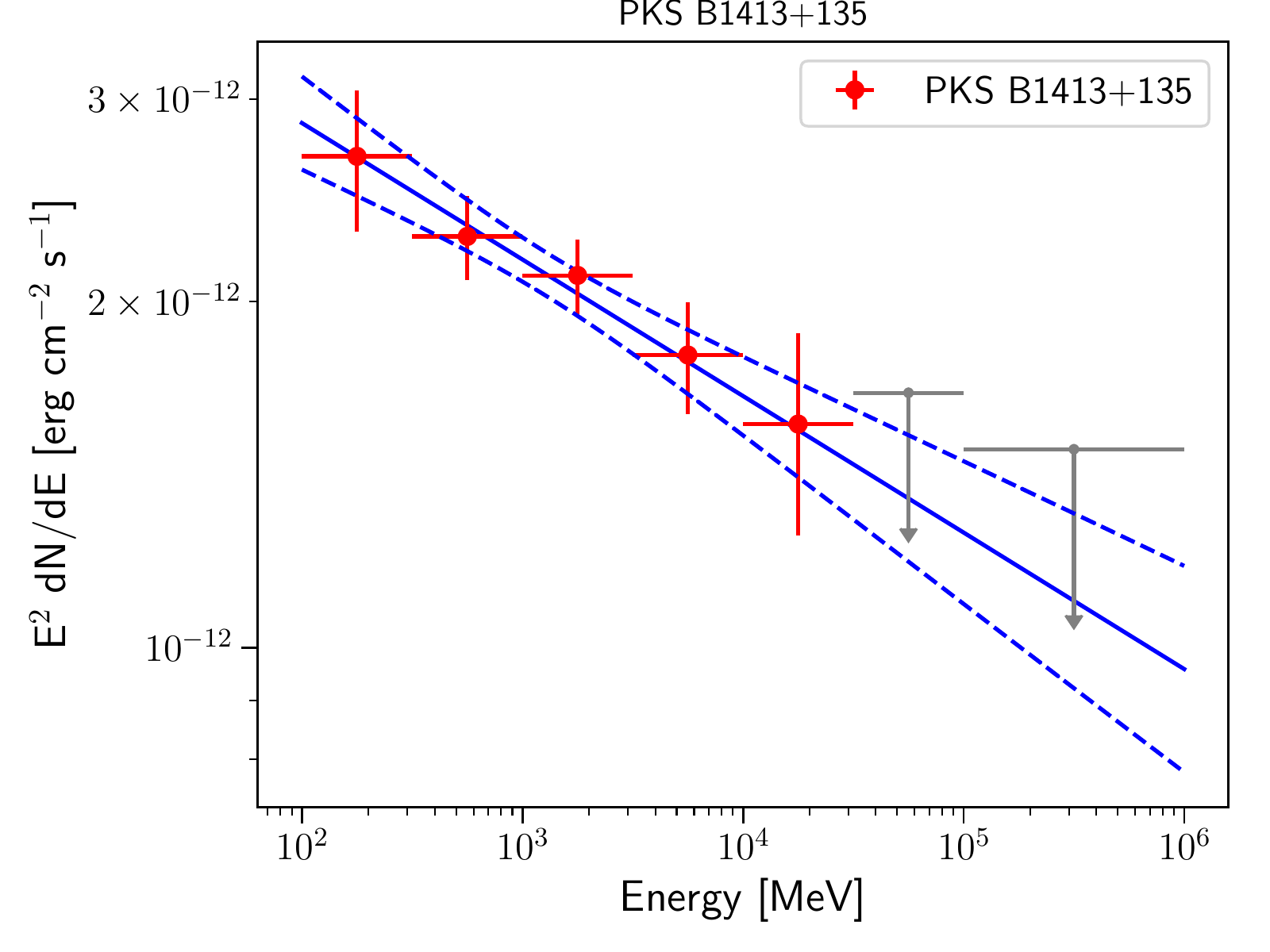}}}
\hspace{0.1cm}
\rotatebox{0}{\resizebox{!}{52mm}{\includegraphics{./plots/PKS_B1413+135_lightcurve.pdf}}}
\\
\rotatebox{0}{\resizebox{!}{52mm}{\includegraphics{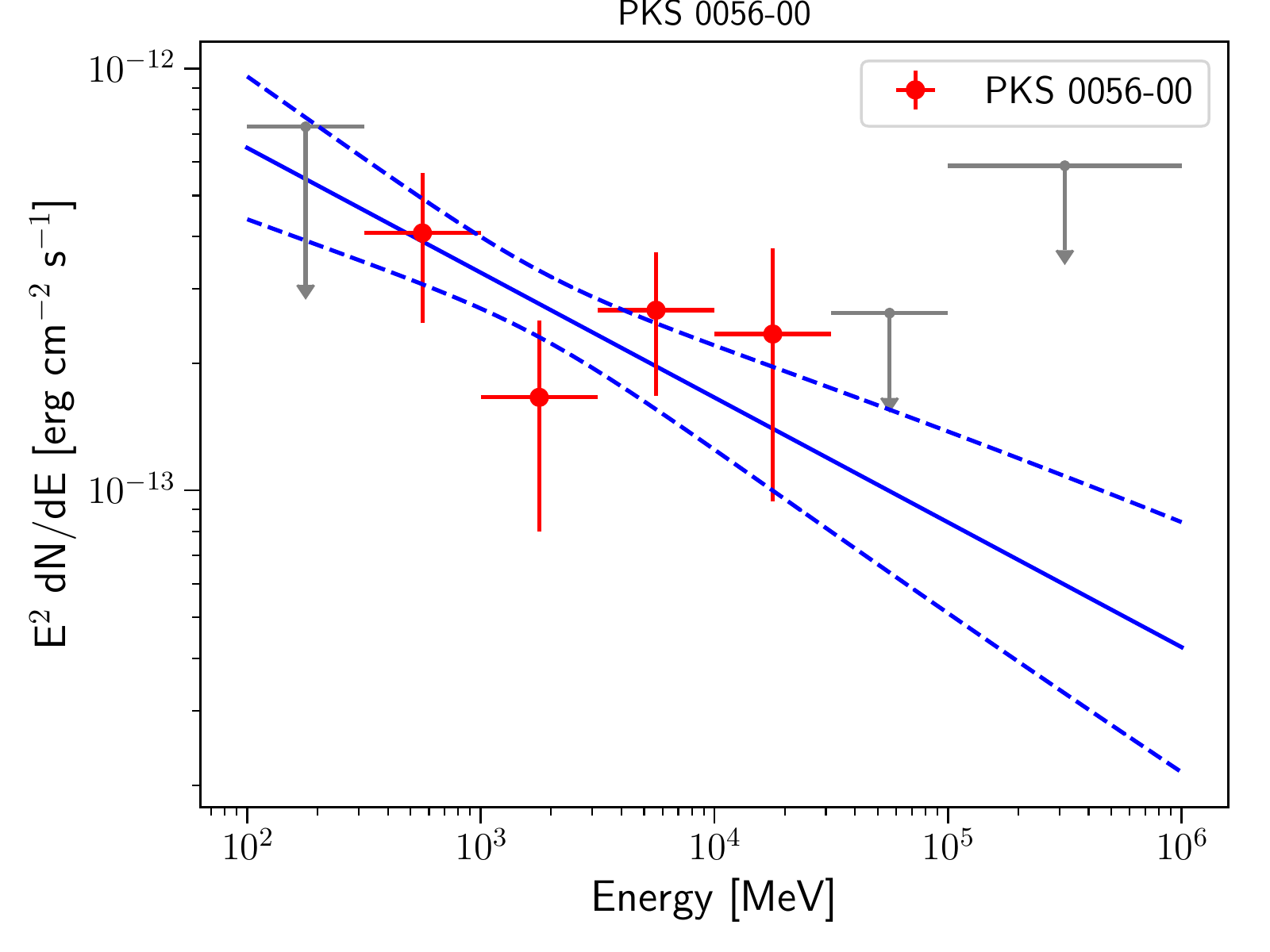}}}
\hspace{0.1cm}
\rotatebox{0}{\resizebox{!}{52mm}{\includegraphics{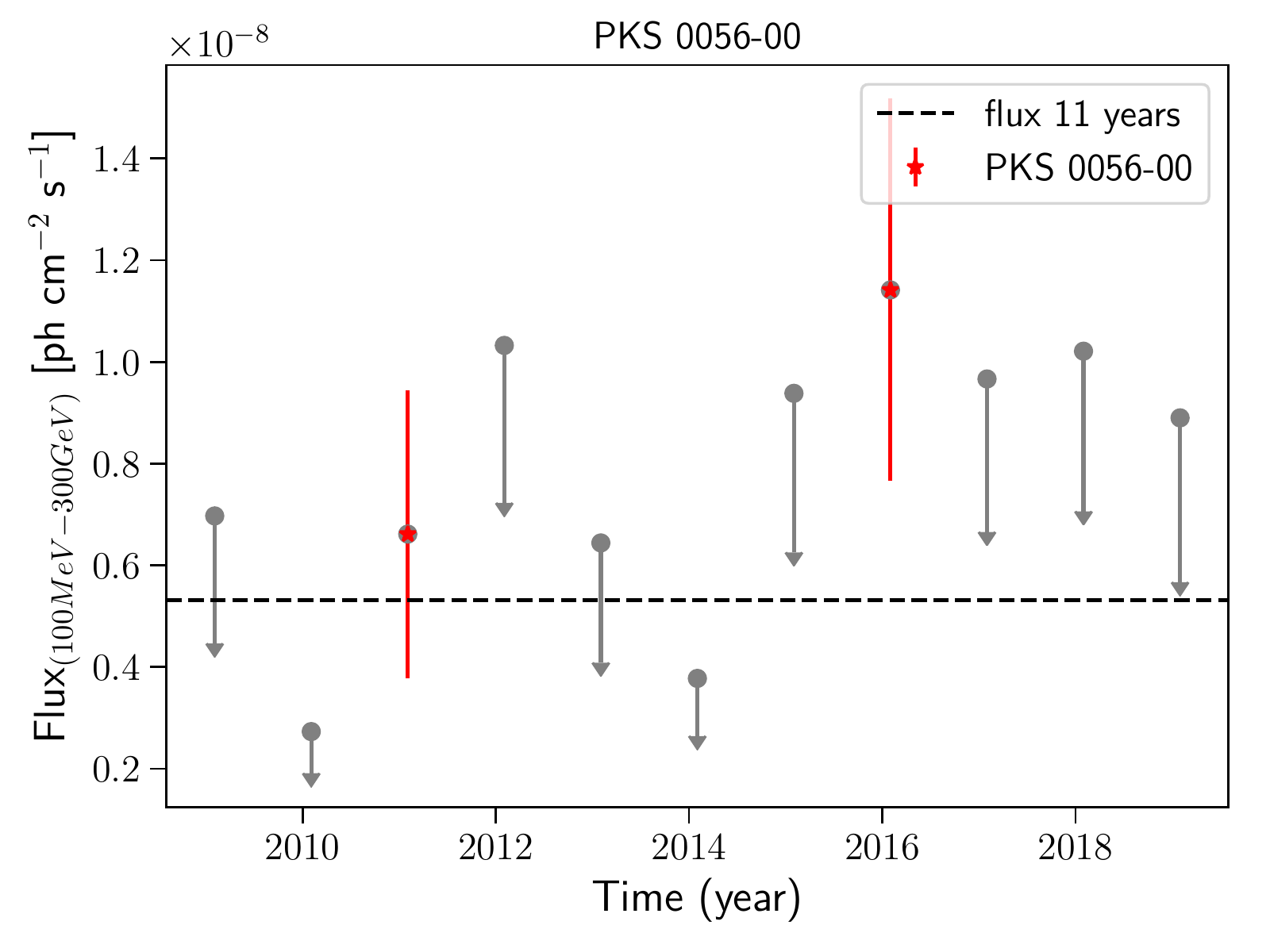}}}
\\
\rotatebox{0}{\resizebox{!}{52mm}{\includegraphics{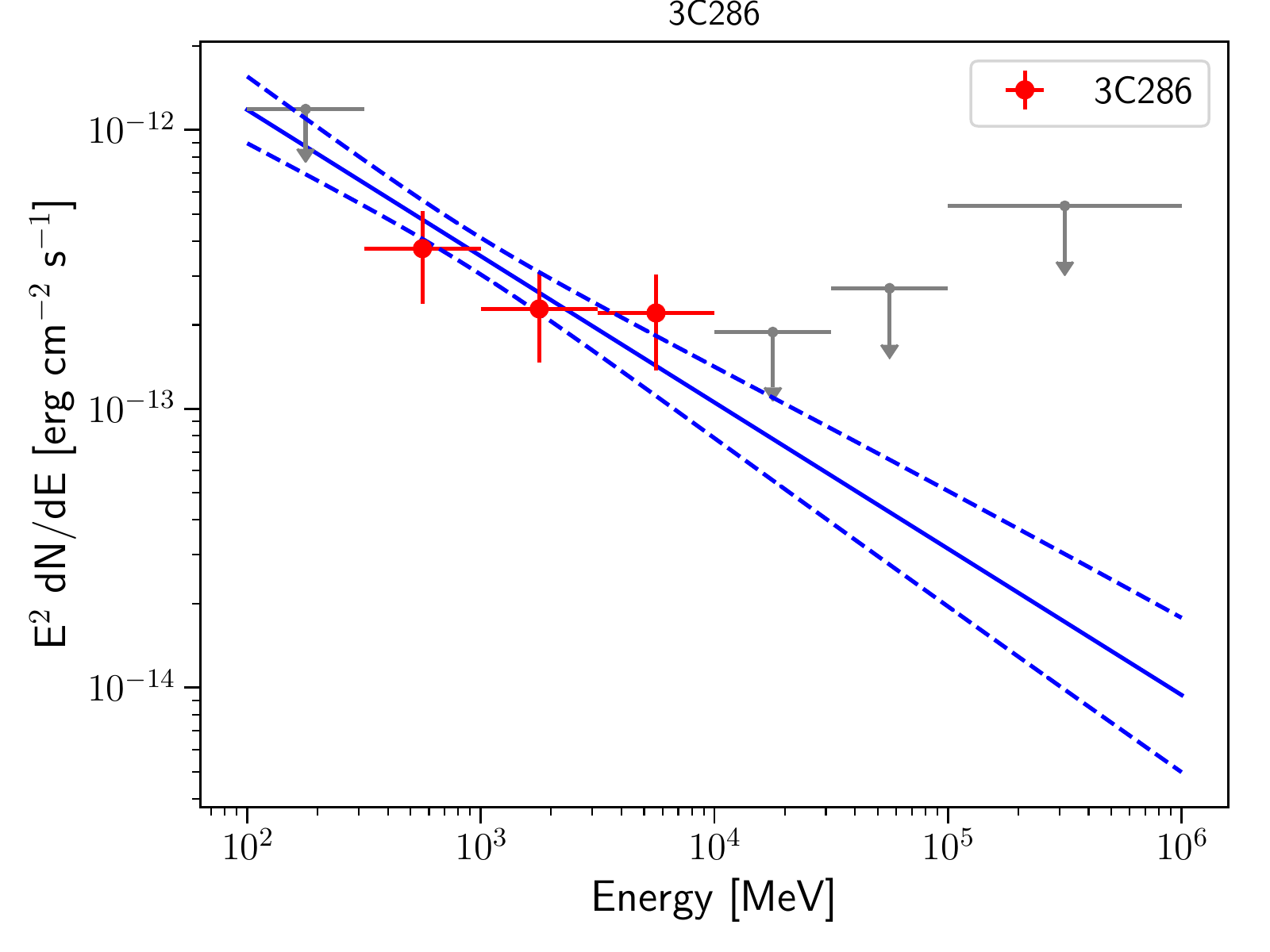}}}
\hspace{0.1cm}
\rotatebox{0}{\resizebox{!}{52mm}{\includegraphics{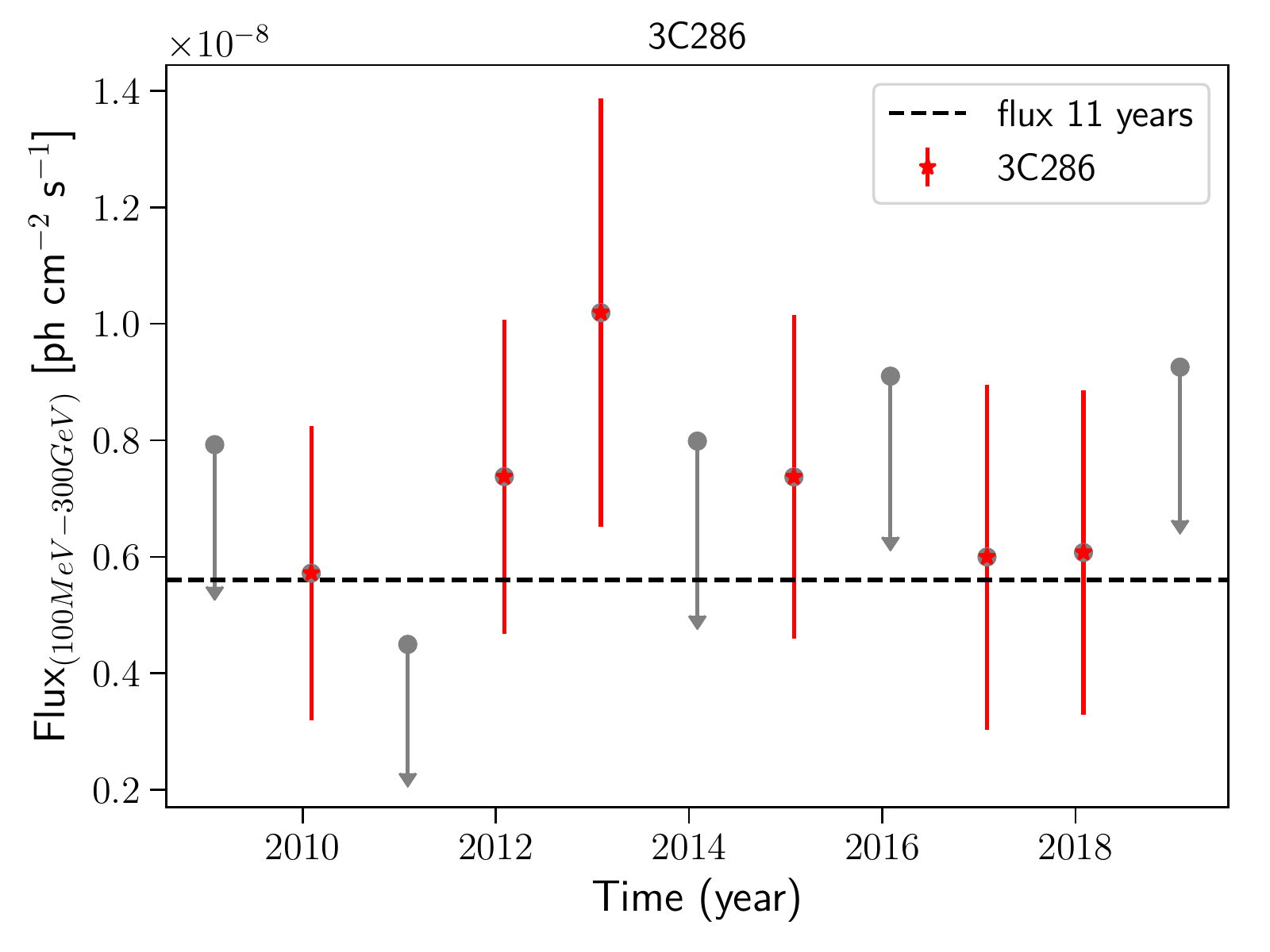}}}
\caption{\small The labels are the same
as in Fig.\ref{fig:sed_lightcurve1}.
The plots are for the quasars: 3C\,380 (first row), PKS\,B1413$+$135 (second row), PKS\,0056-00 (third row) and 3C\,286 (fourth row).}
\label{fig:sed_lightcurve3}
\end{center}
\end{figure*}

During the analysis we found a new gamma-ray source (TS = 26) in the vicinity ($\sim 0.16^{\circ}$) of the quasar B3\,1242$+$410 (see Fig. \ref{fig:map_B3_1242}), included in our sample. The LAT best-fit position of the new gamma-ray source is (R.A., Dec.(J2000)= 191.01$^{\circ} \pm \,0.04^{\circ}$, 40.75$^{\circ} \pm \,0.04^{\circ}$), with a 68\% confidence-level uncertainty R68 = 0.06$^{\circ}$. The gamma-ray source is not significantly associated with the radio object B3\,1242$+$410 (P $<$ 0.8). 
The nearest and most plausible counterpart of the new gamma-ray source is the blazar 5BZQ\,J1243$+$4043, which is located $0.03^{\circ}$ away. When we add a source to the model at the location of the blazar and rerun the likelihood analysis we obtain a TS $=1$ for the quasar B3\,1242$+$410. We report, therefore, only an upper limit for that source.

\onecolumn

\begin{table*}
\caption{\small \label{table_sample} List of all young radio galaxies contained in our sample. We report in this table name, type (galaxy/quasar), redshift, projected linear size (LS) [kpc], radio turnover frequency ($\nu_p$) [GHz], radio luminosity at 5 GHz [W Hz$^{-1}$], reference for radio information, gamma-ray significance (TS), gamma-ray flux ({0.1--1000 GeV}) in units of 10$^{-9}$ ph cm$^{-2}$ s$^{-1}$, power-law photon index ($\gamma$) and gamma-ray luminosity [10$^{44}$ erg s$^{-1}$] of each detected source.
We used a threshold of TS=10 for reporting upper limits on the gamma-ray flux and luminosity.
References:
dV09: \citet{2009A&A...498..641D},
K20: \citet{2020ApJ...897..164K},
L20: \citet{2020MNRAS.491...92L},
Li20: \citet{2020ApJ...899..141L},
O04: \citet{2004A&A...426..463O}
O08: \citet{2008A&A...487..885O},
O14: \citet{2014MNRAS.438..463O}, 
P20: \citet{2020A&A...635A.185P},
R06: \citep{2006A&A...449...49R},
W20: \citet{2020ApJ...892..116W},
Z20: \citet{2020ApJ...899....2Z}.
}
\small
\hspace*{-0.8cm}
\centering
\begin{tabular}{c|cccccc|cccc}
\hline \hline
Name & type & z & LS & $\nu_p$ & log L$_{5\, \textrm{GHz}}$& Ref. & TS & Flux$_{\gamma}$ & $\Gamma$ & Lum$_{\gamma}$\\
 & & & kpc & GHz & W Hz$^{-1}$ & & & 10$^{-9}$ cm$^{-2}$ s$^{-1}$ & & 10$^{44}$ erg s$^{-1}$\\
\hline
OQ\,208 & G & 0.076 & 0.007 & 4.0 & 25.59 & K20 & 0 & <0.27 & 2.00 & <0.050\\
J0111$+$3906 & G & 0.688 & 0.056 & 7.0 & 27.36 & O14 & 0 & <0.25 & 2.00 & <6.8\\
J1335$+$5844 & G & 0.58 & 0.105 & 8.8 & 26.94 & O14 & 0 & <0.11 & 2.00 & <1.9\\
J1735$+$5844 & G & 0.835 & 0.061 & 8.2 & 27.43 & O14 & 0 & <0.24 & 2.00 & <10.5\\
J1511$+$0518 & G & 0.084 & 0.01 & 10.9 & 24.97 & O14 & 0 & <0.20 & 2.00 & <0.045\\
J0428$+$3259 & G & 0.479 & 0.016 & 5.9 & 26.59 & O14 & 0 & <0.22 & 2.00 & <2.5\\
J0951$+$3451 & G & 0.29 & 0.021 & 7.7 & 25.43 & O14 & 0 & <0.14 & 2.00 & <0.49\\
0316$+$161 & G & 0.907 & 4.384 & 1.5 & 27.93 & O14 & 0 & <0.10 & 2.00 & <5.0\\
0404$+$768 & G & 0.598 & 0.866 & 0.55 & 27.53 & O14 & 12 & 2.70 $\pm$ 0.81 & 2.61 $\pm$ 0.29 & 22.0\\
0428$+$205 & G & 0.219 & 0.351 & 1.4 & 26.42 & O14 & 0 & <1.10 & 2.00 & <2.0\\
1323$+$321 & G & 0.369 & 0.305 & 0.68 & 27.07 & O14 & 19 & 1.36 $\pm$ 0.40 & 2.15 $\pm$ 0.23 & 4.8\\
1358$+$624 & G & 0.431 & 0.28 & 0.79 & 27.0 & O14 & 0 & <0.26 & 2.00 & <2.3\\
0710$+$439 & G & 0.518 & 0.136 & 2.88 & 27.22 & O14 & 0 & <0.14 & 2.00 & <1.9\\
2352$+$495 & G & 0.2379 & 0.187 & 0.87 & 26.37 & O14 & 0 & <0.18 & 2.00 & <0.39\\
1943$+$546 & G & 0.263 & 0.181 & 0.75 & 26.27 & O14 & 0 & <0.49 & 2.00 & <1.4\\
B1819$+$6707 & G & 0.22 & 0.112 & 0.976 & 25.28 & O14 & 0 & <0.24 & 2.00 & <0.44\\
B1946$+$7048 & G & 0.1 & 0.087 & 1.98 & 25.18 & O14 & 0 & <0.42 & 2.00 & <0.14\\
B3\,0039$+$391 & G & 1.006 & 2.738 & <0.2 & 26.66 & O14 & 1 & <0.48 & 2.00 & <32.7\\
B3\,0120$+$405 & G & 0.84 & 18.345 & <0.18 & 26.68 & O14 & 0 & <0.27 & 2.00 & <11.7\\
B3\,0213$+$412 & G & 0.515 & 12.358 & 0.2 & 26.3 & O14 & 0 & <0.41 & 2.00 & <5.5\\
B3\,0744$+$464 & G & 2.926 & 11.04 & <0.39 & 27.85 & O14 & 0 & <0.15 & 2.00 & <145.2\\
B3\,0814$+$441 & G & 0.12 & 8.556 & <0.112 & 24.41 & O14 & 0 & <0.21 & 2.00 & <0.10\\
B3\,0935$+$428A & G & 1.291 & 10.969 & <0.46 & 26.97 & O14 & 0 & <0.33 & 2.00 & <42.3\\
B3\,0955$+$390 & G & 0.52 & 29.813 & <0.15 & 26.15 & O14 & 0 & <0.29 & 2.00 & <3.9\\
B3\,1025$+$390B & G & 0.361 & 16.022 & <0.14 & 26.18 & O14 & 1 & <0.41 & 2.00 & <2.4\\
B3\,1027$+$392 & G & 0.56 & 10.323 & 0.23 & 26.24 & O14 & 0 & <0.27 & 2.00 & <4.4\\
B3\,1157$+$460 & G & 0.7428 & 5.852 & 0.44 & 26.89 & O14 & 0 & <0.18 & 2.00 & <6.0\\
B3\,1201$+$394 & G & 0.4448 & 11.951 & <0.14 & 26.08 & O14 & 0 & <0.13 & 2.00 & <1.2\\
B3\,1458$+$433 & G & 0.927 & 12.608 & <0.19 & 26.72 & O14 & 0 & <0.15 & 2.00 & <8.5\\
B3\,2358$+$406 & G & 0.978 & 0.799 & 0.59 & 27.32 & O14 & 0 & <0.30 & 2.00 & <19.1\\
3C\,49 & G & 0.621 & 6.781 & 0.194 & 27.16 & O14 & 1 & <0.55 & 2.00 & <11.5\\
3C\,237 & G & 0.877 & 9.301 & 0.094 & 27.88 & O14 & 0 & <0.30 & 2.00 & <14.8\\
3C\,241 & G & 1.617 & 10.268 & 0.104 & 27.65 & O14 & 0 & <0.13 & 2.00 & <29.8\\
3C\,346 & G & 0.162 & 22.056 & <0.045 & 25.99 & O14 & 13 & 1.23 $\pm$ 0.43 & 2.07 $\pm$ 0.20 & 0.82\\
TXS\,0128$+$554 & G & 0.0365 & 0.012 & 0.66 & 23.695 & Li20 & 178 & 8.03 $\pm$ 1.46 & 2.20 $\pm$ 0.07 & 0.19\\
B3\,1049$+$384 & G & 1.018 & 0.14 & 0 & 25.82 & O04 & 0 & <0.30 & 2.00 & <21.2\\
1345$+$125 & G & 0.122 & 0.152 & 0.67 & 26.06 & O14 & 0 & <0.22 & 2.00 & <0.11\\
B3\,0034$+$444 & G & 2.79 & 11.9 & 0 & 27.9 & R06 & 0 & <0.21 & 2.00 & <179.9\\
0035$+$227 & G & 0.096 & 0.022 & 0 & 24.716 & W20 & 0 & <0.10 & 2.00 & <0.030\\
0116$+$319 & G & 0.0602 & 0.12 & 0.55 & 25.09 & W20 & 3 & <0.74 & 2.00 & <0.083\\
1031$+$567 & G & 0.46 & 0.109 & 1.3 & 26.97 & W20 & 0 & <0.24 & 2.00 & <2.5\\

1245$+$676 & G & 0.107 & 0.01 & 1.42 & 24.716 & W20 & 0 & <0.25 & 2.00 & <0.094\\
1607$+$26 & G & 0.473 & 0.24 & 1.0 & 27.16 & W20 & 0 & <0.37 & 2.00 & <4.0\\

1843$+$356 & G & 0.763 & 0.022 & 2 & 27.32 & W20 & 11 & 0.59 $\pm$ 0.24 & 1.93 $\pm$ 0.24 & 22.6\\

1934-638 & G & 0.183 & 0.085 & 1.4 & 26.75 & W20 & 0 & <0.14 & 2.00 & <0.17\\
2021$+$614 & G & 0.227 & 0.016 & 5 & 26.63 & W20 & 0 & <0.13 & 2.00 & <0.26\\

NGC\,6328 & G & 0.014 & 0.002 & 4 & 24.28 & W20 & 41 & 5.30 $\pm$ 1.86 & 2.54 $\pm$ 0.17 & 0.011\\
NGC\,3894 & G & 0.0108 & 0.01 & 5 & 23.21 & P20 & 95 & 2.03 $\pm$ 0.48 & 2.05 $\pm$ 0.09 & 0.006\\
0402$+$379 & G & 0.0545 & 0.0073 & 10 & 24.88 & K20 & 6 & <3.90 & 2.00 & <0.19\\
\hline \\
\end{tabular}
\end{table*}

\begin{table*}
\caption{\small \label{table_sample2} Continued from Table \ref{table_sample}. $^*$PKS\,1007$+$142 is also known as J100955.51$+$140154.2.}
\small
\hspace*{-0.8cm}
\centering
\begin{tabular}{c|cccccc|cccc}
\hline \hline
Name & type & z & LS & $\nu_p$ & log L$_{5\, \textrm{GHz}}$& Ref. & TS & Flux$_{\gamma}$ & $\Gamma$ & Lum$_{\gamma}$\\
 & & & [kpc] & [GHz] & [W Hz$^{-1}$] & & & [10$^{-9}$ ph\,cm$^{-2}$\,s$^{-1}$] & & [10$^{44}$ erg s$^{-1}$]\\
\hline

J073328$+$560541 & G & 0.104 & 0.09 & 0.46 & 24.68 & dV09 & 1 & <0.36 & 2.00 & <0.13\\
J073934$+$495438 & G & 0.054 & 0.002 & 0.95 & 23.63 & dV09 & 8 & <1.69 & 2.00 & <0.051\\
J083139$+$460800 & G & 0.127 & 0.02 & 2.2 & 24.62 & dV09 & 0 & <0.27 & 2.00 & <0.15\\
J083637$+$440109 & G & 0.054 & 1.7 & <0.15 & 23.66 & dV09 & 0 & <0.35 & 2.00 & <0.031\\
J090615$+$463618 & G & 0.085 & 0.049 & 0.68 & 24.49 & dV09 & 0 & <0.12 & 2.00 & <0.027\\
J102618$+$454229 & G & 0.153 & 0.045 & 0.18 & 24.55 & dV09 & 0 & <0.16 & 2.00 & <0.13\\
J103719$+$433515 & G & 0.023 & 0.009 & <0.15 & 22.96 & dV09 & 0 & <0.20 & 2.00 & <0.003\\
J115000$+$552821 & G & 0.139 & 0.1 & <0.23 & 24.57 & dV09 & 2 & <0.49 & 2.00 & <0.32\\
J120902$+$411559 & G & 0.095 & 0.035 & 0.37 & 24.26 & dV09 & 0 & <0.52 & 2.00 & <0.15\\
J131739$+$411545 & G & 0.066 & 0.005 & 2.3 & 24.37 & dV09 & 5 & <5.49 & 2.00 & <0.098\\

J140051$+$521606 & G & 0.116 & 0.32 & <0.15 & 24.36 & dV09 & 17 & 0.12 $\pm$ 0.05 & 1.64 $\pm$ 0.32 & 0.20\\

J140942$+$360416 & G & 0.148 & 0.07 & 0.33 & 24.45 & dV09 & 0 & <0.32 & 2.00 & <0.25\\
J143521$+$505122 & G & 0.099 & 0.27 & <0.15 & 24.2 & dV09 & 0 & <0.16 & 2.00 & <0.052\\
J150805$+$342323 & G & 0.045 & 0.15 & <0.23 & 23.35 & dV09 & 0 & <0.11 & 2.00 & <0.007\\
J160246$+$524358 & G & 0.106 & 0.35 & <0.15 & 24.75 & dV09 & 2 & <0.52 & 2.00 & <0.19\\
J161148$+$404020 & G & 0.152 & 3.4 & <0.15 & 25.03 & dV09 & 2 & <0.43 & 2.00 & <0.35\\
J171854$+$544148 & G & 0.147 & 0.175 & 0.48 & 24.86 & dV09 & 0 & <0.14 & 2.00 & <0.10\\
J093609$+$331308 & G & 0.076 & 0.002 & 2.2 & 23.84 & dV09 & 0 & <0.08 & 2.00 & <0.015\\
J101636$+$563926 & G & 0.232 & 0.89 & <0.15 & 24.91 & dV09 & 0 & <0.11 & 2.00 & <0.23\\
J105731$+$405646 & G & 0.008 & 0.0001 & 1.25 & 21.59 & dV09 & 1 & <0.41 & 2.00 & <0.001\\
J115727$+$431806 & G & 0.229 & 2.3 & <0.15 & 25.25 & dV09 & 0 & <1.44 & 2.00 & <1.2\\
J132513$+$395552 & G & 0.074 & 0.014 & 1.9 & 23.69 & dV09 & 2 & <0.53 & 2.00 & <0.091\\
J134035$+$444817 & G & 0.065 & 0.0041 & 2.3 & 23.89 & dV09 & 1 & <0.44 & 2.00 & <0.058\\
J155927$+$533054 & G & 0.178 & 4.5 & <1.5 & 24.67 & dV09 & 0 & <0.34 & 2.00 & <0.39\\
J002225.42$+$001456.1 & G & 0.306 & 0.27 & 0.8 & 26.52 & L20 & 1 & <0.57 & 2.00 & <2.2\\
J002833.42$+$005510.9 & G & 0.104 & 3.35 & <0.4 & 24.35 & L20 & 0 & <0.52 & 2.00 & <0.19\\
J002914.24$+$345632.2 & G & 0.517 & 0.2 & 0.8 & 27.13 & L20 & 2 & <0.74 & 2.00 & <10.0\\
J075756.71$+$395936.1 & G & 0.066 & 0.25 & <0.4 & 23.5 & L20 & 0 & <0.31 & 2.00 & <0.042\\
J081323.75$+$073405.6 & G & 0.112 & 2.81 & <0.4 & 24.71 & L20 & 1 & <0.59 & 2.00 & <0.25\\
J082504.56$+$315957.0 & G & 0.265 & 7.75 & <0.4 & 24.82 & L20 & 1 & <0.18 & 2.00 & <0.50\\
J083411.09$+$580321.4 & G & 0.093 & 0.0086 & <0.4 & 24.13 & L20 & 15 & 3.53 $\pm$ 0.96 & 2.66 $\pm$ 0.20 & 0.30\\
J083825.00$+$371036.5 & G & 0.396 & 1.02 & <0.4 & 25.92 & L20 & 0 & <0.28 & 2.00 & <2.0\\
J085408.44$+$021316.1 & G & 0.459 & 3.43 & <0.4 & 25.34 & L20 & 0 & <0.20 & 2.00 & <2.0\\
J090105.25$+$290146.9 & G & 0.194 & 19.33 & 0.06 & 25.96 & L20 & 1 & <0.50 & 2.00 & <0.70\\
J092405.30$+$141021.4 & G & 0.136 & 0.74 & <0.4 & 24.18 & L20 & 13 & 2.15 $\pm$ 0.69 & 2.33 $\pm$ 0.24 & 0.584\\
J093430.68$+$030545.3 & G & 0.225 & 1.62 & <0.4 & 25.25 & L20 & 0 & <0.15 & 2.00 & <0.29\\
PKS 1007$+$142$^{*}$ & G & 0.213 & 3.29 & <0.4 & 25.71 & L20 & 31 & 4.64 $\pm$ 1.55 & 2.56 $\pm$ 0.18 & 2.8\\ 
J101251.77$+$403903.4 & G & 0.506 & 32.8 & <0.4 & 27.01 & L20 & 1 & <0.44 & 2.00 & <5.6\\
J101714.23$+$390121.1 & G & 0.211 & 20.97 & <0.4 & 25.8 & L20 & 0 & <0.26 & 2.00 & <0.43\\
J104029.94$+$295757.7 & G & 0.091 & 3.67 & <0.4 & 24.34 & L20 & 1 & <0.49 & 2.00 & <0.13\\
114339.59$+$462120.4 & G & 0.116 & 17.06 & <0.4 & 24.66 & L20 & 1 & <0.57 & 2.00 & <0.26\\
J115919.97$+$464545.1 & G & 0.467 & 5.0 & <0.4 & 26.4 & L20 & 0 & <0.36 & 2.00 & <3.7\\
131057.00$+$445146.2 & G & 0.391 & 3.87 & <0.4 & 25.26 & L20 & 1 & <0.22 & 2.00 & <1.5\\
J132419.67$+$041907.0 & G & 0.263 & 17.04 & <0.4 & 24.99 & L20 & 0 & <0.29 & 2.00 & <0.80\\
J140416.35$+$411748.7 & G & 0.36 & 1.01 & <0.4 & 25.51 & L20 & 0 & <0.12 & 2.00 & <0.71\\
J141327.22$+$550529.2 & G & 0.282 & 0.81 & <0.4 & 24.92 & L20 & 0 & <0.25 & 2.00 & <0.81\\
J142104.24$+$050844.7 & G & 0.445 & 1.68 & <0.4 & 25.91 & L20 & 0 & <0.21 & 2.00 & <1.9\\
J144712.76$+$404744.9 & G & 0.195 & 26.28 & <0.4 & 25.23 & L20 & 0 & <0.26 & 2.00 & <0.37\\
J152349.34$+$321350.2 & G & 0.11 & 0.4 & <0.4 & 24.2 & L20 & 0 & <0.13 & 2.00 & <0.052\\
J154609.52$+$002624.6 & G & 0.558 & 0.04 & 0.6 & 27.02 & L20 & 0 & <0.27 & 2.00 & <4.3\\
J155235.38$+$441905.9 & G & 0.452 & 6.93 & <0.4 & 25.56 & L20 & 17 & 0.78 $\pm$ 0.26 & 2.07 $\pm$ 0.19 & 6.0\\
J160335.16$+$380642.8 & G & 0.241 & 6.09 & <0.4 & 27.03 & L20 & 0 & <0.15 & 2.00 & <0.35\\
J161823.57$+$363201.7 & G & 0.733 & 0.44 & <0.4 & 26.82 & L20 & 5 & <0.69 & 2.00 & <21.8\\
J165822.18$+$390625.6 & G & 0.425 & 0.97 & <0.4 & 27.1 & L20 & 2 & <0.67 & 2.00 & <5.6\\

\hline
\end{tabular}
\end{table*}

\begin{table*}
\caption{\small \label{table_sample3} List of all young radio quasars contained in our sample. For parameters description and references see Table \ref{table_sample}. $^{*}$PKS\,0056-00 and $^{**}$PKS\,B1413$+$135, are also known as J005905.51$+$000651.6 and J141558.81$+$132023.7, respectively.}
\small
\hspace*{-0.8cm}
\centering
\begin{tabular}{c|cccccc|cccc}
\hline \hline
Name & type & z & LS & $\nu_p$ & log L$_{5\, \textrm{GHz}}$& Ref. & TS & Flux$_{\gamma}$ & $\Gamma$ & Lum$_{\gamma}$\\
 & & & [kpc] & [GHz] & [W Hz$^{-1}$] & & & [10$^{-9}$ ph cm$^{-2}$ s$^{-1}$] & & [10$^{44}$ erg s$^{-1}$]\\
\hline

J0650$+$6001 & Q & 0.455 & 0.04 & 8.0 & 26.93 & O14 & 3 & <0.55 & 2.00 & <5.4\\
1225$+$368 & Q & 1.973 & 0.509 & 5.2 & 28.17 & 014 & 0 & <0.36 & 2.00 & <129\\
B3\,0137$+$401 & Q & 1.62 & 35.939 & <0.26 & 27.09 & O14 & 1 & <0.34 & 2.00 & <76.2\\
B3\,0701$+$392 & Q & 1.283 & 15.167 & 0.34 & 27.13 & O14 & 0 & <0.26 & 2.00 & <32.7\\
B3\,1242$+$410 & Q & 0.813 & 0.34 & 0.65 & 27.35 & O14 & 1 & <0.97 & 2.00 & <24.9\\
B3\,2311$+$469 & Q & 0.745 & 11.575 & <0.17 & 27.18 & O14 & 2 & <0.68 & 2.00 & <22.3\\
B3\,2349$+$410 & Q & 2.046 & 10.146 & <0.3 & 27.44 & O14 & 0 & <0.18 & 2.00 & <70.0\\
3C\,43 & Q & 1.459 & 22.17 & <0.05 & 28.09 & O14 & 0 & <0.32 & 2.00 & <54.1\\
3C\,48 & Q & 0.367 & 6.579 & 0.109 & 27.38 & O14 & 0 & <0.12 & 2.00 & <0.73\\
3C\,119 & Q & 1.023 & 1.617 & 0.303 & 28.23 & O14 & 0 & <0.14 & 2.00 & <9.9\\

3C\,138 & Q & 0.759 & 5.9 & 0.176 & 27.97 & O14 & 34 & 2.09 $\pm$ 0.89 & 2.05 $\pm$ 0.12 & 64.2\\
3C\,147 & Q & 0.545 & 4.454 & 0.231 & 27.92 & O14 & 22 & 6.89 $\pm$ 1.51 & 2.69 $\pm$ 0.16 & 47.1\\

3C\,186 & Q & 1.067 & 17.959 & 0.082 & 27.26 & O14 & 0 & <0.28 & 2.00 & <22.1\\
3C\,190 & Q & 1.1944 & 33.356 & 0.088 & 27.86 & O14 & 2 & <0.56 & 2.00 & <58.6\\
3C\,216 & Q & 0.6702 & 56.12 & 0.066 & 27.23 & O14 & 153 & 7.78 $\pm$ 0.98 & 2.60 $\pm$ 0.09 & 97.4\\
3C\,286 & Q & 0.85 & 24.51 & <0.05 & 28.41 & L20 & 67 & 5.60 $\pm$ 1.10 & 2.52 $\pm$ 0.12 & 111.3\\ 
3C\,309.1 & Q & 0.905 & 17.215 & <0.076 & 28.08 & O14 & 207 & 6.33 $\pm$ 0.74 & 2.47 $\pm$ 0.07 & 178.5\\
3C\,380 & Q & 0.692 & 10.8 & <0.3 & 27.68 & Z20 & 2274 & 36.43 $\pm$ 1.48 & 2.41 $\pm$ 0.02 & 504.5\\
PKS\,0056-00$^{*}$ & Q & 0.719 & 15.099 & <0.14 & 27.5 & L20 & 52 & 5.31 $\pm$ 1.48 & 2.30 $\pm$ 0.15 & 64.6\\ 
J080413.88$+$470442.8 & Q & 0.51 & 6.18 & <0.4 & 26.52 & L20 & 0 & <0.14 & 2.00 & <1.8\\
J080442.23$+$301237.0 & Q & 1.45 & 10.98 & <0.4 & 27.72 & L20 & 0 & <0.26 & 2.00 & <43.3\\
J080447.96$+$101523.7 & Q & 1.968 & 31.02 & <0.02 & 28.12 & L20 & 0 & <0.12 & 2.00 & <43.3\\
J081253.10$+$401859.9 & Q & 0.551 & 7.72 & <0.4 & 26.73 & L20 & 0 & <0.18 & 2.00 & <2.9\\
J084856.57$+$013647.8 & Q & 0.35 & 6.16 & <0.4 & 25.15 & L20 & 0 & <0.33 & 2.00 & <1.7\\
J085601.22$+$285835.4 & Q & 1.084 & 6.53 & <0.4 & 27.09 & L20 & 0 & <0.15 & 2.00 & <12.1\\
J091734.79$+$501638.1 & Q & 0.632 & 4.87 & <0.4 & 25.82 & L20 & 1 & <0.47 & 2.00 & <10.2\\
J092608.00$+$074526.6 & Q & 0.442 & 7.98 & <0.4 & 25.6 & L20 & 0 & <0.32 & 2.00 & <2.9\\
J094525.90$+$352103.6 & Q & 0.208 & 4.45 & <0.4 & 24.72 & L20 & 0 & <0.16 & 2.00 & <0.27\\
J095412.57$+$420109.1 & Q & 1.787 & 16.89 & <0.4 & 27.56 & L20 & 0 & <0.21 & 2.00 & <58.6\\
J105628.25$+$501952.0 & Q & 0.82 & 8.18 & <0.4 & 26.01 & L20 & 0 & <0.61 & 2.00 & <25.4\\
J112027.80$+$142055.0 & Q & 0.363 & 0.4 & 1.0 & 26.65 & L20 & 0 & <0.29 & 2.00 & <1.7\\
J113138.89$+$451451.1 & Q & 0.398 & 4.88 & <0.4 & 26.59 & L20 & 0 & <0.30 & 2.00 & <2.1\\
J114311.02$+$053516.0 & Q & 0.497 & 17.03 & <0.4 & 25.76 & L20 & 0 & <0.32 & 2.00 & <3.9\\
J114856.56$+$525425.2 & Q & 1.632 & 0.0068 & 8.7 & 27.84 & L20 & 1 & <1.35 & 2.00 & <201.0\\
J115618.74$+$312804.7 & Q & 0.417 & 4.96 & 0.1 & 26.8 & L20 & 0 & <0.31 & 2.00 & <2.5\\
J120321.93$+$041419.0 & Q & 1.224 & 0.6 & 0.4 & 27.74 & L20 & 1 & <0.81 & 2.00 & <90.2\\

J120624.70$+$641336.8 & Q & 0.372 & 6.98 & <0.08 & 26.72 & L20 & 0 & <0.63 & 2.00 & <3.9\\

J125325.72$+$303635.1 & Q & 1.314 & 4.62 & <0.4 & 27.31 & L20 & 0 & <0.35 & 2.00 & <46.7\\
J130941.51$+$404757.2 & Q & 2.907 & 0.01 & 2.0 & 27.97 & L20 & 0 & <0.20 & 2.00 & <182.9\\
J131718.64$+$392528.1 & Q & 1.563 & 0.29 & <0.4 & 27.55 & L20 & 0 & <0.19 & 2.00 & <38.9\\
J133037.69$+$250910.9 & Q & 1.055 & 0.39 & <0.04 & 28.28 & L20 & 0 & <0.36 & 2.00 & <27.5\\
J134536.94$+$382312.5 & Q & 1.852 & 0.93 & <0.4 & 27.98 & L20 & 0 & <0.40 & 2.00 & <125.0\\
J140028.65$+$621038.5 & Q & 0.429 & 0.39 & 0.6 & 27.08 & L20 & 0 & <0.25 & 2.00 & <2.2\\
J140319.30$+$350813.3 & Q & 2.291 & 10.21 & <0.4 & 26.12 & L20 & 3 & <0.51 & 2.00 & <263.1\\
J141414.83$+$455448.7 & Q & 0.458 & 0.17 & 1.4 & 26.22 & L20 & 0 & <0.40 & 2.11 & <3.1\\

PKS\,B1413$+$135$^{**}$ & Q & 0.247 & 0.03 & 10 & 26.19 & L20 & 1193 & 14.72 $\pm$ 1.02 & 2.10 $\pm$ 0.03 & 27.8\\ 
J144516.46$+$095836.0 & Q & 3.541 & 0.15 & 0.9 & 29.14 & L20 & 0 & <0.12 & 2.00 & <179.4\\
J150426.69$+$285430.5 & Q & 2.285 & 0.35 & <0.4 & 28.02 & L20 & 0 & <0.44 & 2.00 & <227.0\\
J152005.47$+$201605.4 & Q & 1.572 & 8.9 & <0.02 & 28.11 & L20 & 4 & <0.65 & 2.00 & <133.7\\
J153409.90$+$301204.0 & Q & 0.929 & 35.86 & <0.4 & 26.03 & L20 & 0 & <0.21 & 2.00 & <12.1\\
J154349.50$+$385601.3 & Q & 0.553 & 8.1 & <0.4 & 25.77 & L20 & 0 & <0.31 & 2.00 & <5.0\\
J154525.48$+$462244.3 & Q & 0.525 & 6.59 & <0.4 & 25.79 & L20 & 0 & <0.16 & 2.00 & <2.2\\
J162111.27$+$374604.9 & Q & 1.271 & 6.28 & <0.4 & 27.27 & L20 & 1 & <0.35 & 2.00 & <42.8\\
J163402.95$+$390000.5 & Q & 1.083 & 6.53 & <0.4 & 27.38 & L20 & 0 & <0.27 & 2.00 & <22.2\\
J164311.34$+$315618.4 & Q & 0.587 & 10.58 & <0.4 & 25.76 & L20 & 2 & <0.43 & 2.00 & <7.8\\
J213638.58$+$004154.2 & Q & 1.941 & 0.02 & 5.2 & 29.36 & L20 & 0 & <0.02 & 2.00 & <6.2\\
J225025.34$+$141952.0 & Q & 0.235 & 0.75 & <0.178 & 26.3 & L20 & 1 & <0.64 & 2.00 & <1.4\\
\hline
\end{tabular}
\end{table*}

\label{lastpage}
\end{document}